\documentclass[a4paper,fleqn,usenatbib]{mnras}

\pdfoutput=1

\usepackage{newtxtext,newtxmath}

\usepackage[T1]{fontenc}
\usepackage{ae,aecompl}

\usepackage{graphicx}	
\usepackage{amsmath}	
\usepackage{amssymb}	

\usepackage{cleveref}
\usepackage{subfig}
\usepackage{tabularx}

\newcommand{\K}{\mathrm{K}}
\newcommand{\sy}{\Msun {\rm yr}^{-1}}
\newcommand{\Gyr}{\mathrm{Gyr}}
\newcommand{\pc}{\rm pc}

\newcommand{\Msun}{\mathrm{M}_{\odot}}
\newcommand{\kms}{\rm km s^{-1}}
\newcommand{\Tv}{T_{\mathrm{v}}}
\newcommand{\Mv}{M_{\mathrm{v}}}
\newcommand{\Rv}{R_{\mathrm{v}}}
\newcommand{\Vv}{V_{\mathrm{v}}}
\newcommand{\tv}{t_{\mathrm{v}}}
\newcommand{\PIt}{\citetalias{Mandelker2016b}}
\newcommand{\PIp}{\citepalias{Mandelker2016b}}
\newcommand{\gtsima}{$\; \buildrel > \over \sim \;$}
\newcommand{\gsim}{\lower.5ex\hbox{\gtsima}}
\newcommand{\ltsima}{$\; \buildrel < \over \sim \;$}
\newcommand{\lsim}{\lower.5ex\hbox{\ltsima}}
\newcommand{\gammab}{\gamma_{\rm b}}
\newcommand{\gammas}{\gamma_{\rm s}}
\newcommand{\Tb}{T_{\rm b}}
\newcommand{\Ts}{T_{\rm s}}
\newcommand{\Tbs}{T_{\rm b,s}}
\newcommand{\rhob}{\rho_{\rm b}}
\newcommand{\rhos}{\rho_{\rm s}}
\newcommand{\rhobs}{\rho_{\rm b,s}}

\newcommand{\qs}{q_{\rm s}}

\newcommand{\eb}{e_{\rm b}}
\newcommand{\es}{e_{\rm s}}
\newcommand{\ebs}{e_{\rm b,s}}
\newcommand{\cb}{c_{\rm b}}
\newcommand{\cs}{c_{\rm s}}
\newcommand{\cbs}{c_{\rm b,s}}
\newcommand{\Vb}{V_{\rm b}}
\newcommand{\Vs}{V_{\rm s}}
\newcommand{\Vbs}{V_{\rm b,s}}
\newcommand{\Mb}{M_{\rm b}}
\newcommand{\Ms}{M_{\rm s}}

\newcommand{\Mtot}{M_{\rm tot}}
\newcommand{\Pic}{\Pi_{\rm c}}
\newcommand{\unpert}{(\Mb,\delta)}
\newcommand{\xint}{x_{\rm int}}
\newcommand{\Rs}{R_{\rm s}}
\newcommand{\tsc}{t_{\rm sc}}
\newcommand{\tRs}{t_{\rm \Rs}}
\newcommand{\Tbox}{T_{\rm box}}
\newcommand{\Mcrit}{M_{\rm crit}}
\newcommand{\Oi}{\omega_{_{\rm I}}}
\newcommand{\Imag}[1]{{\rm Im}(#1)}

\newcommand{\vpi}{\varpi_{_{\rm I}}}
\newcommand{\tg}{t_{\rm growth}}
\newcommand{\tkh}{t_{\rm KH}}

\newcommand{\Nef}{N_{\rm e\:folding}}
\newcommand{\kj}{k_{\rm j}}
\newcommand{\phij}{\phi_{\rm j}}

\newcommand{\lambdaj}{\lambda_{\rm j}}
\newcommand{\hj}{h_{\rm j}}
\newcommand{\jmin}{j_{\rm min}}
\newcommand{\jmax}{j_{\rm max}}
\newcommand{\Dj}{\Delta j}
\newcommand{\Hj}{H_{\rm j}}
\newcommand{\Wj}{W_{\rm j}}
\newcommand{\hs}{h_{\rm s}}

\newcommand{\hb}{h_{\rm b}}

\newcommand{\hbs}{h_{\rm b,s}}

\newcommand{\Vc}{V_{\rm c}}

\newcommand{\Ev}{E_{\rm v}}
\newcommand{\Evspat}{E_{\rm v,spat.}}
\newcommand{\vcm}{v_{\rm z,cm}}
\newcommand{\tausurf}{\tau_{\rm surface}}
\newcommand{\fpure}{f_{\rm pure}}
\newcommand{\Tpure}{T_{\rm pure}}
\newcommand{\rhopure}{\rho_{\rm pure}}
\newcommand{\xix}{\xi_{\rm x}}
\newcommand{\ux}{u_{\rm x}}
\newcommand{\tNL}{t_{\rm NL}}
\newcommand{\tNLcrit}{t_{\rm NL,crit}}
\newcommand{\hNL}{h_{\rm NL}}
\newcommand{\taubody}{\tau_{\rm body}}
\newcommand{\lambdacrit}{\lambda_{\rm crit}}
\newcommand{\ncrit}{n_{\rm crit}}
\newcommand{\vbreak}{v_{\rm break}}
\newcommand{\Rsbsurface}{R_{\rm s,break}^{\rm surface}}
\newcommand{\Rsbbody}{R_{\rm s,break}^{\rm body}}
\newcommand{\Rsb}{R_{\rm s,break}}
\newcommand{\tbreakbody}{t_{\rm break}^{\rm body}}
\newcommand{\Rsisurface}{R_{\rm s,inflow}^{\rm surface}}
\newcommand{\Rsibody}{R_{\rm s,inflow}^{\rm body}}
\newcommand{\Rsi}{R_{\rm s,inflow}}
\newcommand{\dx}{{\rm dx}}
\newcommand{\dz}{{\rm dz}}
\newcommand{\mb}{m_{\rm b}}
\newcommand{\ms}{m_{\rm s}}
\newcommand{\dmb}{{\rm d\mb}}
\newcommand{\dms}{{\rm d\ms}}
\newcommand{\dV}{{\rm dV}}
\newcommand{\vx}{v_{\rm x}}
\newcommand{\vz}{v_{\rm z}}
\newcommand{\vorty}{\omega_{\rm y}}
\newcommand{\epsener}{\varepsilon_{\rm ener}}

\setcitestyle{notesep={ }}
\defcitealias{Mandelker2016b}{Paper I}
\numberwithin{equation}{section}

\title[Nonlinear KH Instability in Cold Streams]{Instability of Supersonic Cold Streams Feeding Galaxies II. Nonlinear Evolution of Surface and Body Modes of Kelvin-Helmholtz Instability}

\author[D. Padnos et al.]{
\parbox[t]{\textwidth} 
{ 
Dan Padnos$^{1}$\thanks{E-mail: dan.padnos@mail.huji.ac.il},
Nir Mandelker$^{2,3}$,
Yuval Birnboim$^{1,4}$,
Avishai Dekel$^{1,5}$,
Mark R. Krumholz$^{4}$
and Elad Steinberg$^{1}$
}
\\
\\
$^{1}$Centre for Astrophysics and Planetary Science, Racah Institute of Physics, The Hebrew University, Jerusalem 91904, Israel\\
$^{2}$Department of Astronomy, Yale University, PO Box 208101, New Haven, CT, USA \\
$^{3}$Heidelberg Institut f{\"u}r Theoretische Studien, Schloss-Wolfsbrunnenweg 35, 69118 Heidelberg, Germany \\
$^{4}$Research School of Astronomy \& Astrophysics, Australian National University, Cotter Road, Weston, ACT 2611, Australia \\
$^{5}$Santa Cruz Institute for Particle Physics, University of California, Santa Cruz, CA 95064, USA 
}

\date{Accepted 2018 March 22. Received 2018 March 13; in original form 2018 January 22}

\pubyear{2018}

\begin{document}
\label{firstpage}
\pagerange{\pageref{firstpage}--\pageref{lastpage}}
\maketitle

\begin{abstract}
As part of our long-term campaign to understand how cold streams feed massive galaxies at high redshift, we study the Kelvin-Helmholtz instability (KHI) of a supersonic, cold, dense gas stream as it penetrates through a hot, dilute circumgalactic medium (CGM). A linear analysis (Paper I) showed that, for realistic conditions, KHI may produce nonlinear perturbations to the stream during infall. Therefore, we proceed here to study the \emph{nonlinear} stage of KHI, still limited to a two-dimensional slab with no radiative cooling or gravity. Using analytic models and numerical simulations, we examine stream breakup, deceleration and heating via surface modes and body modes. The relevant parameters are the density contrast between stream and CGM ($\delta$), the Mach number of the stream velocity with respect to the CGM ($\Mb$) and the stream radius relative to the halo virial radius ($\Rs/\Rv$). We find that sufficiently thin streams disintegrate prior to reaching the central galaxy. The condition for breakup ranges from $\Rs<0.03\Rv$ for $(\Mb \sim 0.75,\delta\sim10)$ to $\Rs<0.003\Rv$ for $(\Mb\sim2.25,\delta\sim 100)$. However, due to the large stream inertia, KHI has only a small effect on the stream inflow rate and a small contribution to heating and subsequent Lyman-$\alpha$ cooling emission. 
\end{abstract}




\section{Introduction}
\label{sec:introduction}

\subsection{General}
\label{sec:introduction-general}

According to the standard $\Lambda {\rm CDM}$ model of cosmology, the most massive galactic dark-matter halos at any epoch lie at the intersections of cosmic web filaments \citep{Zeldovich1970,Bond96,Springel05}. During the peak phase of star-formation, at redshift $z\sim2$, these include halos with a virial mass exceeding $\Mv\sim 10^{12}\Msun$, above the critical mass for shock heating \citep{bd03,db06}, which contain hot gas at the virial temperature, $\Tv \sim 10^6\K$. However, the gas flowing along the filaments that feed such massive halos is significantly denser than the halo gas, allowing it to cool more rapidly and preventing the formation of a stable virial shock within the filaments. Instead, these gas streams are expected to penetrate through the hot circumgalactic medium (CGM) onto the central galaxy, all the while remaining cold and dense \citep{db06,Dekel09}, with a typical temeperture of $\Ts\gsim 10^4\K$, set by the steep drop in the cooling rate below that temperature \citep{Sutherland1993}.

In cosmological simulations, cold streams with diameters of a few to ten percent of the virial radius are seen to penetrate deep into the halos of massive star-forming galaxies (SFGs), confirming the theoretical picture described above \citep{Keres05,Ocvirk08,Dekel09,CDB,FG11,vdv11}. The gas accretion rate via cold streams in simulations is on the order of $\sim 100\sy$, comparable to both the theoretical cosmological gas accretion rate \citep{Dekel09} and the observed star-formation rate (SFR) in SFGs \citep{Genzel06,Forster06,Elmegreen07,Genzel08,Stark08}. This suggests that cold streams must bring a significant fraction of the cosmological gas inflow into the central galaxy \citep{Dekel09}, as is the case in cosmological simulations \citep{Dekel2013}. 

The simulated streams maintain roughly constant inflow velocities when traversing the CGM, instead of accelerating as expected due to free-fall into the halo gravitational potential \citep{Dekel09,Goerdt15a}. This indicates that a dissipation process, as yet unidentified, acts upon the streams along the way. The associated loss of kinetic energy may be observed as Lyman-$\alpha$ emission \citep{Dijkstra09,Goerdt10,FG10}, possibly accounting for Lyman-$\alpha$ ``blobs" (LABs) observed at $z>2$ \citep{Steidel00,Matsuda06,Matsuda11}. Since cold streams consist of mostly neutral Hydrogen, they may also be visible in Lyman-$\alpha$ absorption and may explain some observed systems \citep{Fumagalli11,Goerdt12,vdv12,Bouche13}. Recent observational evidence reveals massive extended cold components in the CGM of high-redshift SFGs, with kinematic properties consistent with predictions for cold streams \citep{Cantalupo14,Martin2014,Martin2014a,Danovich15,Bouche16,Borisova2016,Fumagalli2017}.

Despite the growing evidence that cold streams are a fundamental part of galaxy formation at high redshift, several important questions regarding their evolution remain unanswered. Do the streams break up as they traverse the CGM? How much of their energy is dissipated in the journey? Should the dissipation be observable in emitted radiation? How does it affect the mass inflow rate? What is the state of the gas when it accretes onto the galaxy? How does this affect the angular momentum and SFR in the disk?

Most attempts to address these questions have used cosmological simulations. The current generation of cosmological simulations, such as the VELA suite of zoom-in simulations \citep{Ceverino14,Zolotov15,Tacchella2016}, reach a resolution of $\sim100\pc$ within streams at large radii. This is comparable to the stream width itself, so hydrodynamic and other instabilities at smaller scales are not captured properly, although the global stream properties such as its radius and mean density may be resolved. Thus, cosmological simulations are presently ill-suited to investigate the ultimate fate of cold streams. This difficulty may be the cause of apparent contradictions between properties of cold streams predicted by different simulations. For example, based on moving mesh \texttt{AREPO} \citep{Springel10} simulations, \citet{Nelson13} argue that cold streams heat up and become diffuse in the inner halo, contrary to comparable Eulerian AMR simulations, where the streams remain cold and collimated. The interpretation of these results is uncertain due to insufficient resolution \citep[see also][]{Nelson2016}, motivating a more careful study of cold stream evolution in the CGM.

As an alternative approach to full cosmological simulations, in this series of papers we use analytic models and idealized simulations, progressively increasing the complexity of our analysis by adding fundamental physical processes one-by-one. In parallel to this methodical approach, cosmological zoom-in simulations tailored to resolve stream instabilities will be presented in a separate paper (Roca-Fabrega et al. in preparation). 

\subsection{Main Results of \PIt}

\citet{Mandelker2016b}, hereafter \PIt, took the first step in our long-term campaign, by studying the linear phase of Kelvin-Helmholtz Instability (KHI) of a cold, dense stream confined in a hot, dilute background, under fully compressible conditions\footnotemark with no radiative cooling or gravity.
\footnotetext{We use ``compressible" to refer to flows with arbitrary Mach number: supersonic ($M>1$), transonic ($M\approx 1$) and subsonic ($M<1$). We use ``incompressible" to refer to the limit $M\rightarrow 0$.} Three geometrical variants of the problem were considered:
\begin{itemize}
	\item A \emph{planar sheet}, where two semi-infinite fluids are initially separated by a single planar interface at $x=0$. 
	\item A \emph{planar slab}, where the stream fluid is initially confined to a slab of finite thickness, $-\Rs<x<\Rs$, surrounded by the background fluid. 
	\item A \emph{cylindrical stream}, where the stream fluid is initially confined to a cylinder of finite radius  $\Rs$,  surrounded by the background fluid.
\end{itemize}
The fluids are characterized by their respective densities and speeds of sound, $\rhobs$ and $\cbs$, and are assumed to be in pressure equilibrium. The reference frame is chosen so that the background is initially stationary, $\Vb=0$, while the stream has velocity $\Vs=V$ parallel to the stream/background interface. 

The chosen setup, with $\Vb=0$, is a reasonably accurate representation of the conditions of our astrophysical scenario, where cold streams flow through hydrostatic gas that has been shocked to the virial temperature \citep{bd03,db06}. While there are indications that some turbulent motion is present in hydrostatic halos, the associated turbulent pressure is expected to be small compared to the thermal pressure. An upper limit on the turbulent pressure can be inferred from measurements of velocities under $100\kms$ in OVI absorbing objects in galaxies observed with the Cosmic Origins Spectrograph \citep{Green2012}. This is roughly half of the virial velocity, limiting the turbulent pressure to $\lsim1/4$ of the thermal pressure. For galaxy clusters this ratio is even smaller, estimated to be $10\%$ at most \citep{Churazov08,Rebusco2006}. These estimates are also supported by current cosmological simulations that should, in principle, correctly capture the large scale motions in gaseous halos.

The stability of the configurations described above, and the linear growth rate in the unstable regime, is determined by two dimensionless parameters: the density contrast, $\delta \equiv \rhos\slash\rhob$, and the Mach number with respect to the background speed of sound, $\Mb \equiv V\slash\cb$. Due to pressure equilibrium, the temperatures and speeds of sound satisfy $\Tb/\Ts=\delta$ and $\cb/\cs=\sqrt{\delta}$. The Mach number with respect to the stream speed of sound satisfies $\Ms \equiv V\slash\cs=\sqrt{\delta}\Mb$.

Linear stability analysis for each of the three problems considered in \PIt~yields an eigenvalue equation for the perturbations, which is further reduced to a dispersion relation, $\omega(k)$, where $\omega$ is the frequency, $k=2\pi/\lambda$ is the perturbation wavenumber and $\lambda$ is its wavelength along the stream. The growth rate\footnotemark~is given by the imaginary part of the frequency, $\Oi \equiv \Imag{\omega}$. The Kelvin-Helmholtz time, i.e. the characteristic time for perturbation growth in the linear regime, is therefore $\tkh\equiv1/\Oi(k)$. If $\Oi>0$ for some eigenmode, it grows exponentially with time as $\exp{(t\slash\tkh)}$.

\footnotetext{Generally, there are two approaches to linear stability analysis; \emph{temporal} and \emph{spatial}. In the former, the wavenumber $k$ is real while the frequency $\omega$ is complex. This represents seeding the entire system with a spatially-oscillating perturbation and studying its \emph{temporal growth}. In the latter, $\omega$ is real while $k$ is complex. This represents seeding a temporally-oscillating  perturbation at the stream origin and studying its downstream \emph{spatial growth}. The analysis performed in \PIt~was temporal.}

The planar sheet admits unstable eigenmodes which decay exponentially away from the initial interface and are therefore called \emph{surface modes} \citepalias[see Figure 2 in][]{Mandelker2016b}. Linear stability analysis shows that instability occurs if and only if the Mach number is below a critical value,
\begin{equation}
\label{eq:sheet-instability-condition}
\Mb < \Mcrit = \left(1 + \delta^{-1/3}\right)^{3/2},
\end{equation}
In cases where \Cref{eq:sheet-instability-condition} is satisfied, perturbations at all wavelength are unstable with
\begin{align}
\label{eq:sheet-tkh}
\tkh = \frac{1}{\vpi kV} = \frac{1}{2\pi\vpi}\frac{\lambda}{V}, &&\text{(surface modes)}
\end{align}
where $\varpi\equiv\omega/(kV)$, and $\vpi=\Imag{\varpi}$ is a dimensionless function of $\unpert$. Note that $\tkh$ scales as the time associated with the perturbation wavelength, $\lambda/V$, so smaller wavelength modes experience faster growth.

The slab and cylinder also admit surface mode solutions, which are equivalent to the planar sheet eigenmodes and converge to the same dispersion relation in the incompressible ($\Mb\ll 1$), short wavelength ($\lambda\lsim\Rs$) limit. As the Mach number is increased, surface modes become stable and are replaced by another class of unstable solutions, called \emph{body modes}. These are associated with waves reverberating back and forth between the stream boundaries, forming a pattern of nodes inside the stream, much like standing waves propagating through a waveguide \citepalias[see Figure 5 in][]{Mandelker2016b}. A necessary condition for body modes to grow is
\begin{equation}
\label{eq:body-instability-condition}
\Mtot \equiv \frac{V}{\cs+\cb} = \frac{\sqrt{\delta}}{1+\sqrt{\delta}}\Mb>1,
\end{equation}
which is roughly the opposite of \Cref{eq:sheet-instability-condition}. Therefore, the $\unpert$ parameter space is divided into a surface-mode-dominated regime and a body-mode-dominated regime, with a narrow range of parameters allowing coexistence.

For a given wavelength, body modes can appear in infinitely many \emph{orders}, forming a discrete set with different growth rates, $\{\Oi^{n}(k)\}_{n=-1}^{\infty}$, where $n+1$ corresponds to the number of transverse nodes in the pressure perturbation within the stream. In a planar slab, even values of $n$ correspond to \emph{sinusoidal-modes} (S-modes), which cause a symmetric  displacement of the slab boundaries, whereas odd values are \emph{pinch-modes} (P-modes), characterized by an antisymmetric displacement. The \emph{effective} Kelvin-Helmholtz time, determined by the mode with the largest growth rate at a fixed wavelength, is approximately
\begin{align}
\label{eq:body-tkh}
\tkh \simeq \frac{1}{\ln{\left(2\pi\vpi\frac{\Rs}{\lambda}\right)}} \tsc,  &&\text{(body modes)}
\end{align}
where $\tsc=2\Rs\slash\cs$ is the sound crossing time of the stream and $\vpi\unpert$ is some dimensionless factor\footnotemark. While the slab admits only two symmetry modes, S- and P-modes, a cylinder admits infinitely many symmetry modes, corresponding to the azimuthal mode number, $m\geq0$, which represents the number of nodes on the circumference of the cylinder. Despite this and other qualitative differences between a planar slab and a cylindrical stream, the effective growth rates of body modes are similar and \Cref{eq:body-tkh} represents both cases with reasonable accuracy. 
\footnotetext{This is a slight divergence from the notation in \PIt, where we used $\vpi \equiv \Oi\slash(kV)$ throughout.}

Comparing \Cref{eq:sheet-tkh} and \Cref{eq:body-tkh}, we find two notable differences between the KH time for surface and body modes: first, for body modes $\tkh$ scales as $\Rs/\cs$ instead of $\lambda/V$; second, although in both cases smaller wavelength perturbations have larger growth rates, the dependence on $\lambda$ is weaker for body modes (logarithmic instead of linear).

The importance of KHI in the evolution of cosmic cold streams was assessed in \PIt~by estimating the number of e-foldings achieved by a perturbation,
\begin{equation}
\label{eq:num-e-folding}
\Nef=\frac{\tg}{\tkh},
\end{equation}
where $\tg$ is the time available for growth before the stream joins the central galaxy, on the order of the virial crossing time, $\tv$. In \PIt~we estimate $\tg\simeq\tv$ for surface modes and $\tg\simeq\tv-\tsc$ for body modes. By substituting $\tkh$ from  either \Cref{eq:sheet-tkh} or \Cref{eq:body-tkh}, we can evaluate \Cref{eq:num-e-folding} as a function of four parameters: $\Mb$, $\delta$, $\Rv\slash\Rs$ and $\Rs\slash\lambda$, where the virial radius, $\Rv$, is introduced through $\tv$. \PIt~included a rough estimation\footnotemark~for the allowed range of parameters for cosmic cold streams, namely $\Mb\sim 1-2$, $\delta\sim 10-100$, $\Rs\slash\Rv\sim 0.005-0.05$ and $\Rs\slash\lambda\gsim 1$. These estimates are repeated more accurately here in \Cref{sec:application-parameters}, including a mutual constraint on $\Mb$, $\delta$ and $\Rv\slash\Rs$. For the aforementioned parameter range, \PIt~found $\Nef$ between $0.1$ and $100$, showing that KHI could in principle have an important role in the evolution of cold streams. In general, for both surface modes and body modes, the instability is attenuated as either the Mach number or the density contrast are increased. When surface modes dominate, the stream is highly unstable and even very small perturbations are expected to become nonlinear. However, only a very minor change in velocity pushes the stream into the regime where body modes dominate, the growth rate is slower and the outcome of the instability is expected to depend largely on the stream width. These conclusions motivate the study of the subsequent nonlinear instability in cold streams feeding massive galaxies at high redshifts.
\footnotetext{Since growth rates diverge as $\lambda\to 0$, the cited values $\Rs\slash\lambda$ deserve particular scrutiny. In a realistic stream, various physical processes are expected to suppress the growth of small-wavelength perturbations, resulting in a finite fastest-growing mode. The assumed range of $\Rs\slash\lambda$ is a crude estimation \citepalias[see section 5 of][]{Mandelker2016b} and awaits more rigorous study in the future.}

\subsection{Validity of the 2D Adiabatic Slab Model}

In this paper, we address the nonlinear evolution of both surface and body modes in 2D planar sheet and slab geometries, still with no radiative cooling or gravity, using 2D hydrodynamic simulations and simple analytic models. The use of a 2D slab model instead of a 3D cylinder is justified by analysis, experiments and simulations as follows. Linear analysis shows the aforementioned agreement in growth rates of the slab and cylinder problems. Specifically for slab surface modes, experiments show that the large-scale nonlinear evolution is dominated by nearly two-dimensional structures with little spanwise variability \citep{Brown1974,Papamoschou1989}. Numerical studies comparing perturbation growth and fragmentation in 2D and 3D simulations of body modes reach similar conclusions in both cases, despite differences in morphology \citep[e.g][]{Bassett1995,Bodo1998}. 

Nevertheless, the 2D model has some limitations. First, as will be noted below, cylindrical streams are likely to decelerate faster than 2D planar slabs, raising a caveat to our findings regarding stream inflow rates. Second, although the large-scale behavior of KHI is captured in 2D, a proper treatment of turbulence requires 3D simulations, and is therefore beyond the scope of this work. Third, the strict separation between a surface-mode- and a body-mode-dominated regime is an accurate description only in 2D; in 3D, we expect to find unstable surface modes at high Mach numbers as well\footnotemark. In future communication, we will address these caveats using 3D simulations of cylinders and investigate the effects of additional physical processes absent from the current treatment, such as cooling, thermal conduction, the external potential of the host halo, self-gravity and magnetic fields.
\footnotetext{The Mach number determining surface mode stability in \Cref{eq:sheet-instability-condition} corresponds to the velocity component parallel to the perturbation wave-vector, $V_{\rm k}={\vec {V}}\cdot {\hat{k}}$. Therefore, perturbations at sufficiently oblique angles relative to the flow can be considered effectively subsonic, even in highly supersonic streams. For slabs, this means that surface modes with sufficiently large $k_{\rm y}$ are unstable even if $\Mb>\Mcrit$. For cylinders, surface modes with sufficiently high azimuthal order $m$ are unstable even if $\Mb>\Mcrit$.}

\subsection{Outline}

This paper is organized as follows. In \Cref{sec:methods} we discuss the numerical simulations and techniques used for their analysis. In \Cref{sec:surface} we address the nonlinear evolution of surface modes in the planar sheet and slab. In \Cref{sec:body} we discuss the nonlinear evolution of body modes in the planar slab. In \Cref{sec:application} we estimate the relevant parameter range for KHI in cosmic cold streams and apply the results of our idealized models to the astrophysical scenario to obtain estimations of the potential reduction of inflow rates, fragmentation and Lyman-$\alpha$ emission due to KHI in cold streams. We summarize our conclusions in \Cref{sec:conclusions}.


\section{Numerical Methods}
\label{sec:methods}

\subsection{Hydrodynamic Code}
\label{sec:methods-cod}

We use the Eulerian AMR code \texttt{RAMSES} \citep{Teyssier02}, with a piecewise-linear reconstruction using the MonCen slope limiter \citep{vanLeer77} and a HLLC approximate Riemann solver \citep{Toro94}, identical to \PIt. All of our simulations are two-dimensional (2D).

\subsection{Unperturbed Initial Conditions}
\label{sec:methods-unperturbed}

\begin{figure*}
	\centering
	\subfloat{
		\includegraphics[trim={0cm 0.75cm 0.75cm 0.75cm}, clip, height=0.35\textheight]{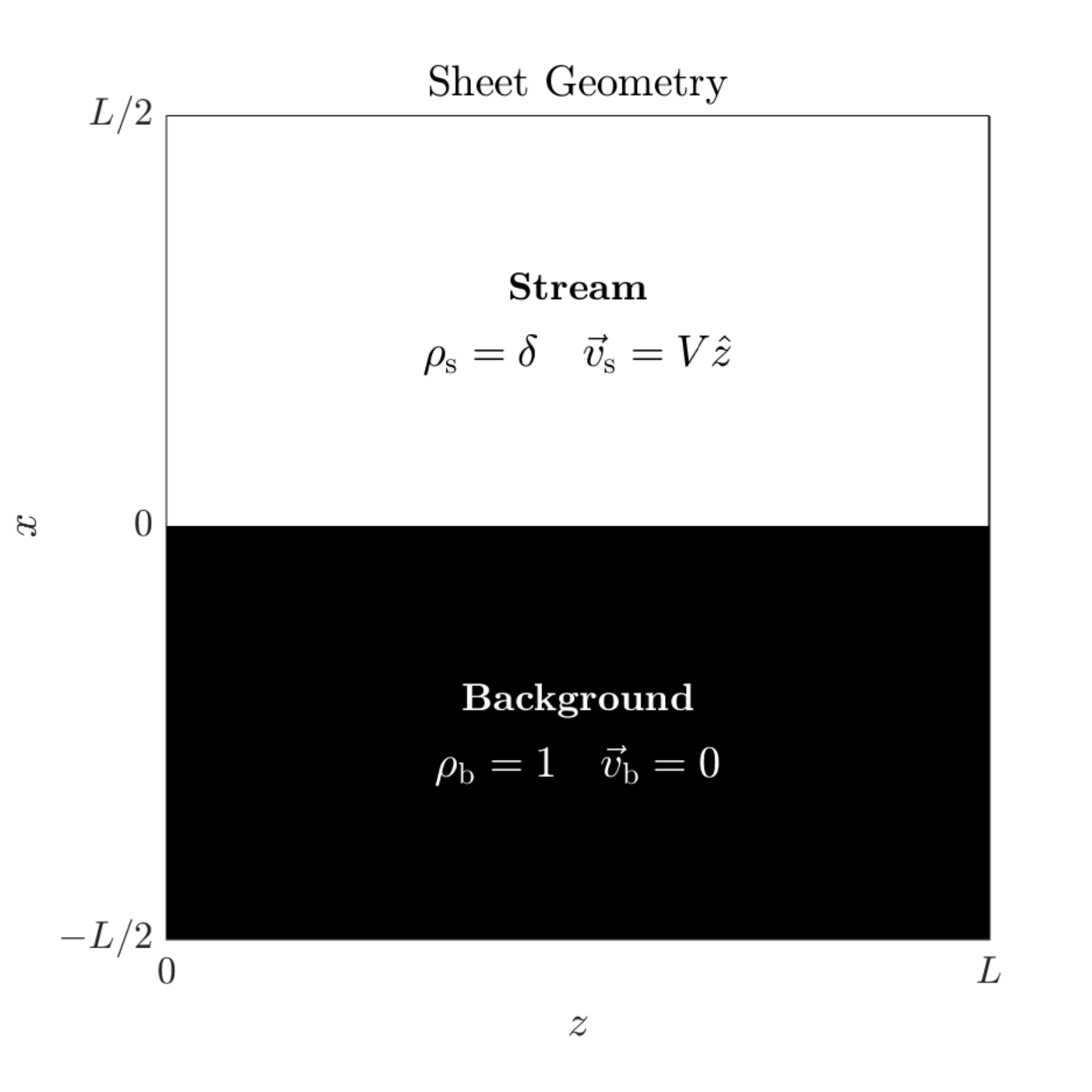}
	}
	\subfloat{
		\includegraphics[trim={0.75cm 0.75cm 0.75cm 0.75cm}, clip, height=0.35\textheight]{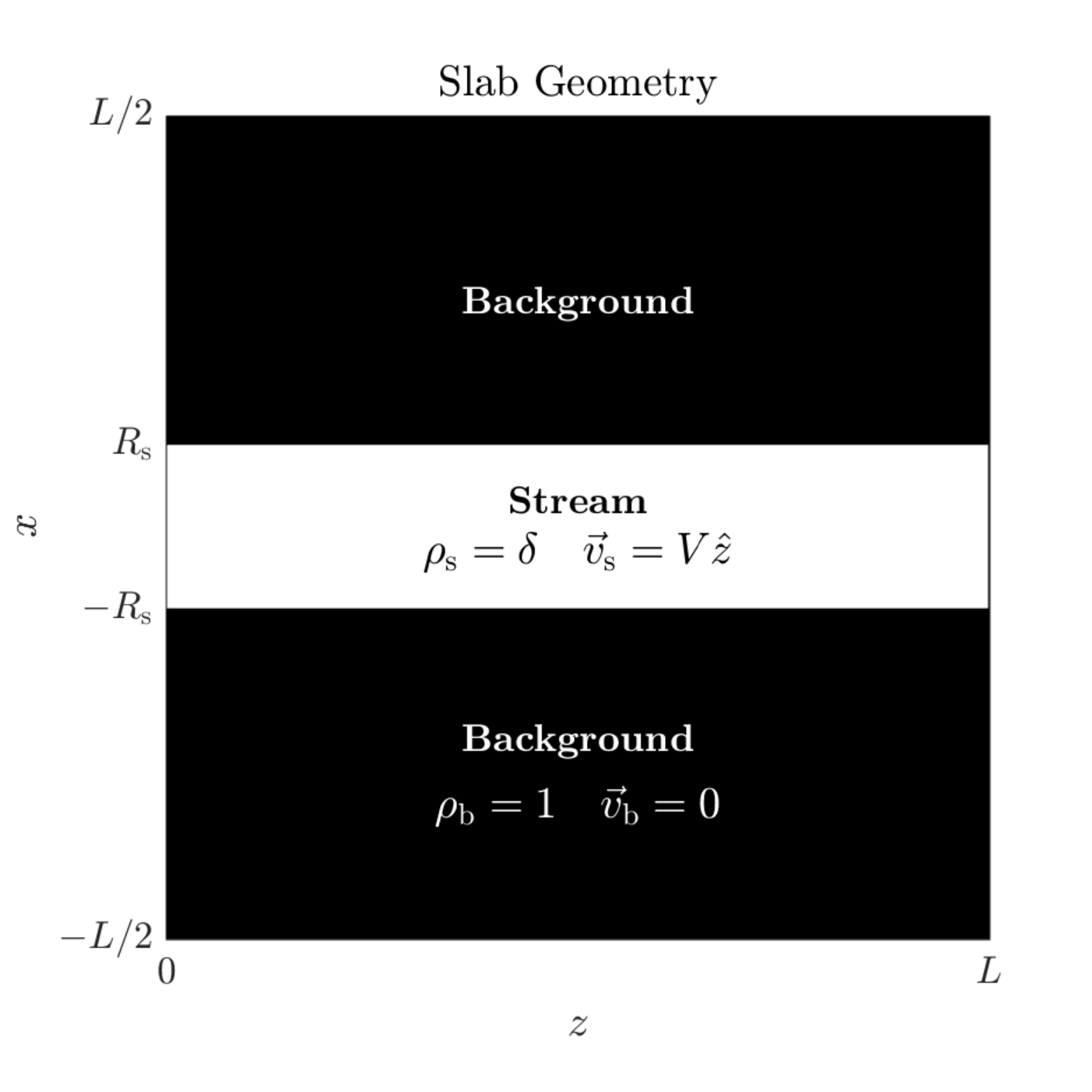}
	}
	\caption{Unperturbed initial conditions for planar sheet and planar slab geometry. In planar sheet simulations, the instability involves one interface at $x=0$. In planar slab simulation, the instability involves two interfaces at $x=\pm\Rs$.}
	\label{fig:unperturbed-ic}
\end{figure*}

The simulation domain is a square of side $L=1$, representing the $xz$ plane, extending from $0$ to $1$ in the $z$ direction and from $-0.5$ to $0.5$ in the $x$ direction. The unperturbed initial conditions are illustrated in \Cref{fig:unperturbed-ic}. For sheet geometry, the interface between the fluids is centered at $x=0$, with the background fluid occupying $x<0$ and the stream fluid occupying $x>0$. For slab geometry, the slab is centered around $x=0$, such that the stream fluid occupies $-\Rs<x<\Rs$ and the background fluid fills the rest of the domain. Both fluids are ideal gases with adiabatic index $\gamma=5/3$, and initial uniform pressure $P_0=1$. The background is initialized with density $\rhob=1$ and velocity ${\vec {v}}_{\rm b}=0$. The stream is initialized with $\rhos=\delta$ and ${\vec {v}}_{\rm s}=V{\hat {z}}=\Mb\cb{\hat {z}}$, where $\cb=\sqrt{5/3}$ is the background sound speed in simulation units. 

In the setup described above, the density and velocity are discontinuous at the interfaces, either $\xint=0$ in sheet geometry or $\xint=\pm\Rs$ in slab geometry. This generates numerical perturbations at the grid scale, which grow faster than the intended perturbations in the linear regime, and may dominate the instability, depending on their initial amplitude. This is remedied by smoothing the unperturbed density and velocity around each interface using the same ramp function as in \PIt,
\begin{equation}
\label{eq:smoothing}
f(x) = f_{\rm b} + \frac{f_{\rm s}-f_{\rm b}}{2}
\times\left[1\pm\tanh\left(\frac{x-\xint}{\sigma}\right)\right].
\end{equation}
where $f$ stands for either $\rho$ or $\vec{v}$ and the sign is chosen for each interface separately such that $f=f_{\rm s}$ inside the stream (positive for $\xint\leq 0$ and negative otherwise). Different values of the width parameter $\sigma$ are used in surface mode and body mode simulations. In all cases, $\sigma$ is sufficiently small, on the order of a few cells, to guarantee that our conclusions do not depend on its exact value. 

\subsection{Boundary Conditions}
\label{sec:methods-boco}

We use periodic boundary conditions at $z=0$ and $z=1$, and outflow boundary conditions at $x=\pm 0.5$, such that gas crossing the boundary is lost from the simulation domain. The boundary conditions at $x=\pm 0.5$ may affect the interface region once sound waves have propagated back and forth between the interface and the boundary. For an interface at $x=0$, the minimal time for this interaction to occur is $\Tbox \simeq L/\cb \simeq 0.775$ in simulation units\footnotemark. Our \emph{planar sheet} simulations end at $t \leq \Tbox$, guaranteeing that the results are not influenced by the outflow boundary conditions. Much longer simulations were performed in order to observe the late time evolution of \emph{planar slabs} with high density contrasts. The influence of the boundary conditions on these simulations was estimated by comparing runs with the fiducial domain size, $L=1$, to runs with a larger domain, $L=2$, and found to be negligible (see \Cref{sec:surface-results-deceleration,sec:body-results-deceleration}).

\footnotetext{Any outgoing shocks in our simulations are sufficiently weak such that their velocity is roughly $\cb$, so the estimated value of $\Tbox$ represents sound waves and shock waves alike.}

\subsection{Computational Grid}
\label{sec:methods-grid}

We used a statically refined grid throughout all runs. The innermost region, $|x|<0.2$, has the highest resolution with cell size $\Delta=2^{-13}=1/8192$ in simulation units. The cell size increases by a factor of $2$ every $0.05$ simulation units in the $x$-direction, up to a maximal cell size of $2^{-9}$ in runs with $L=1$. The grid used in runs with $L=2$ is exactly identical to $L=1$ in the region $|x|<0.5$, which is padded with a uniform cell size of $2^{-8}$ for $|x|>0.5$. 

Overall, our results are converged in terms of the computational grid. For planar sheet simulations, convergence is tested by doubling the resolution of the entire grid, reaching a minimal cell size of $\Delta=2^{-14}$, with negligble effect on our conclusions. This is reported in \Cref{app:convergence}. For slabs, the important parameter for convergence is the number of cells across the slab width, $\Rs/\Delta$. Convergence is therefore tested by varying the stream radius and all other physical length scales by the same factor, while keeping the computational grid the same. The effect on our results is small for slabs as well. These tests are reported in \Cref{sec:surface,sec:body} where appropriate.

\subsection{Tracing the Two Fluids}
\label{sec:methods-analysis}

Our simulations include a passive scalar field, denoted by $\psi(x,z,t)$. The passive scalar is initialized such that $\psi=1$ in the stream and $\psi=0$ in the background. Since this field is advected with the flow, it serves as a Lagrangian tracer for the fluid in the simulation (which is Eulerian). An element characterized by passive scalar value $\psi$, density $\rho$ and volume $\dV$ contains a mass of stream and background fluid given by
\begin{equation}
\label{eq:mass-stream-background}
\dms = \psi \rho \dV \qquad \text{and} \qquad \dmb = (1-\psi) \rho \dV.
\end{equation}

The volume-weighted line average of $\psi$ along the $\hat{z}$-axis is given by 
\begin{equation}
\label{eq:volume-averaged-color}
\overline{\psi}(x,t) = \frac{\int_{0}^{L}{\psi(x,z,t) \dz}}{\int_{0}^{L} \dz}.
\end{equation}
In the unperturbed state, each stream/background interface is characterized by a sharp transition\footnotemark~between $\overline{\psi}=0$ in the background and $\overline{\psi}=1$ in the stream. Hence, $\overline{\psi}(x,t=0)$ is a step function for planar sheet geometry and a top-hat function for planar slab geometry. 
\footnotetext{Neglecting the smoothing introduced in \Cref{eq:smoothing}.}

Due to the nonlinear growth of surface modes, the initial discontinuities of $\overline{\psi}(x)$ are smeared over a finite width around each interface. In the neighborhood of each interface, we assume that $\overline{\psi}(x,t>0)$ remains monotonic and reaches its asymptotic values $\overline{\psi}=0$ and $\overline{\psi}=1$. 
Therefore, we can use the inverse $x(\overline{\psi})$ to define the edges of the perturbed region around an interface, $x(\overline{\psi}=\varepsilon)$ on the background side and $x(\overline{\psi}=1-\varepsilon)$ on the stream side, where $\varepsilon$ is an arbitrary threshold. The background-side thickness of the perturbed region in our simulation is then defined by
\begin{equation}
\label{eq:hb}
\hb \equiv
\begin{cases}
-x(\overline{\psi}=\varepsilon) & \quad\text{sheet} \\
\max{x(\overline{\psi}=\varepsilon)}-\Rs & \quad\text{slab (top)} \\
-\Rs-\min{x(\overline{\psi}=\varepsilon)}  & \quad\text{slab (bottom)}
\end{cases},
\end{equation}
whereas the stream-side thickness is
\begin{equation}
\label{eq:hs}
\hs \equiv
\begin{cases}
x(\overline{\psi}=1-\varepsilon) & \quad\text{sheet} \\
\Rs-\max{x(\overline{\psi}=1-\varepsilon)} & \quad\text{slab (top)} \\
\min{x(\overline{\psi}=1-\varepsilon)}+\Rs  & \quad\text{slab (bottom)}
\end{cases},
\end{equation}
and the overall thickness of the perturbed region is the sum, 
\begin{equation}
\label{eq:h}
h \equiv \hb+\hs
\end{equation}
The different cases in \Cref{eq:hb,eq:hs} correspond to our various possible initial interface positions, namely $x=0$ for planar sheet geometry and $x=\pm\Rs$ for planar slab geometry. The signs are chosen such that $\hbs\geq 0$. In the following sections, the position of the interface in question will be either irrelevant or made clear by the context.

A related metric is used to estimate the total width of a planar slab perturbed by either surface or body modes. We define the time-dependent ``stream width'', $w$, as the extent of the region satisfying $\overline{\psi}\geq\varepsilon$,
\begin{equation}
\label{eq:stream-width}
w \equiv
\max{x(\overline{\psi}=\varepsilon)}-\min{x(\overline{\psi}=\varepsilon)}.
\end{equation}

The metrics presented above depend on $\varepsilon$ and are expected to converge for $\varepsilon\to 0^{+}$. The exact dependence of $h$ and $w$ on $\varepsilon$ depends on the shape of $\overline{\psi}(x)$, which varies somewhat with $\unpert$, introducing minor systematic differences between estimations in different cases. Through experimentation we find that the thickness with $\varepsilon$ of a few percent traces the limits of the perturbed region reasonably well in all runs. Unless stated otherwise, we use $\varepsilon=0.02$. This is analogous to the common experimental approach of estimating the ``visual thickness'' by tracing the visible limits of the largest perturbations in high-speed photography (e.g. \citealt{Brown1974,Papamoschou1989,Rossmann2002}), and is therefore useful for comparison with experiments. 

\subsection{Numerical Mixing}

In our simulations, the stream and the background fluid eventually become tangled in complex, small-scale structures. Therefore, a non-negligible fraction of the stream mass might occupy cells with a mixed composition of stream and background fluid. Depending on the mixing ratio, the temperature\footnotemark~of a mixed cell can be dominated by the hot background, creating the false impression that the cold fluid has been heated considerably. When attempting to numerically evaluate the heating experienced by the stream, it is important to take into account only cells with sufficiently pure composition, so as not to be mislead by this numerical artifact.

\footnotetext{Temperature and specific internal energy are closely related by $e=1/(\gamma-1)\cdot kT/(\mu m_{\rm p})$, where $k$ is the Boltzmann constant and $\mu m_{\rm p}$ is the particle mass, and are therefore used interchangeably.}

Consider a cell containing masses $\rm dm_{\rm b,s}$ of background fluid and stream fluid respectively. Assuming both components maintain their initial thermodynamic state, the specific internal energies are 
\begin{equation}
\ebs \equiv \frac{1}{\gamma-1}\frac{P}{\rhobs},
\end{equation}
for the background and the stream respectively. The specific internal energy of the mixed cell, $e$, must be equal to the mass-weighted mean of its components, 
\begin{equation}
\label{eq:energy-mix}
e = \frac{\es\dms+\eb\dmb}{\dms + \dmb} = \left[1+(\delta-1)(1-\psi)\right] e_{\rm s}.
\end{equation}
For $\delta>1$, we see that an impure composition, corresponding to $0<\psi<1$, leads to an over-estimation of the internal energy of the stream when using the cell-averaged energy, $e>\es$. We require that the associated relative error be less than some threshold, 
\begin{equation}
\frac{e-\es}{\es} < \epsener.
\end{equation}
Substituting \Cref{eq:energy-mix} for $e$ yields the condition
\begin{equation}
\label{eq:pure-stream}
\psi > 1 - \frac{\epsener}{\delta-1} > 1-\frac{\epsener}{\delta},
\end{equation}
where the last inequality holds for any $\delta>1$. Note that the numeric heating discussed above is of no concern for $\delta=1$. Nevertheless, for consistency, we use \Cref{eq:pure-stream} to define ``pure'' or ``unmixed'' stream fluid for all $\delta$. The results show little sensitivity to the exact value of $\epsener$ (see \Cref{sec:surface-results-heating,sec:body-results-heating}).

The mass fraction of stream fluid in unmixed cells is given by 
\begin{equation}
\fpure = \frac{\int \Theta\left(\psi-1+\frac{\epsener}{\delta}\right) \dms}{\int \dms},
\end{equation}
where $\Theta$ is the Heaviside step function and the integration is done over the entire simulation domain. The mass-weighted mean density and temperature of the unmixed stream fluid are therefore
\begin{equation}
\Tpure = \frac{\int \Theta\left(\psi-1+\frac{\epsener}{\delta}\right) T \dms}{\int \Theta\left(\psi-1+\frac{\epsener}{\delta}\right) \dms},
\end{equation}
\begin{equation}
\rhopure = \frac{\int \Theta\left(\psi-1+\frac{\epsener}{\delta}\right) \rho \dms}{\int \Theta\left(\psi-1+\frac{\epsener}{\delta}\right) \dms}.
\end{equation}


\section{The Nonlinear Evolution of Surface Modes}
\label{sec:surface}

This section extends the study of surface modes into the nonlinear stage for both the planar sheet and planar slab. In \Cref{sec:surface-qual} we present a qualitative description of the nonlinear evolution to guide the intuition. The numerical simulations used for this section are described in \Cref{sec:surface-simulations}. The results are presented in \Cref{sec:surface-results}.


\subsection{Qualitative Description}
\label{sec:surface-qual}

It is instructive to begin our discussion by considering two thought experiments. We first present these experiments and then discuss their implications.

In the first thought experiment, let a \emph{single-mode} perturbation of wavelength $\lambda$ grow in a KH-unstable planar sheet. Assume that the initial displacement amplitude is small, $h(t=0) \ll \lambda$, where $h(t)$ is the peak-to-valley amplitude of the perturbation at the interface\footnotemark. As long as the limit $h(t) \ll \lambda$ holds, linear analysis yields $h(t) = h(0) \exp{\left[t/\tkh(\lambda)\right]}$. At some finite time the perturbation is bound to reach an amplitude comparable to its wavelength, $h(t) \approx \lambda$, rendering the assumptions of linear  analysis invalid. How will the single-mode perturbation evolve from this point and on, outside the linear regime?
	
In the second thought experiment, let a \emph{multi-mode} initial perturbation comprising a set of wavelengths $\left\{\lambdaj\right\}$ with initially small amplitudes $\left\{h_{\rm j}(t=0)\right\}$ grow in a KH-unstable planar sheet. As long as $\hj \ll \lambdaj$ holds for all $j$, each mode grows independently of the rest, according to its own respective linear growth rate, $\hj(t) = h_{\rm j0} \exp{\left[t/\tkh(\lambdaj)\right]}$. How will the system evolve once some of the modes leave the linear regime?

\footnotetext{The peak-to-valley amplitude refers to the distance along the $x$-axis between the top (peak) point of the perturbed interface to the bottom (valley) point. This differs from \PIt, where $h$ was used for the one-sided amplitude. Since the eigenmodes of the linearized problem displace the interface an equal amount in both directions, this corresponds to replacing $h$ by $h/2$. Note that the displacement is not necessarily symmetric in the nonlinear regime, i.e. the one-sided amplitudes can differ from $h/2$. This is discussed in \Cref{sec:surface-results-entrainment}.}

One can seek solutions to these problems by applying dimensional arguments. There are two different ways of constructing a length scale for the peak-to-valley amplitude. One is
\begin{equation}
\label{eq:saturated-solution}
h \sim \lambda,
\end{equation}
where the growth of a given perturbation mode saturates at an amplitude comparable to its wavelength, $\lambda$, and the other is  
\begin{equation}
\label{eq:self-similar-solution}
h \sim Vt,
\end{equation}
where the perturbed region grows indefinitely in a self-similar manner.

If we assume that the initial conditions dominate the evolution of the instability indefinitely, we must prefer the saturated solution in \Cref{eq:saturated-solution}. Conversely, if we require that initial conditions are ``forgotten'' at late times, the self-similar solution in  \Cref{eq:self-similar-solution} must hold. As we demonstrate shortly, while the saturated behavior may appear as a transient, a practical system will inevitably converge to the self-similar form.

A single-mode initial perturbation is known to evolve into a row of identical singular point vortices, often called ``eddies" \citep[e.g.][]{Corcos1976}. The vortices add more windings with time but their height reaches a finite asymptotic value proportional to the wavelength, as predicted in \Cref{eq:saturated-solution}. In particular, for an  incompressible flow, the evolution of a row of vortices can be solved using methods of complex potential, yielding an asymptotic value of $h(t\to\infty)=\left[\arcsin{(1)}/\pi\right] \lambda \cong 0.56\lambda$ for the height of an individual eddy \citep{Corcos1976, Rikanati2003}. This analysis assumes a perfect single-wavelength perturbation in an infinite or periodic domain.

The growth of a multi-mode initial perturbation is driven by an attractive interaction between saturated co-rotating vortices, leading to vortex mergers, sometimes referred to as ``pairing'' or ``amalgamation''. This was first observed in the pioneering experiments by \citet{Brown1974} and \citet{Winant1974}. The result of a merger between two vortices with heights $h_1$ and $h_2$ is the formation of a larger vortex with its height approximately equal to the sum of the component heights, $h \simeq h_1+h_2$ \citep{Winant1974, Rikanati2003}. These interactions cause the vortex population to increase in size and decrease in number with time, resulting in a self-similar growth of the perturbed region with $h(t)=\alpha Vt$ as predicted in \Cref{eq:self-similar-solution}. The coefficient
\begin{equation}
\label{eq:alpha}
\alpha \equiv \frac{h}{Vt}
\end{equation}
is referred to as the \emph{dimensionless growth rate} and the flow confined within the perturbed region is often called a \emph{``shear layer''} or \emph{``mixing layer''}. This behavior was demonstrated in numerous experiments and numerical simulations (see references in \Cref{sec:surface-simulations}).

\begin{figure*}
	\centering
	\subfloat{
		\includegraphics[trim={1cm 1.4cm 1.25cm 1.75cm}, clip,width=0.7\textwidth]{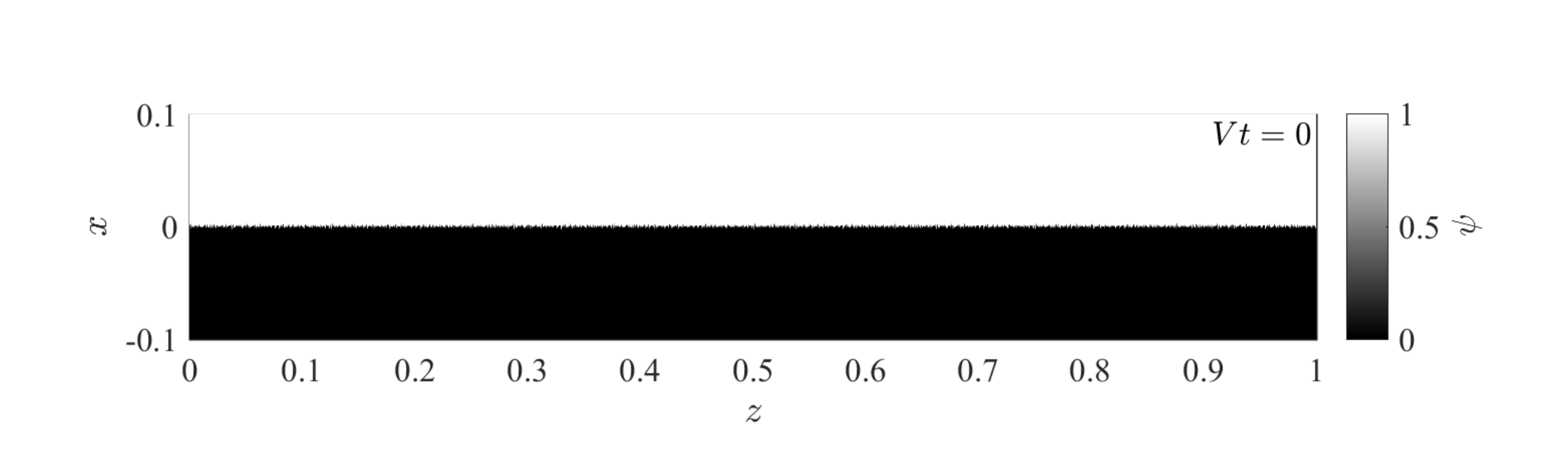}
    }\\
	\subfloat{
		\includegraphics[trim={1cm 1.4cm 1.25cm 1.75cm}, clip,width=0.7\textwidth]{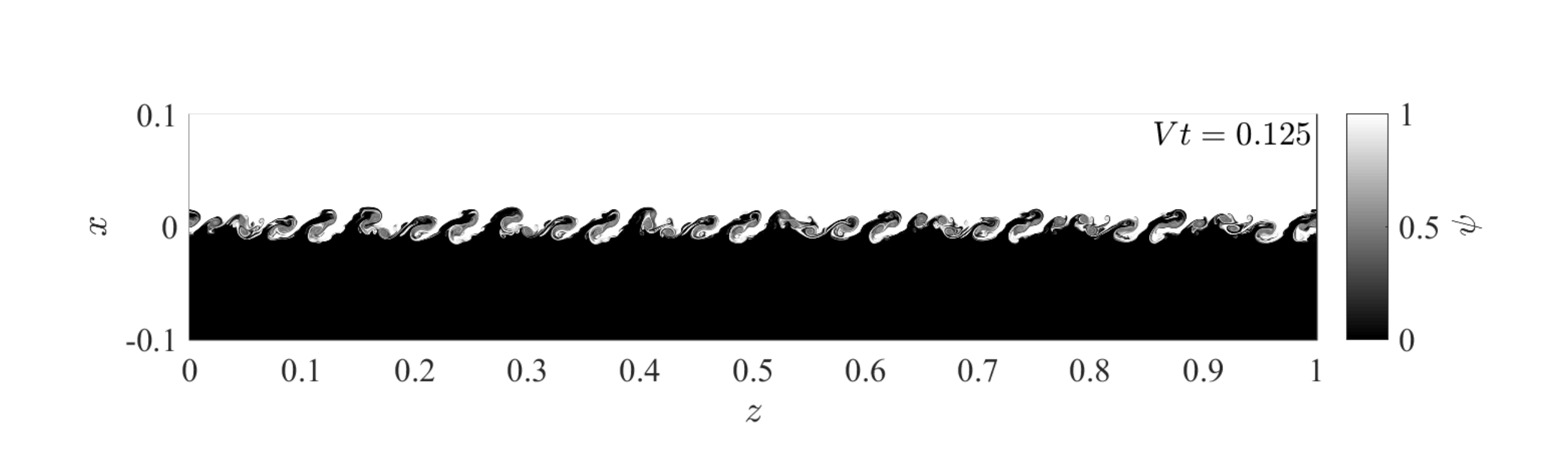}
	}\\
	\subfloat{
		\includegraphics[trim={1cm 1.4cm 1.25cm 1.75cm}, clip,width=0.7\textwidth]{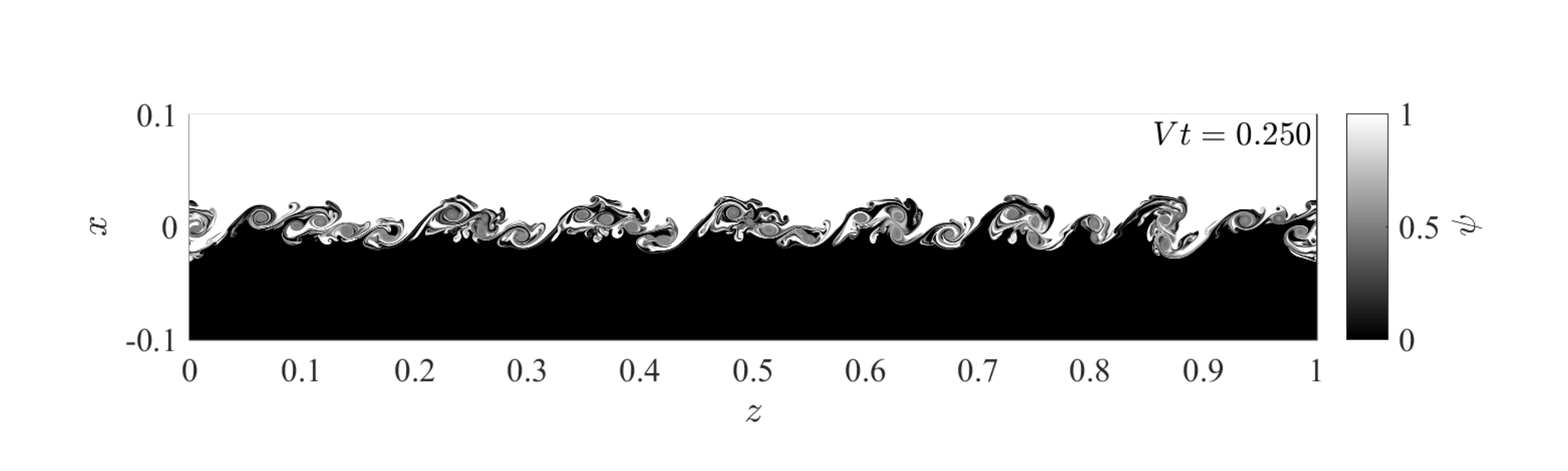}
	}\\
	\subfloat{
		\includegraphics[trim={1cm 1.4cm 1.25cm 1.75cm}, clip,width=0.7\textwidth]{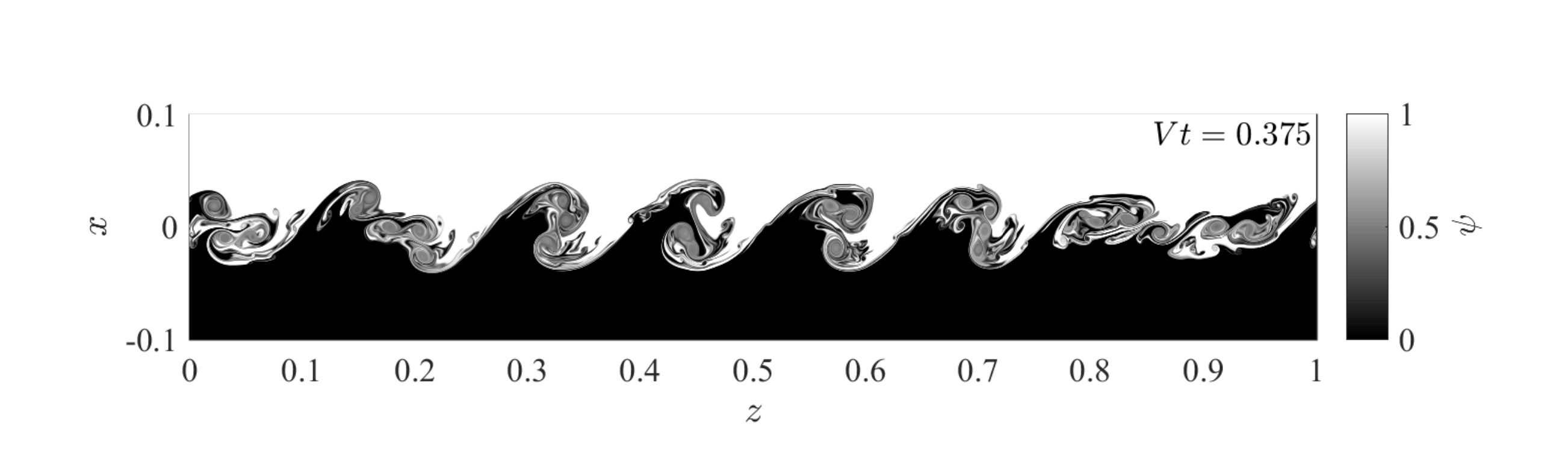}
	}\\
	\subfloat{
		\includegraphics[trim={1cm 0.8cm 1.25cm 1.75cm}, clip,width=0.7\textwidth]{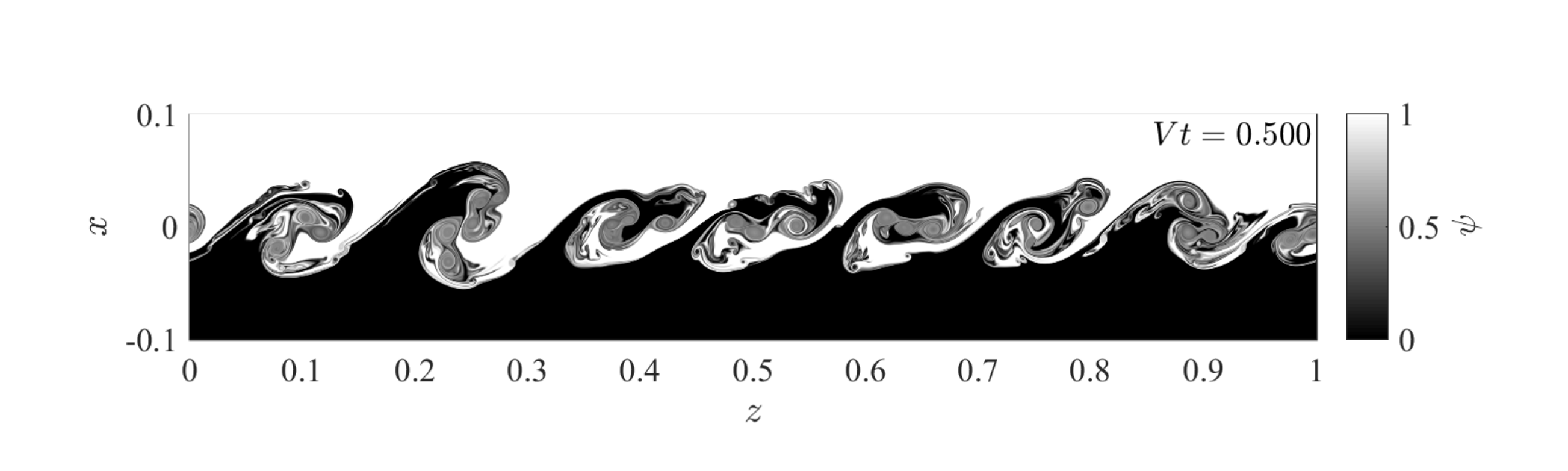}
	}
	\caption{Snapshots of $\psi(x,z)$, a passive scalar field used as a lagrangian tracer for the stream fluid (see \Cref{sec:methods-analysis}), taken from a planar sheet simulation. The values $\psi=0$ (background) and $\psi=1$ (stream) correspond to cells with a pure fluid composition, while cells with $0<\psi<1$ contain a mixture of both fluids. The different times are represented by the self-similar coordinate, $Vt$. The numerical setup is described in \Cref{sec:surface-simulations}. The unperturbed initial conditions were $(\Mb=0.5,\delta=1)$ and the initial perturbations were full eigenmodes with a sparse white noise spectrum of type A, as described in \Cref{sec:surface-simulations}.}
	\label{fig:demo-eddies-color}
\end{figure*}

\begin{figure*}
	\centering
	\subfloat{
		\includegraphics[trim={1cm 1.4cm 1.25cm 1.75cm}, clip,width=0.7\textwidth]{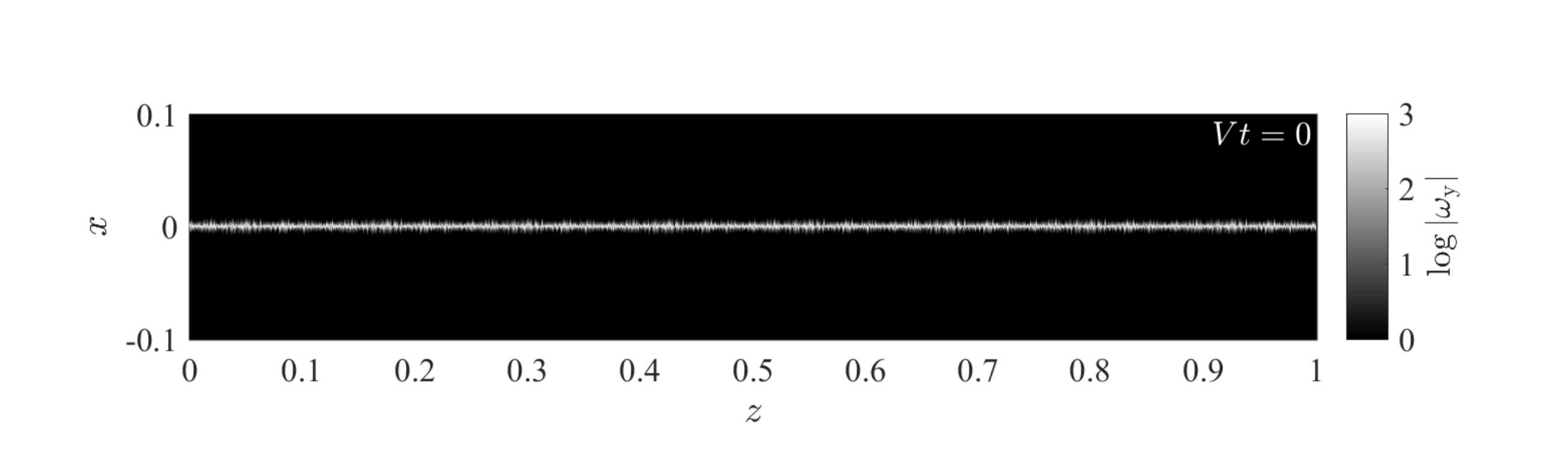}
	}\\
	\subfloat{
		\includegraphics[trim={1cm 1.4cm 1.25cm 1.75cm}, clip,width=0.7\textwidth]{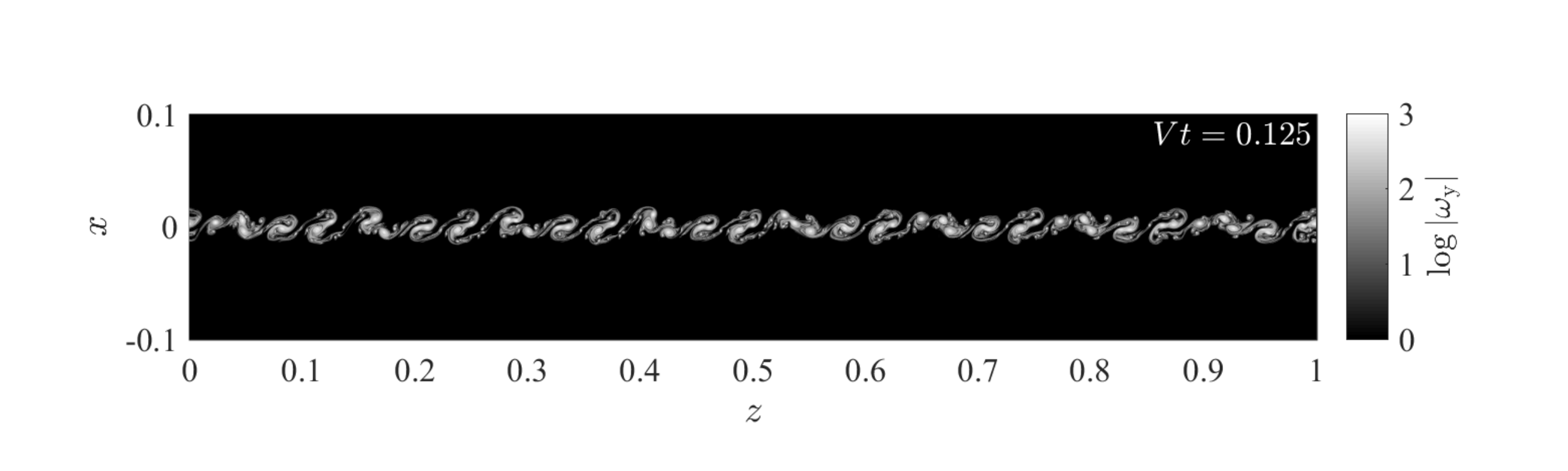}
	}\\
	\subfloat{
		\includegraphics[trim={1cm 1.4cm 1.25cm 1.75cm}, clip,width=0.7\textwidth]{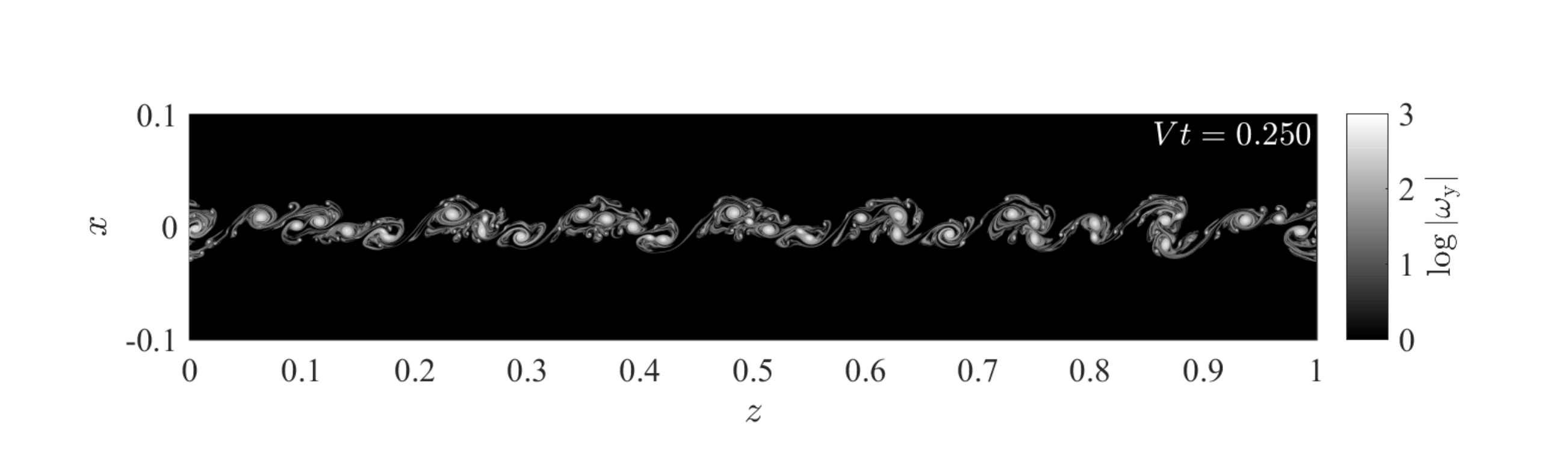}
	}\\
	\subfloat{
		\includegraphics[trim={1cm 1.4cm 1.25cm 1.75cm}, clip,width=0.7\textwidth]{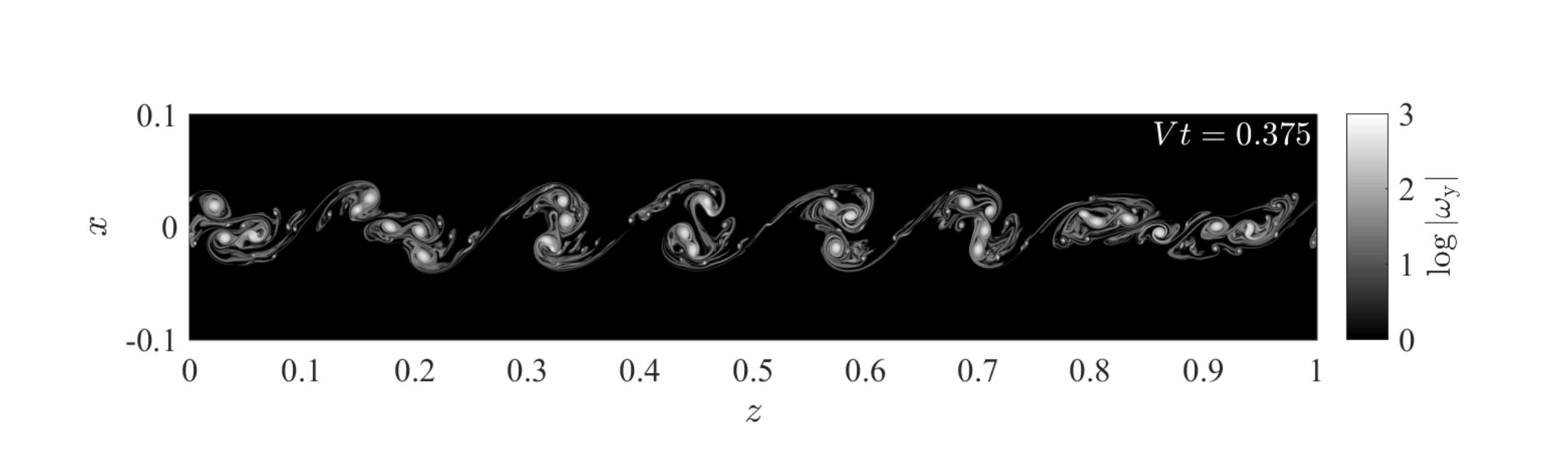}
	}\\
	\subfloat{
		\includegraphics[trim={1cm 0.8cm 1.25cm 1.75cm}, clip,width=0.7\textwidth]{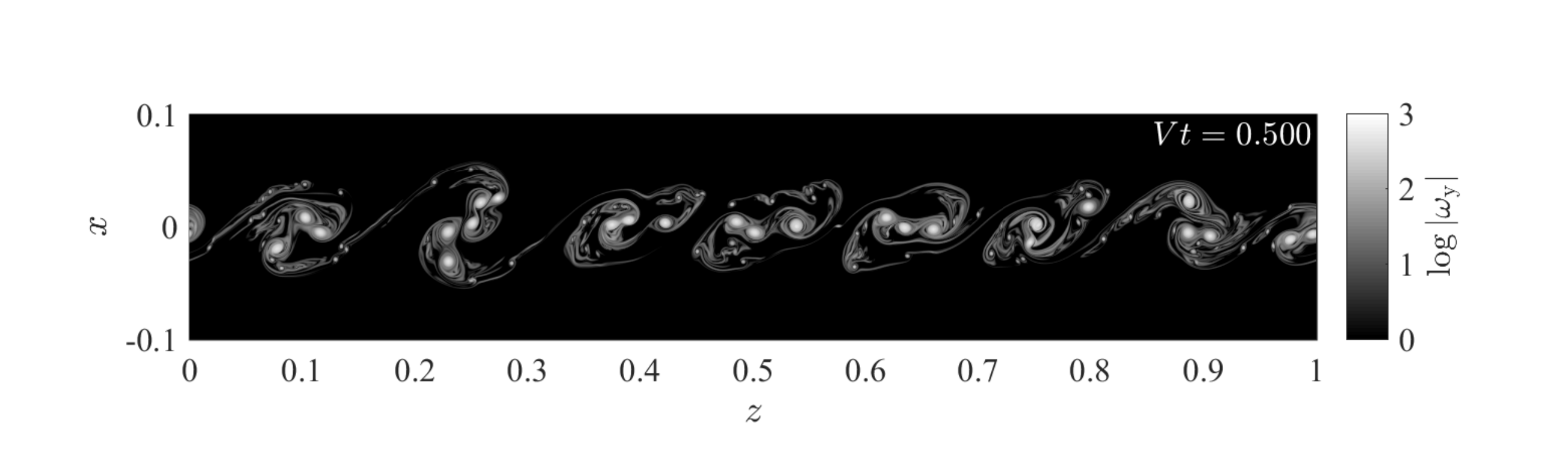}
	}
	\caption{Snapshots of vorticty, $\vorty$, at different times, represented by the self-similar coordinate, $Vt$. The snapshots are taken from the same run as in \Cref{fig:demo-eddies-color}.}
	\label{fig:demo-eddies-vorticity}
\end{figure*}

\Cref{fig:demo-eddies-color} shows the evolution of a multi-mode initial perturbation via vortex pairing in a sequence of snapshots from one of our planar sheet simulations. The small initial perturbations quickly develop into a row of saturated eddies, which then undergo repeated mergers with each other. Each ``generation'' of eddies is characterized by a typical size, which increases with time, while their number decreases. \Cref{fig:demo-eddies-vorticity} shows the vorticity, defined as $\vorty \equiv \frac{\partial \vx}{\partial z}-\frac{\partial \vz}{\partial x}$ (not to be confused with the frequency), in the same snapshots. High-vorticity peaks in \Cref{fig:demo-eddies-vorticity} correspond to the centers of eddies in \Cref{fig:demo-eddies-color}. 

\begin{figure}
	\centering
	\includegraphics[trim={0cm 1.5cm 0.5cm 1.5cm}, clip, width=0.475\textwidth]{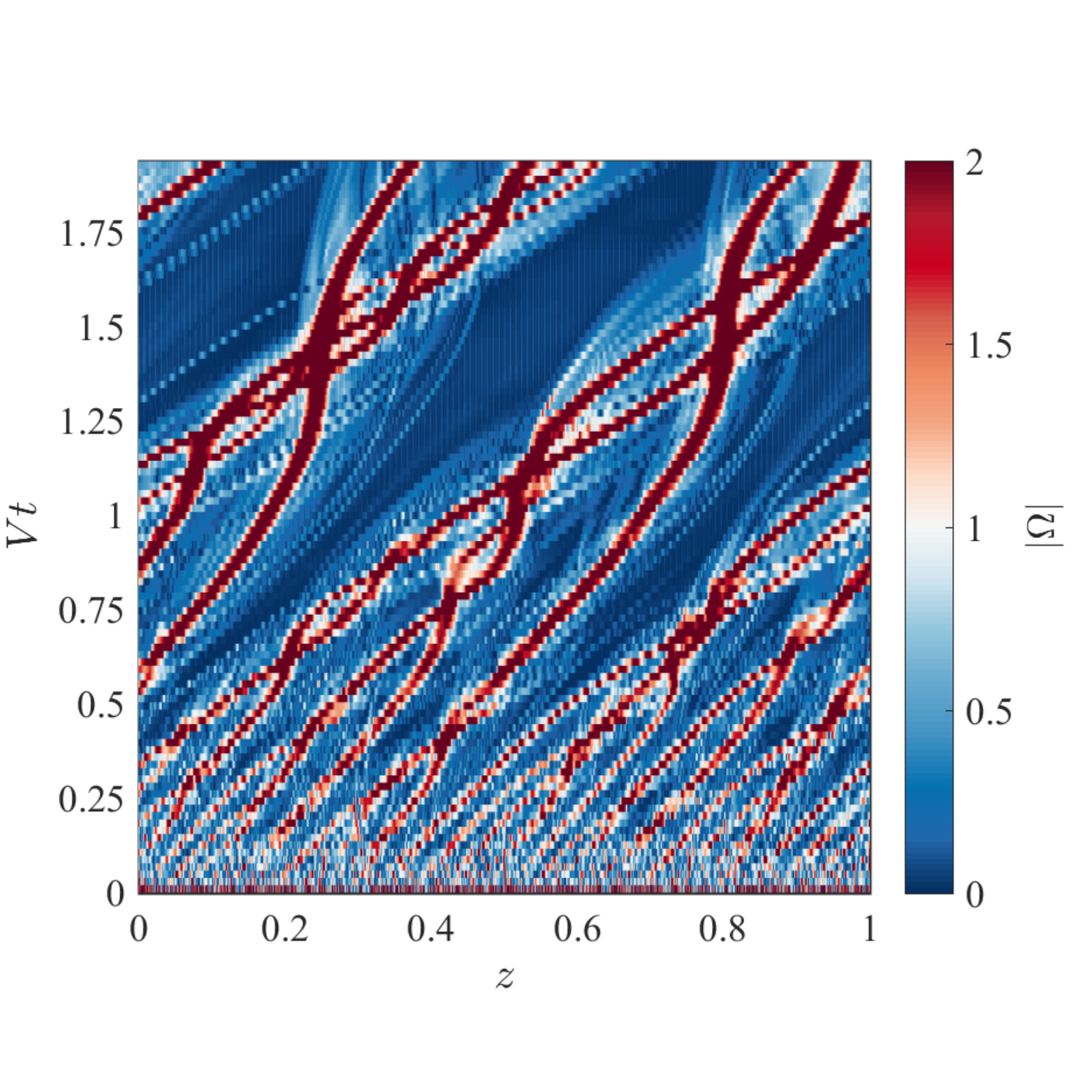}
	\caption{Average vorticity profiles as a function of time, represented by the self-similar coordinate, $Vt$, and the streamwise position, $z$, in a planar sheet simulation. The color corresponds to magnitude of the average vorticity, $|\Omega|$, where $\Omega$ is the volume average of $\vorty$ along the $x$ axis. The red streaks mark the trajectory of the dominant eddies, producing a ``merger tree''-like plot. The initial conditions are identical to \Cref{fig:demo-eddies-color,fig:demo-eddies-vorticity}, but the resolution was reduced by a factor of 4 to reduce run time.}
	\label{fig:eddie_merger_tree}
\end{figure}

In \Cref{fig:eddie_merger_tree}, we plot the average vorticity along the $x$ axis, $\Omega \equiv \int{\vorty(x,z,t)\dx} \;\big\slash \int{\dx}$, as a function of the $z$-coordinate and of time. By tracing the trajectory of high-vorticity peaks along the $z$ axis, we can follow the evolution of the vortex population in a ``vortex merger tree'', similar to the plots extracted from experiments by \citet{Brown1974} and from models by \citet{Rikanati2003}. This representation highlights the decrease in the number of vortices with time. In addition,  \Cref{fig:eddie_merger_tree} shows that, on average, the eddies ``drift'' downstream at a characteristic velocity, corresponding to the ``convection velocity'' discussed in \Cref{sec:surface-results-convection}.

\begin{figure*}
	\centering
	\subfloat{		
		\includegraphics[trim={1cm 1.4cm 1.25cm 1.75cm}, clip,width=0.7\textwidth]{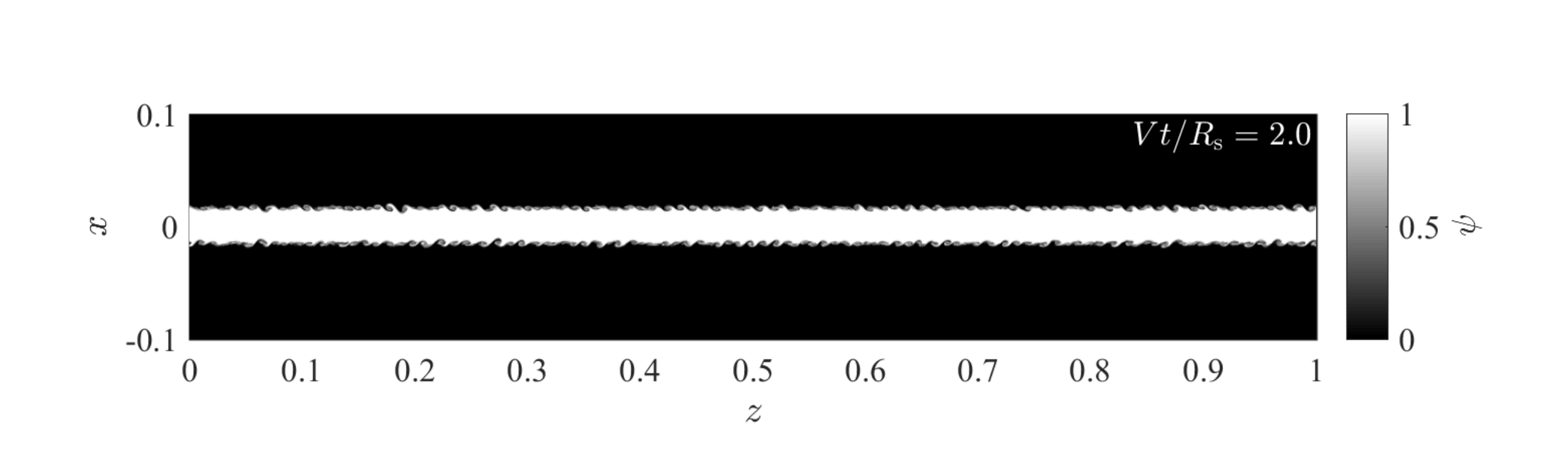}
		\label{fig:demo-eddies-slab0}
	}\\
	\subfloat{
		\includegraphics[trim={1cm 1.4cm 1.25cm 1.75cm}, clip,width=0.7\textwidth]{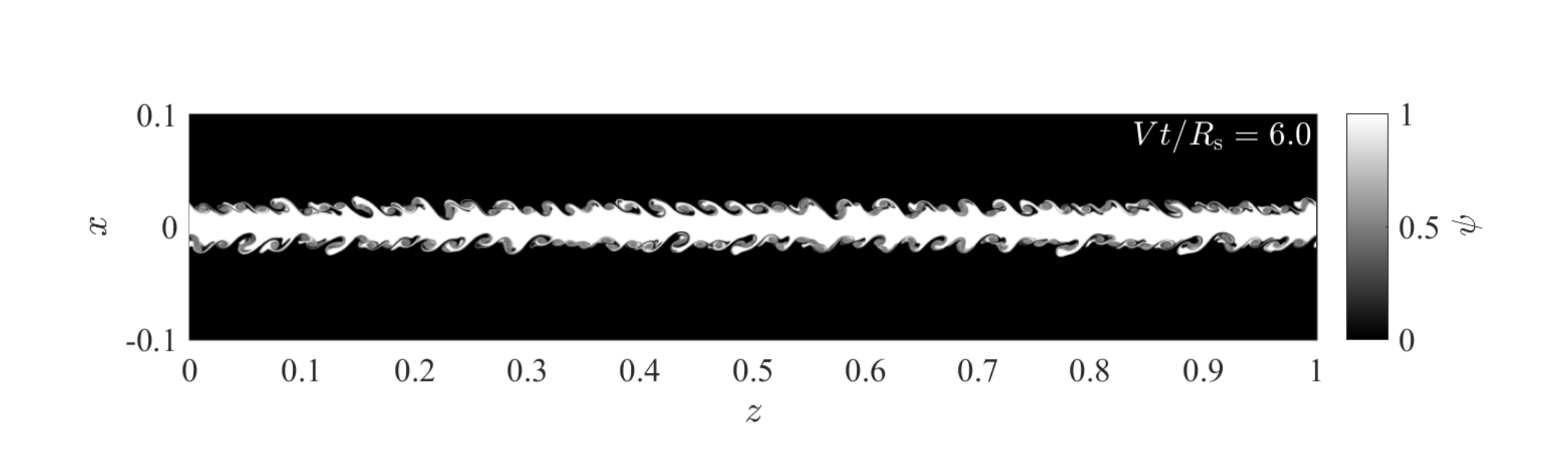}
		\label{fig:demo-eddies-slab1}
	}\\
	\subfloat{
		\includegraphics[trim={1cm 1.4cm 1.25cm 1.75cm}, clip,width=0.7\textwidth]{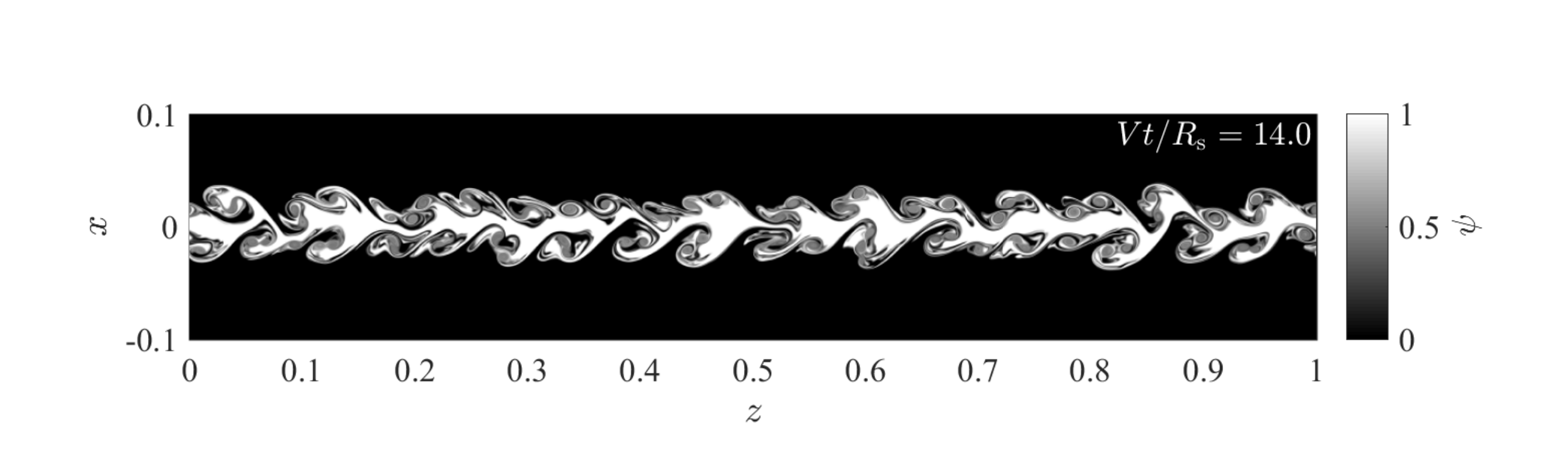}
		\label{fig:demo-eddies-slab2}
	}\\
	\subfloat{
		\includegraphics[trim={1cm 1.4cm 1.25cm 1.75cm}, clip,width=0.7\textwidth]{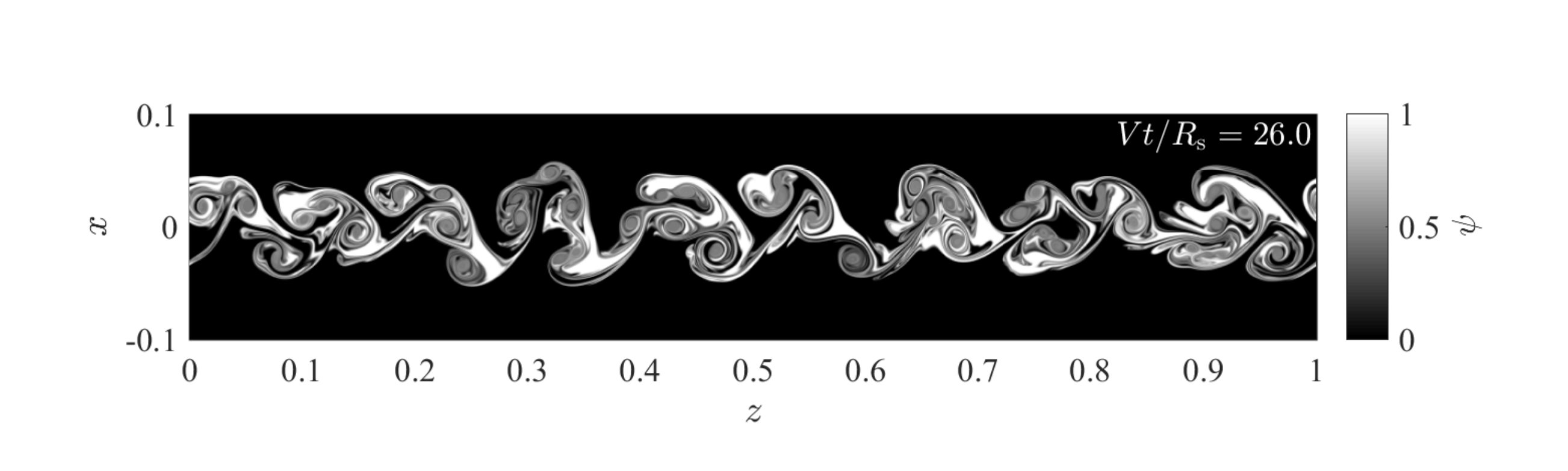}
		\label{fig:demo-eddies-slab3}
	}\\
	\subfloat{
		\includegraphics[trim={1cm 0.8cm 1.25cm 1.75cm}, clip,width=0.7\textwidth]{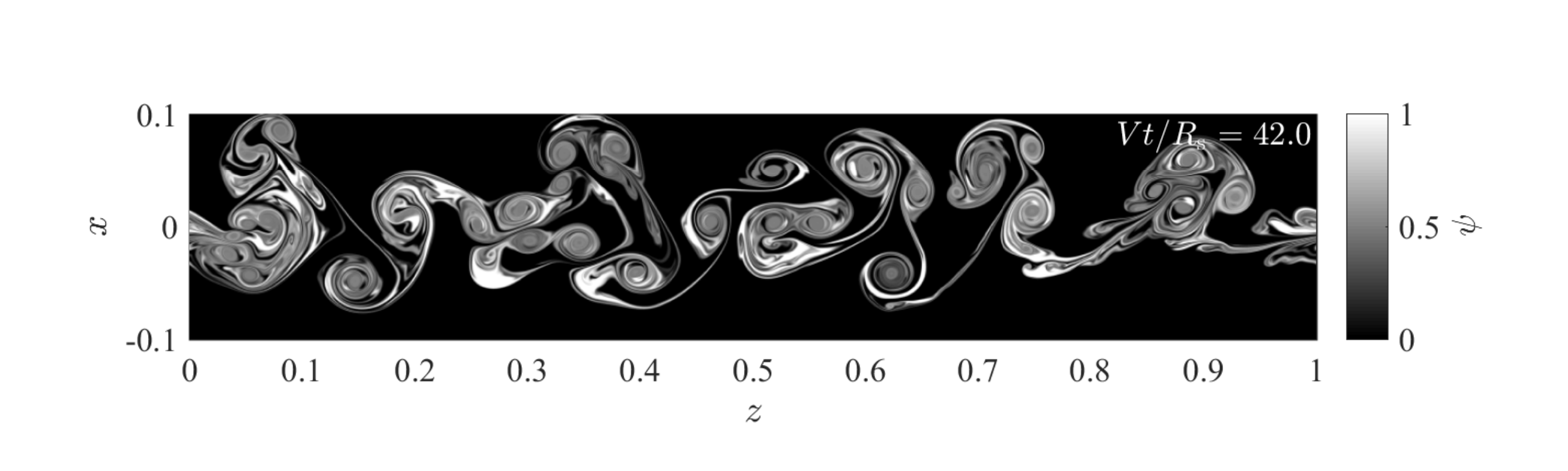}
		\label{fig:demo-eddies-slab4}
	}
	\caption{As in \Cref{fig:demo-eddies-color}, but for a planar slab simulation where the times are rescaled by the dynamic time $\tRs=\Rs/V$. The unperturbed initial conditions are $(\Mb=0.5,\delta=1)$, placing the stream in the surface-mode-dominated regime. The initial stream radius is $\Rs=1/64$. The initial perturbations are interface-only with a broadband white noise spectrum (see \Cref{sec:surface-simulations}).}
	\label{fig:demo-eddies-slab}
\end{figure*}

In a periodic or infinite row of identical vortices with equal spacing, symmetry dictates that any vortex experiences zero net attraction to its neighbors. Namely, the attraction induced on some vortex by another vortex in the row is always canceled out due to an identical vortex at an equal distance in the opposite direction. Therefore, in an ideal single-mode scenario, the eddies remain stationary and never merge, producing the saturated solution,  \Cref{eq:saturated-solution}. An immediate corollary is that any slight deviation from perfect symmetry will result in net attraction between the  vortices, triggering a transition to growth via vortex mergers \citep{Rikanati2003} at a finite time and eventually achieving the self-similar solution of \Cref{eq:self-similar-solution}. In this sense, self-similar growth can be seen as the only long-lived solution possible in a non-ideal system.

The inflow of gas in cold streams into galactic halos occurs in a noisy environment, riddled with perturbations at various scales, due to minor mergers, galactic winds, filament-filament interactions and other sources. It is therefore reasonable to assume that nonlinear KHI in cold streams lacks the fine-tuning required for maintaining the saturated solution for any significant length of time. Hence, we focus our attention on the problem of \emph{self-similar shear layer growth}, characterized by \Cref{eq:self-similar-solution}.

The discussion so far assumed a planar sheet geometry. We now extend our argumentation to the problem of nonlinear surface modes in a planar slab. Consider a planar slab in the surface-mode-dominated regime, perturbed with a multi-mode initial perturbation at both interfaces. Since the linear growth rate of surface modes decreases as the wavelength is made larger\footnotemark, nonlinear evolution begins with the smallest wavelength perturbations. For sufficiently short wavelengths, $\lambda\ll\Rs$, the slab geometry can be ignored altogether, because the additional boundary of the stream should not affect perturbations at much smaller scales. Therefore, the description of shear layer growth in a planar sheet can, by and large, also be applied to a planar slab at early times, $h\ll\Rs$. We expect to see shear layers growing independently on opposite sides of the slab, subject to the self-similar scaling \Cref{eq:self-similar-solution} with the same growth rate as in a planar sheet. This is demonstrated in the top two panels of \Cref{fig:demo-eddies-slab}, showing how a multi-mode initial perturbation develops into two independent shear layers on both sides of a planar slab unstable to surface modes. 

\footnotetext{Both the slab and the sheet growth rates vanish as $\lambda\to\infty$. A careful analysis of the long wavelength limit in slab geometry, $\lambda\gg\Rs$, reveals that the decay is more rapid than implied by \Cref{eq:sheet-tkh}, with $\tkh \propto \lambda^{1.5}$ instead of $\tkh \propto \lambda$ \citepalias[see Section 2.3 in][]{Mandelker2016b}. This only serves to reinforce the argument above.}

In principle, the argument above breaks down when the shear layer thickness is on the order of the slab thickness, $h\approx\Rs$. Therefore, we may expect a deviation from the planar sheet growth rates as $h\to\Rs$. Moreover, since eddies on opposite sides of the slab are counter-rotating, they repel rather than attract each other \citep{Paterson1984}. This leads to more complex vortex trajectories, due to the combination of attractive interactions between neighbors and repulsive interactions between opposite counterparts. Nevertheless, further vortex mergers are inevitable, driving the continued growth of the perturbed region as shown in the bottom three panels of \Cref{fig:demo-eddies-slab}. This process continues until the initial kinetic energy of the stream has been fully converted into turbulent motion and heat, with subsequent mixing absorbing the stream into the background fluid. Dimensional analysis shows that the relevant timescale for the terminal evolution of the stream is the typical turnover time for an eddy at the scale of the stream radius,

\begin{equation}
\label{eq:slab-surface-timescale}
\tRs = \frac{\Rs}{V}.
\end{equation}


\subsection{Simulations}
\label{sec:surface-simulations}

The range of parameters studied in the numerical simulations presented in this section is listed in \Cref{tab:parameters-surface}. These span the range of density contrast and Mach number relevant to cosmic cold streams with $0.5<\Mb<1.5$ and $1<\delta<100$, and include a few effectively incompressible cases with $\Mb=0.05$. In most cases, we simulated both a planar sheet and a planar slab. The combination $(\Mb=1.5,\delta=100)$ is skipped because it is stable to surface modes by virtue of \Cref{eq:sheet-instability-condition}.
 
\begin{table} 
	\centering
	\caption{Parameters of simulations studying the nonlinear evolution of surface modes. The first four entries correspond to effectively incompressible planar sheet runs. The next eight entries are compressible combinations of $\unpert$ for which we performed both planar sheet and planar slab simulations. The last three entries are planar slab simulations which study the scaling with $\Rs$, the boundary effect and convergence with respect to $\Rs/\Delta$. We performed three simulation runs with different realizations of broadband white noise interface-only perturbations (see \Cref{tab:spectra-surface}) in all cases, except for the last three entries where only one realization was tested. In addition, we repeated the case $(\Mb=0.5,\delta=1)$ with different functional forms of initial perturbations, as described in \Cref{sec:surface-results-initial-perturbations}. The smoothing width parameter in \Cref{eq:smoothing} was $\sigma=\Delta$ in all cases. }
	\label{tab:parameters-surface}
	\begin{tabularx}{\columnwidth}{XXXXXX}
		\hline
		$\Mb$ &$\delta$ &$\Mtot$ &$\Rs$ &$L$ &$\Rs/\Delta$ \\
		\hline
		0.05 &1   &0.025 &N/A   &1.0 &N/A \\
		0.05 &5   &0.034 &N/A   &1.0 &N/A \\ 
		0.05 &20  &0.041 &N/A   &1.0 &N/A \\ 
		0.05 &50  &0.044 &N/A   &1.0 &N/A \\
		\hline
		0.50 &1   &0.250 &1/64  &1.0 &128 \\ 
		0.50 &10  &0.380 &1/64  &1.0 &128 \\ 
		0.50 &100 &0.454 &1/64  &1.0 &128 \\ 
		1.00 &1   &0.500 &1/64  &1.0 &128 \\ 
		1.00 &10  &0.760 &1/64  &1.0 &128 \\ 
		1.00 &100 &0.909 &1/64  &1.0 &128 \\ 
		1.50 &1   &0.750 &1/64  &1.0 &128 \\
		1.50 &10  &1.140 &1/64  &1.0 &128 \\
		\hline
		1.00 &10  &0.760 &1/128 &1.0 &64  \\ 	  	
		1.00 &10  &0.760 &1/64  &2.0 &128 \\
		1.00 &10  &0.760 &1/32  &2.0 &256 \\ 	  	 	  	
		\hline
	\end{tabularx}
\end{table}

\begin{table} 
	\centering
	\caption{Parameters of initial perturbation spectra. In all cases we used a fixed amplitude for all wavelengths. For planar sheet simulations with a broadband or a sparse noise spectrum, the displacement amplitude was $\Hj=\Delta$ and the velocity amplitude was $\Wj = (2\Hj/L) V = (2\Delta/L)V$. For narrowband white noise we used $\Hj=2\Delta$. For planar slab simulations, all amplitudes were reduced by a factor of $4$ in order to keep the total displacement as small as possible compared to the slab width, $h=\sum_{j} \hj \ll \Rs$.} \label{tab:spectra-surface}
	\begin{tabularx}{\columnwidth}{Xcccc}
		\hline
		Name & $\#$ Modes & $\jmin$ & $\jmax$ & $\Dj$ \\
		\hline
		Broadband White Noise & 897 & 128 & 1024 & 1 \\
		Narrowband White Noise & 129 & 128 & 256 & 1 \\ 
		Sparse White Noise A & 113 & 128 & 1024 & 8 \\ 
		Sparse White Noise B & 97 & 256 & 1024 & 8 \\ 
		Sparse White Noise C & 49 & 128 & 512 & 8 \\  
		\hline
	\end{tabularx}
\end{table}

Each simulation run was initialized with a random realization of a set of periodic perturbations. We used different types of initial perturbations, varying by functional form and by power spectrum. To comply with periodic boundary conditions, all wavelengths were harmonics of the box length, $\kj = 2\pi j$, with $j\geq1$ an index representing an individual perturbation mode. To generate a specific realization we assigned each mode a random phase, $\phij\in\left[0,2\pi\right)$.

For a given run, we used one of three functional forms for the perturbations:
\begin{itemize}
	\item \emph{``Interface-Only''} - each mode is a cosine perturbation to the $x$-position of the interface, $\hj(z) = \Hj \cos{\left(\kj z+\phij\right)}$, where $\Hj$ is the displacement amplitude.
	\item \emph{``Velocity-Only''} - each mode is a cosine perturbation to the transverse velocity decaying exponentially away from the interface, ${\vec {v}}_{\rm j}(x,z) = \Wj\cos{\left(\kj z+\phij\right)}\exp{\left(-\kj |x|\right)}\hat{z}$, where $\Wj$ is the velocity amplitude.
	\item \emph{``Full Eigenmodes''} - each mode includes the same cosine perturbation to the interface as in the first case, as well as self-consistent perturbations to the density, pressure and velocities according to the eigenmode solution of the linear problem \PIp.
\end{itemize}

The different spectra we used appear in \Cref{tab:spectra-surface}. In each case, some subset of modes $j \in \left\{\jmin,\jmin+\Dj,\jmin+2\Dj,\ldots,\jmax\right\}$ was given a nonzero amplitude, identical for all $j$'s. If $\Dj=1$ all modes in the range $\jmin\leq j \leq\jmax$ are included, representing white noise within this band, whereas $\Dj>1$ corresponds to a sparse subset of modes. 

In \Cref{sec:surface-results-initial-perturbations} we demonstrate that the key results of this work exhibit little sensitivity to the chosen type of initial perturbations. Subsequently, most of our results will be obtained using realizations of interface-only perturbations with a broadband white noise spectrum.

Numerical simulations were previously used to study shear layer growth in various other contexts. Particular attention was devoted to studying the effect of compressibility on temporal shear layer growth in 2D and 3D simulations  \citep[e.g.][]{Sandham1994,Vreman1996,Pantano2002,Mahle2007,Foysi2010}. Other work presented simulations of spatial growth in planar and cylindrical geometry \citep[e.g.][]{Wilson1994,Freund2000,Laizet2010,Bogey2011,Zhou2012}. The validity of these investigations notwithstanding, this work improves upon them in a number of ways. We explore cases with very high density contrast, up to $\delta=100$, whereas previous studies concentrated on $\delta\leq10$ and are therefore of lesser relevance to cosmic cold streams. In addition, we present the first comprehensive study of the deceleration induced on a slab due to shear layer growth and the first comparison of shear layers in sheet and slab geometries. 

Shear layer growth was also studied extensively in laboratory experiments (e.g. \citealt{Browand1966}; \citealt{Brown1974}; \citealt{Winant1974}; \citealt{Dimotakis1976}; \citealt{Konrad1977}; \citealt{Chinzei1986}; \citealt{Papamoschou1988}; \citealt{Hall1993}; \citealt{Slessor1998}; \citealt{Rossmann2002}; additional references in \citealt{Freeman2014}). These experiments are limited to $\delta\leq7$, so their conclusions are not directly applicable to the astrophysical application in mind, further motivating our present work. Nevertheless, since we compare our numerical results to the findings of these investigations, we briefly describe the experimental systems and how they differ from our simulations. In a typical experiment, two flows of unequal velocities are separated by a wall or a wedge, parallel to the flow direction. The wall ends at some point, bringing the flows into contact and thus initiating the growth of the instability. A perturbed region is observed to extend downstream from this point of origin, its width growing linearly with the downstream coordinate, $h \propto z$. The spatial growth rate is defined as
\begin{equation}
\label{eq:h'}
h' = \frac{{\rm d}h}{\dz}
\end{equation}
Experiments therefore typically study \emph{spatial} evolution, in contrast to most theoretical work, including our own, which studies \emph{temporal} evolution. While these two classes of problems can be related using the proper transformation (see \Cref{sec:surface-results-convection}), allowing us to compare our results to available experimental data, one important difference cannot be bridged: while temporal systems are invariant to Galilean transformations, spatial systems are not, because the laboratory frame of reference is unique in being stationary with respect to the origin of the perturbations. This introduces an additional governing parameter to the spatial case, on top of the parameters controlling the temporal case. A typical choice for this parameter \citep[e.g.][]{Dimotakis1991} is the ratio of background velocity to stream velocity, evaluated in the laboratory reference frame, 
\begin{equation}
\label{eq:velocity-ratio}
r \equiv \frac{V_{\rm b}}{V_{\rm s}}.
\end{equation}
Spatial growth experiments are thus parametrized by the Mach number, $\Mb$, the density contrast, $\delta$, and the velocity ratio, $r$. Temporal shear layer growth in our numerical simulations depends on the former two quantities alone.  


\subsection{Results}
\label{sec:surface-results}

Motivated by the qualitative picture presented in \Cref{sec:surface-qual} and using the set of simulations described in \Cref{sec:surface-simulations}, this section addresses the following quantitative questions:
\begin{itemize}
	\item How sensitive are shear layer growth rates to the detailed properties of the initial perturbations? Is there a ``universal'' growth rate?
	\item How does the growth rate depend on the unperturbed conditions, namely density contrast, $\delta$, and Mach number, $\Mb$, in the parameter range relevant to cold streams?
	\item Does the growth of the shear layer differ in the stream and the background? How does the ratio $\hs/\hb$ (see \Cref{sec:methods-analysis}) depend on $\unpert$?
	\item How does the growth rate in a planar slab diverge from the planar sheet solution as the shear layer thickness becomes comparable to the slab width, $h\approx\Rs$?
	\item For slabs, what is the rate of deceleration of the stream fluid due to shear layer growth as a function of  $\unpert$?
	\item For slabs, how much of the stream kinetic energy is converted into stream internal energy (which can subsequently be emitted as Lyman-$\alpha$ radiation)? How does this depend on $\unpert$?
\end{itemize}
The answers to these questions, presented below, are used in \Cref{sec:application} to predict the outcome of KHI in cold streams feeding massive SFGs at high redshift.

\subsubsection{Weak Dependence on Initial Perturbations}
\label{sec:surface-results-initial-perturbations}

\begin{figure}
	\centering
	\includegraphics[trim={0cm 0.5cm 1.0cm 1.0cm}, clip, width=0.475\textwidth]{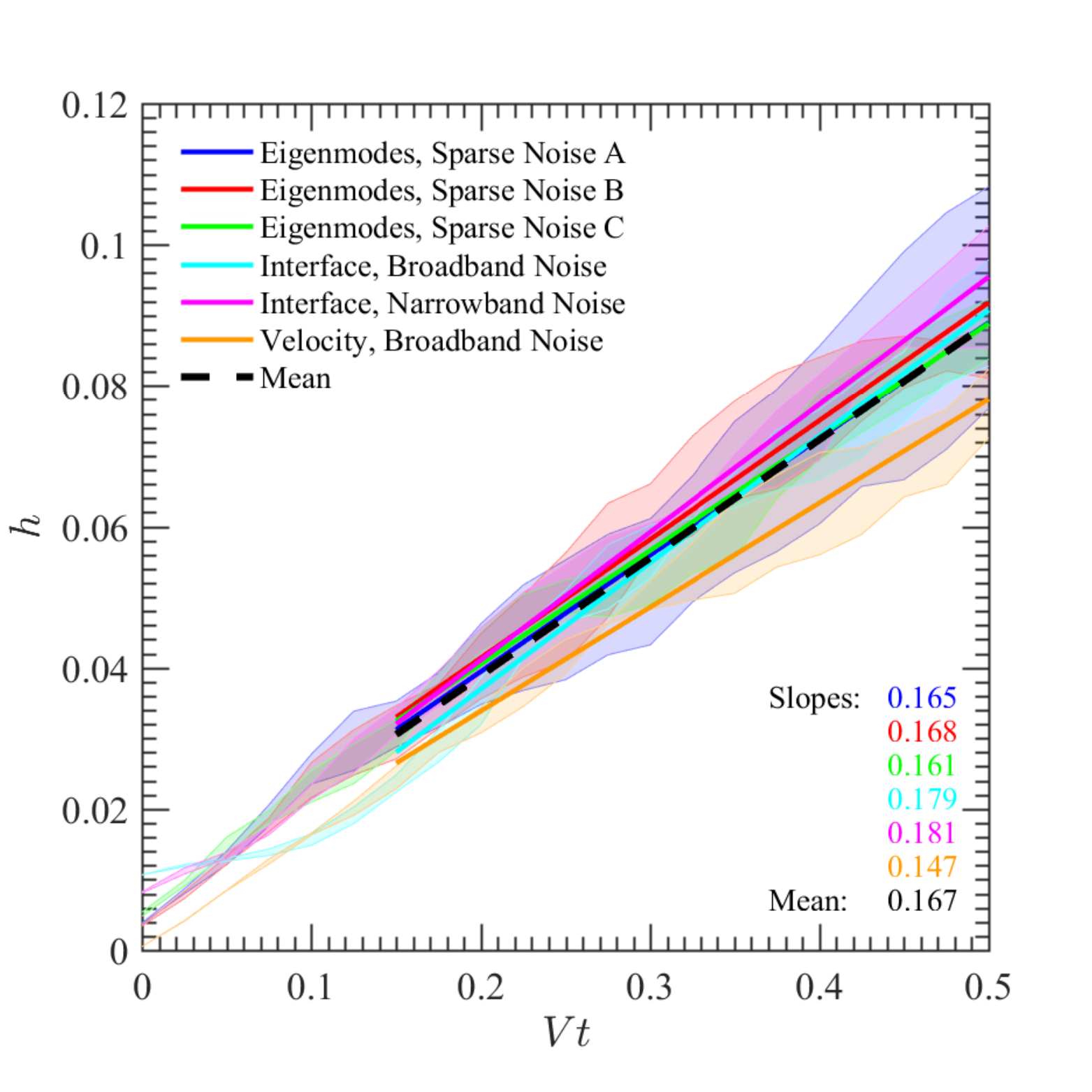}
	\caption{Shear layer thickness, $h$, defined in \Cref{eq:h}, for different initial perturbations in planar sheet simulations with $\Mb=0.5$ and $\delta=1$. The colors denote different spectra and functional forms for the initial perturbations (see \Cref{sec:surface-simulations}). For each of these cases we ran three realizations, spanning the shaded areas of different colors. A linear curve was fitted to the mean thickness of each three realization at times $Vt \geq 0.15$, plotted in solid lines. The dashed black line is a linear fit to the overall mean, obtained by averaging over the colored lines. The slopes of all the linear curves, representing the dimensionless growth rates, $\alpha$, are printed in the plot.}
	\label{fig:ic_independence}
\end{figure}

Self-similar shear layer growth is expected to become independent of the initial perturbations at late times. We demonstrate this behavior in \Cref{fig:ic_independence}, which shows the shear layer thickness as a function of $Vt$ in planar sheet simulations with equal densities, $\delta=1$, and relatively low Mach number, $\Mb=0.5$, seeded with various types of initial perturbations. The results for other values of $\unpert$ show a similarly weak dependence on the details of the initial perturbations.

The different initial perturbations shown in \Cref{fig:ic_independence} have a variety of initial thicknesses and evolve differently at early times. For example, interface-only perturbations initially grow at a slower rate than both eigenmode and velocity-only perturbations, since the former lack a disturbance to the bulk flow at $t=0$. The perturbations to velocity and pressure take finite time to develop \citepalias[see discussion in Section 3.3 of][]{Mandelker2016b}, thus delaying the growth of an interface-only perturbation at early times.

At later times, $Vt \gsim 0.15$, the different cases converge to a universal linear curve. The mean growth rates obtained for different types of initial perturbations are within $\pm 10\%$ of the overall mean. While there are systematic differences between different types of initial perturbations, these are small and comparable to the span of growth rates among different realizations of the same kind. This $\pm 10\%$ scatter can be considered as the typical uncertainty of our numerical analysis.  It must be stressed that this scatter is largely a result of the finite size of the simulation domain; at late times, the number of eddies in the simulation is rather small, typically $10$ to $20$, so the growth rate becomes dominated by stochastic variations. For an infinite domain, the scatter is expected to vanish. 

In the following discussion, we assume that shear layers in cold streams reach their asymptotic, ``universal'' growth rate early during the infall. This assumption makes the outcome of the instability independent of whatever complex conditions seed the initial perturbations. 

\subsubsection{Density Contrast and Mach Number Scaling}
\label{sec:surface-results-density-mach}

\begin{figure*}
	\centering
	\subfloat{
		\includegraphics[trim={0cm 0.5cm 1cm 1.25cm}, clip, height=0.35\textheight]{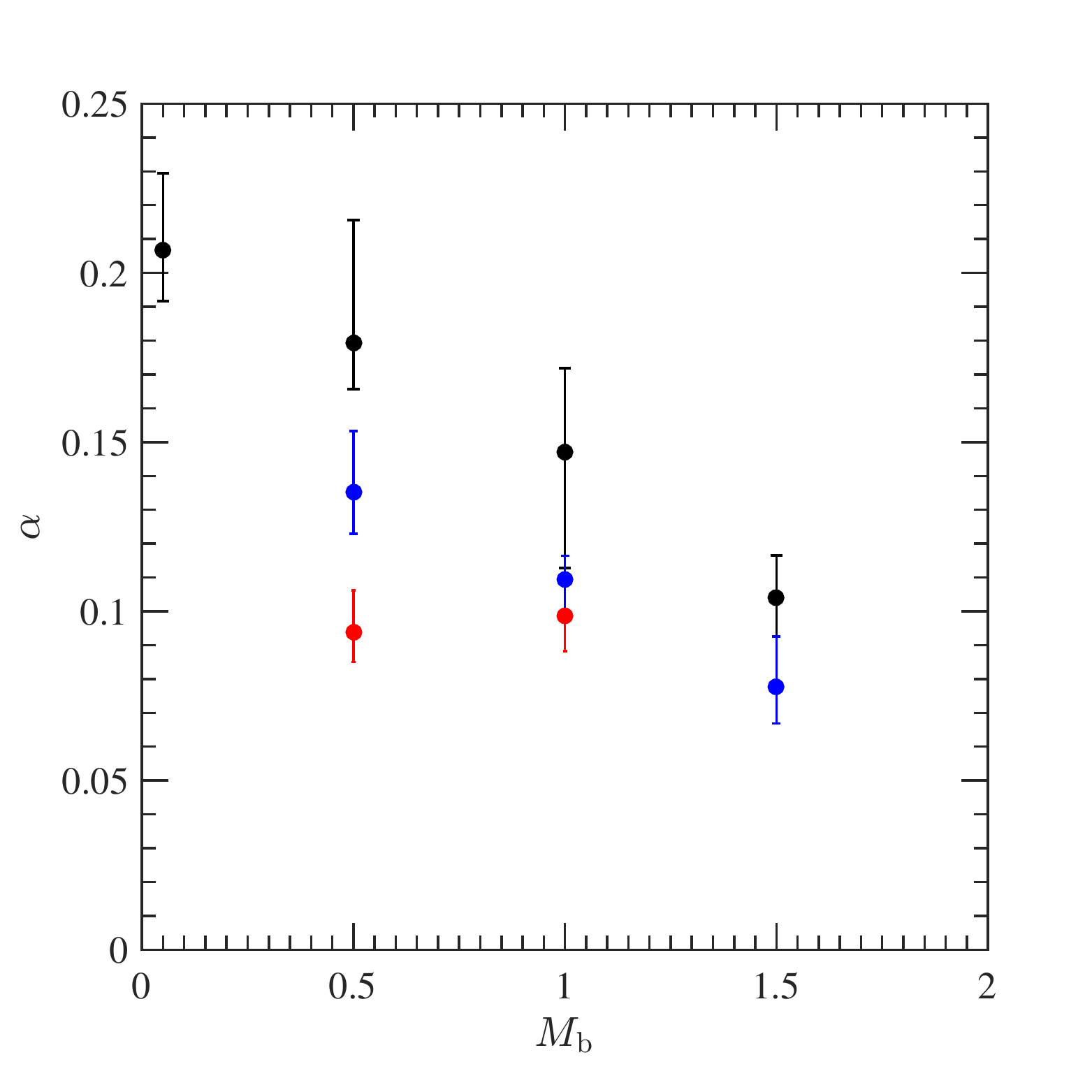}
	}
	\subfloat{
		\includegraphics[trim={0.75cm 0.5cm 1cm 1.25cm}, clip, height=0.35\textheight]{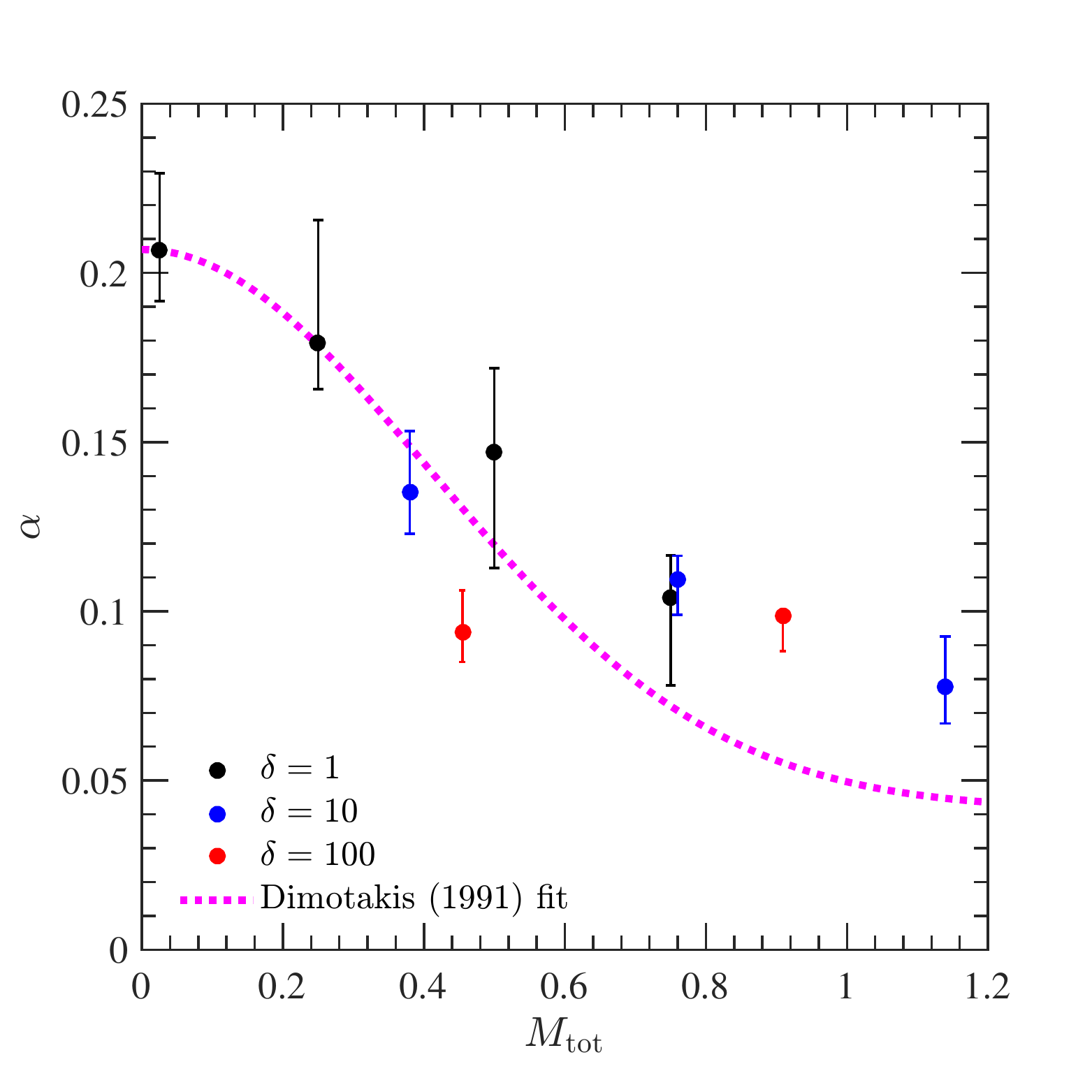}
	}
	\caption{The dimensionless growth rate, $\alpha$, in planar sheet simulations with different unperturbed initial conditions parametrized by $\unpert$. Both panels show the same data, plotted with different horizontal axis: $\Mb$ in the left-hand panel and $\Mtot$ in the right-hand panel. The colors denote different values of $\delta$. One incompressible case is shown  $(\Mb=0.05,\delta=1)$ as well as eight compressible cases: $\Mb=0.5,1.0,1.5$ for $\delta=1,10$ and $\Mb=0.5,1.0$ for $\delta=100$. For each of these cases we ran three realizations of broadband white noise interface-only perturbations. The circles mark the mean growth rate $\alpha$ while the errorbars bound the best fit values of different realizations. The dotted magenta line on the right panel is a curve fitted to experimental data, \Cref{eq:alpha-dimotakis}.}
	\label{fig:alpha}
\end{figure*}

Encouraged by the fact that the growth rate depends weakly on the specific initial perturbations, we turn our attention to its dependence on the unperturbed conditions. The left-hand panel in \Cref{fig:alpha} shows the dimensionless growth rate $\alpha$, introduced in \Cref{eq:alpha}, as a function of $\Mb$ for different values of $\delta$. Qualitatively, the scaling with both parameters is similar to the linear regime \citepalias[see Figure 1 in][]{Mandelker2016b}. For a fixed background Mach number, $\alpha$ decreases with the density contrast. For a fixed density contrast, $\alpha$ decreases with the background Mach number, with the exception of $\delta=100$, where the mean growth rate at $\Mb=1.0$ is higher than at $\Mb=0.5$, similar to the results of the linear regime. The incompressible case $(\Mb=0.05,\delta=1)$ is consistent with the known incompressible temporal growth rate, $\alpha \approx 0.2$, observed in experiments  \citep{Brown1974} and statistical models \cite{Rikanati2003}. Note that the errorbars are consistent with the previously cited $\pm 10\%$.

In the incompressible limit, \emph{temporal} shear layer growth is independent of density contrast \citep{Rikanati2003}, i.e. $\alpha(\Mb \to 0, \delta)\simeq 0.2$ is a constant. On the other hand, the \emph{spatial} growth rate, \Cref{eq:h'}, does depend on the density contrast in the incompressible limit \citep{Konrad1977,Dimotakis1986}. Many authors \citep[e.g.][]{Dimotakis1991,Slessor2000,Freeman2014} observed that the \emph{spatial} growth rate can be expressed as the product of a density contrast dependent factor and a compressibility dependent factor. If this observation is also correct for \emph{temporal} shear layers, then the temporal growth rate must be completely independent of density contrast, and can be expressed as a function of a single parameter quantifying compressibility effects. The parameter typically used to scale the compressibility dependence of spatial growth rates is the total Mach number, defined in \Cref{eq:body-instability-condition}.

The right-hand panel in \Cref{fig:alpha} shows the same growth rates as the left-hand panel, plotted against $\Mtot$, defined in \Cref{eq:body-instability-condition}, instead of $\Mb$. Our results support the claim that $\alpha$ depends on compressibility alone, parametrized by $\Mtot$. This behavior is unique to the \emph{nonlinear} phase of temporal KHI, since the dispersion relation for the planar sheet does depend on $\delta$ at the incompressible limit and in general it can not be expressed as function of $\Mtot$ only \citepalias[see equations (18)-(19) in][]{Mandelker2016b}. The subject of compressibility scaling is discussed in further detail in \Cref{app:compressibility-scaling}, including a comparison of the results of this work with previous publications. Our results are roughly consistent with the empirical fit proposed by \citet{Dimotakis1991},
\begin{equation}
\label{eq:alpha-dimotakis}
\alpha_{\rm Dim.} = \alpha_0 \times \bigg[ 0.8 \exp{(-3\Mtot^2)} + 0.2 \bigg],
\end{equation}
using $\alpha_0=\alpha(\Mtot\to0)\simeq 0.21$. As can be seen in \Cref{fig:alpha}, \Cref{eq:alpha-dimotakis} underestimates the growth rate in our simulations by $\sim50\%$ for the parameter range appropriate for cosmic cold streams, $\Mtot \gsim 0.8$. For these values of $\Mtot$, the growth rate varies slowly and can be crudely approximated by a constant value, $\alpha \approx 0.1$, approximately half the incompressible growth rate.

\subsubsection{Convection Velocity}
\label{sec:surface-results-convection}

In a planar sheet, the mean downstream velocity transitions from $\vz(x\to -\infty)=\Vb$ to $\vz(x\to \infty)=\Vs$ over the thickness of the shear layer. The shear layer is populated by large coherent structures (eddies), which drift downstream at approximately the center of mass velocity of the shear layer,
\begin{equation}
\label{eq:Vc_definition}
\Vc \equiv
\frac{\int_{-\hb}^{\hs}\int_{0}^{L} \rho\vz \dz\dx}{\int_{-\hb}^{\hs}\int_{0}^{L} \rho \dz\dx},
\end{equation} 
often called the ``convection velocity''. For incompressible flow, an expression for $\Vc$ can be obtained by recognizing that stagnation points exist between each pair of eddies and requiring continuity in total pressure at these points, i.e. Bernoulli's principle  \citep{Coles1985,Dimotakis1986}. The resulting expression is
\begin{equation}
\label{eq:Vc_general}
\Vc = \frac{\Vb+\Vs}{2} + \left[\frac{\sqrt{\delta}-1}{\sqrt{\delta}+1}\right] \frac{\Vs-\Vb}{2},
\end{equation}
where the first term is the average velocity of the stream and the background and the second term implies the eddies tend to drift more with the dense stream as $\delta$ is increased. In our simulations $\Vb=0$, hence
\begin{equation}
\label{eq:Vc_delta}
\Vc = \frac{\sqrt{\delta}}{\sqrt{\delta}+1}V.
\end{equation}
Intuitively, this satisfies $\Vc(\delta=1)=V/2$ and $\Vc(\delta\to\infty)=V$. 

\begin{figure}
	\centering
	\includegraphics[trim={0.25cm 0.5cm 1cm 1cm}, clip, width=0.475\textwidth]{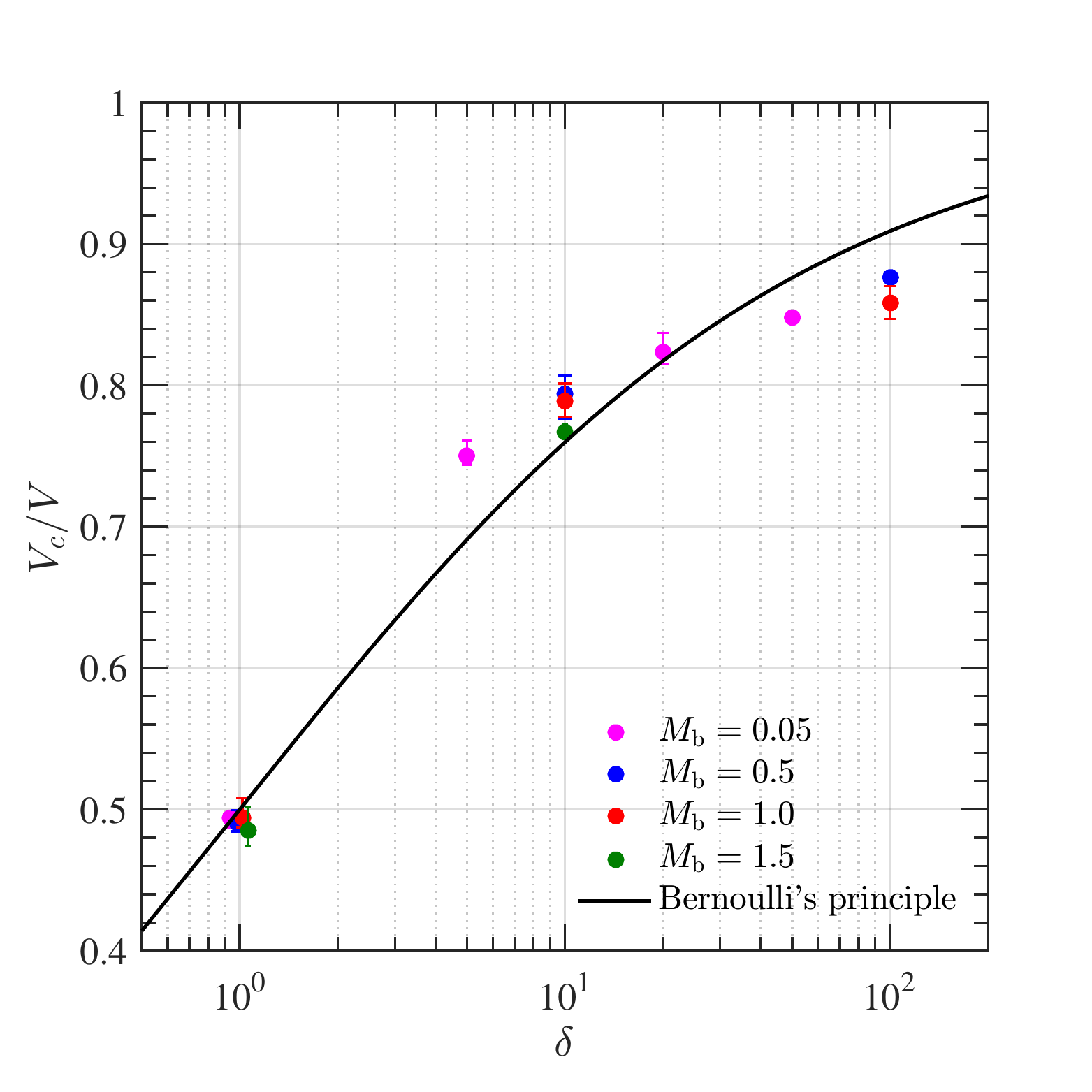}
	\caption{The convection velocity, $\Vc$, estimated using thickness $\hbs$ with $\varepsilon=0.02$ to define the edges of the shear layer in \Cref{eq:Vc_definition}. The colors correspond to different values of $\Mb$. The solid black line corresponds to the incompressible or isentropic prediction, based on Bernoulli's princple \Cref{eq:Vc_delta}. For each combination $\unpert$ we ran three realizations of broadband white noise interface-only perturbations. Circles mark the mean convection velocity. The errorbars are due to the span of velocities for different realizations. The points at $\delta=1$ were shifted slightly along the horizontal axis to make them distinguishable. The values in the figure represent the convection velocity after shear layer growth has become independent of the detailed properties of the initial perturbations (see  \Cref{sec:surface-results-initial-perturbations}). At this stage, the fluctuations in convection velocity are small. } 
	\label{fig:convection-velocity}
\end{figure}

The derivation of $\Vc$ was extended for compressible isentropic (adiabatic and reversible) flow by \citet{Bogdanoff1983} and \citet{Papamoschou1988}. If the adiabatic indices of both fluids are identical, $\gammab=\gammas$ (as they are in our case), the isentropic derivation recovers the incompressible result in \Cref{eq:Vc_delta}. Despite the high Mach numbers, the heating observed in our simulations is quite small (see \Cref{sec:surface-results-heating}), implying nearly isentropic flow, so \Cref{eq:Vc_delta} is expected to predict $\Vc$ reasonably well. This expectation is validated by the results shown in \Cref{fig:convection-velocity}. The convection velocity was deduced from the planar sheet simulations by applying \Cref{eq:Vc_definition} for a few snapshots in which the shear layer is already well-developed and taking the mean value. The fluctuations around the mean are small. For a wide range of Mach numbers, the relation $\Vc(\delta)$ in our simulations is well-approximated by \Cref{eq:Vc_delta}. We note a systematic underestimation of $<5\%$ for $\delta \gsim 20$.

It must be noted that compressible experiments show a significant deviation from \Cref{eq:Vc_delta}, even at Mach numbers comparable to or smaller than those considered in this work \citep{Papamoschou1989}. \citet{Dimotakis1991} proposed a semi-empirical model that reproduced this phenomena by taking into account the formation of shocks. The model is not Galilean-invariant, hinting that the effect observed in experiments may be inherent to spatial, rather than temporal systems. Consequently, the correction to \Cref{eq:Vc_delta} ought to depend on the velocity ratio, $r=\Vb/\Vs$ (see discussion of laboratory experiments in \Cref{sec:surface-simulations}). We are interested specifically in the convection velocity for cold streams feeding galaxies, where the background gas is nearly stationary compared to the cold stream, $\Vb \ll \Vs$, i.e. $r \simeq 0$. Our simulations, which correspond to $r=0$, show good agreement with \Cref{eq:Vc_delta}. We therefore conclude that compressibility-related corrections for \Cref{eq:Vc_delta} are small in the case of our astrophysical scenario, despite the caveat cited here. 

\subsubsection{Transformation to Spatial Growth}
\label{sec:surface-results-spatial}

The convection velocity, obtained in the previous section, can be used to transform back and forth between temporal growth rate and spatial growth rate. Consider a spatial system with velocity ratio $r$ and density contrast $\delta$. Viewed in the frame of reference moving at the velocity of the large structure, $\Vc$, every individual eddy undergoes temporal growth. Returning to the laboratory reference frame, the time it takes an eddy to reach position $z$ is $t=z/\Vc$. Therefore, the predicted amplitude at position $z$ is
\begin{equation}
\label{eq:temporal-to-spatial}
h(z)=\alpha \frac{V}{\Vc} z, 
\end{equation}
and the spatial growth rate, \Cref{eq:h'}, is 
\begin{equation}
\label{eq:spatial-growth-rate}
h' = \alpha \frac{V}{\Vc} = \frac{\sqrt{\delta}+1}{\sqrt{\delta}}\alpha,
\end{equation}
where \Cref{eq:Vc_delta} is used to obtain the last equality, assuming $\Vb=0$ (i.e. $r=0$). This reasoning was used by \citet{Brown1975} and \citet{Rikanati2003} to successfully reproduce the experimental dependence of spatial growth on the density contrast and the velocity ratio\footnotemark.
\footnotetext{\citet{Dimotakis1986} introduced an additional multiplicative factor to \Cref{eq:spatial-growth-rate} accounting for the increasing size of vortices along the downstream coordinate in spatial shear layers. The resulting correction is small for the conditions relevant to our problem, namely $\delta \gsim 10$ and $r\simeq 0$, and is therefore not incorporated into our analysis.}

\begin{figure}
	\centering
	\includegraphics[trim={0.25cm 0.5cm 1cm 1cm}, clip, width=0.475\textwidth]{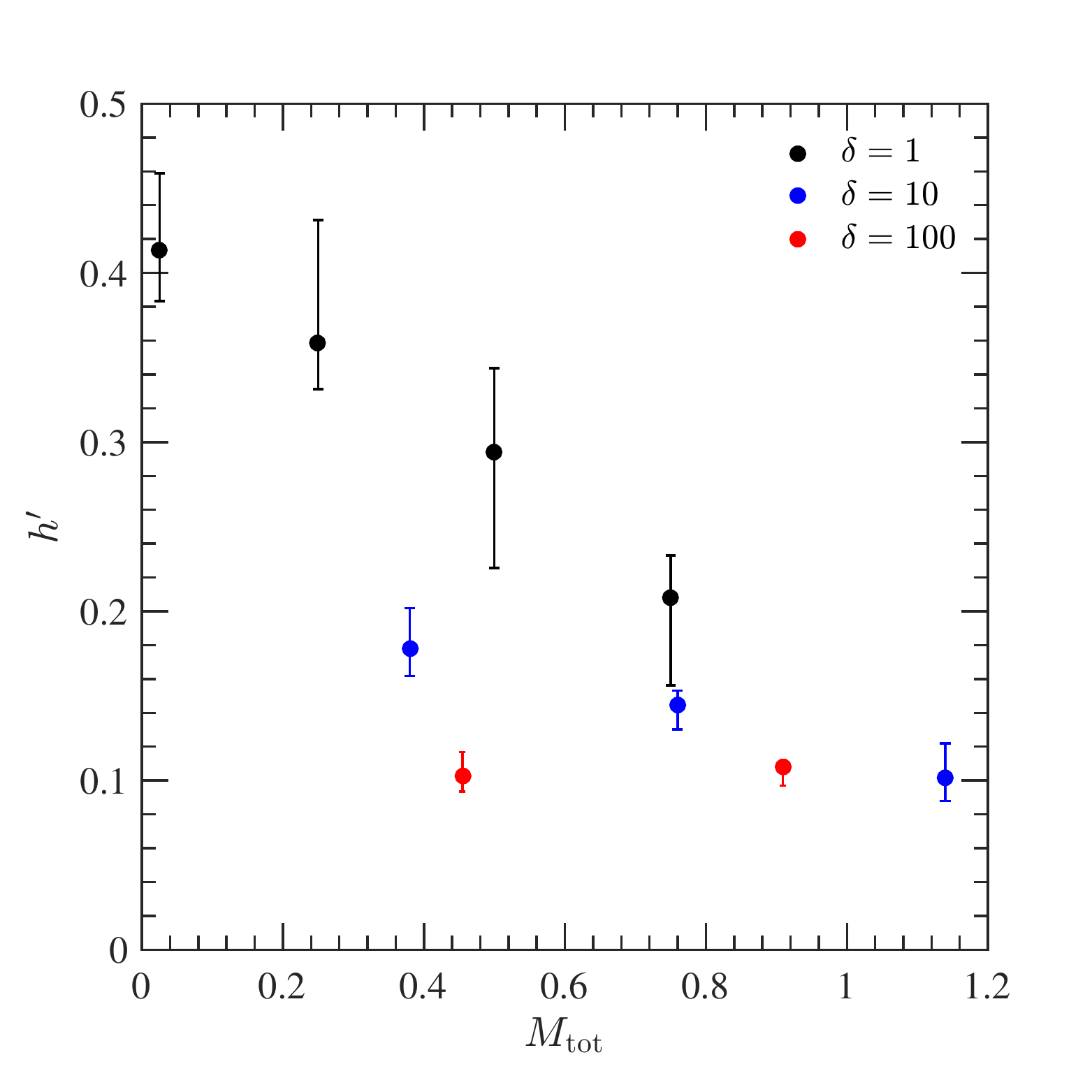}
	\caption{Spatial growth rate, $h'$, based on the temporal growth rates, $\alpha$ in \Cref{fig:alpha}. The transformation from temporal to spatial growth rates was achieved by using \Cref{eq:spatial-growth-rate}.}
	\label{fig:spatial-Mtot}
\end{figure}

\Cref{fig:spatial-Mtot} translates the temporal growth rates from \Cref{fig:alpha} to spatial growth rates, using \Cref{eq:spatial-growth-rate}. The large convection velocity at high density contrast, $\Vc(\delta \gg 1) \approx V$, translates to a factor of $\sim 2$ decrease in the spatial growth rate relative to the $\delta=1$ runs, on top of the decrease in temporal growth rate due to compressibility effects shown in \Cref{fig:alpha}. This brings the predicted spatial growth rate for the conditions of cosmic cold streams down to $h' \approx 0.1$, a factor of $\sim 4$ less than the $(\delta=1,\Mtot\to 0)$ value.

\subsubsection{Entrainment Ratio}
\label{sec:surface-results-entrainment}

As the shear layer grows, either temporally or spatially, it entrains an increasing amount of fluid from the unperturbed regions on both sides. So far, we have essentially discussed the total rate of entrainment (inclusion in the shear layer) from both fluids, without addressing the relative contribution of each component, i.e. the \emph{entrainment ratio}. Experiments show that shear layer growth can be asymmetric, $\hs\ne\hb$, implying an entrainment ratio far from unity \citep{Brown1974,Konrad1977}. Since, in the astrophysical context, we are specifically interested in the disruption of the stream rather than the background, it is important to quantify this asymmetry.

The entrainment ratio can be derived using simple arguments. Consider two fluids with unperturbed densities $\rhobs$ and velocities $\Vbs$,  initially separated by an interface at $x=0$. At some later time, a shear layer has developed around the interface, extending a distance $\hbs$ into the two fluids respectively. Assume only the region $-\hb<x<\hs$ is perturbed, while the flow outside the shear layer maintains its initial conditions.  Conservation of mass between the initial and final state then reads
\begin{equation}
\label{eq:entrain-mass-cons}
\int_{-\hb}^{\hs}\int_{0}^{L} \rho \dz\dx = \rhob\hb L+\rhos\hs L,
\end{equation}
where $L$ is the extent of the computational domain in either direction. Conservation of $z$-momentum yields 
\begin{equation}
\label{eq:entrain-mom-cons}
\int_{-\hb}^{\hs}\int_{0}^{L} \rho\vz \dz\dx = \rhob\hb L\Vb + \rhos\hs L\Vs.
\end{equation}
Substituting equation \Cref{eq:Vc_definition} into \Cref{eq:entrain-mom-cons} and then using \Cref{eq:entrain-mass-cons} we have 
\begin{equation}
\label{eq:entrain-Vc}
\Vc = \frac{\rhos\hs\Vs + \rhob\hb\Vb}{\rhos\hs+\rhob\hb},
\end{equation}
The volume entrainment ratio, $\Ev$, is defined as 
\begin{equation}
\label{eq:entrain-definition}
\Ev \equiv \frac{\hs}{\hb}.
\end{equation}
By substituting \Cref{eq:Vc_general} for $\Vc$ in \Cref{eq:entrain-Vc}, we obtain the entrainment ratio as a function of the dimensionless parameters,
\begin{equation}
\label{eq:entrain-delta}
\Ev = \frac{1}{\sqrt{\delta}}.
\end{equation}
Since \Cref{eq:entrain-mass-cons,eq:entrain-mom-cons} are Galilean invariant, it is not surprising to find that the final result \Cref{eq:entrain-delta} is independent of $r$. Interestingly, \citet{Brown1975} reached the same conclusion based on scaling arguments and empirical data. This prediction is consistent with experimental entrainment ratios reported therein, with fixed $r=0.38$ and varying $\delta$. 

Experiments \citep[e.g.][]{Konrad1977} also indicate a dependence of $\Ev$ on $r$ in spatial systems. \citet{Dimotakis1986} explained this observation and derived an expression for the  entrainment ratio in spatial, incompressible shear layers, $\Evspat(r,\delta)$. Using a modified expression for compressible flow presented in \citet{Dimotakis1991}, we find
\begin{equation}
\label{eq:entrain-dimotakis}
\Evspat(r,\delta \gsim 10,\Mtot \gsim 0.5) \simeq \frac{1}{\sqrt{\delta}}\left(1+0.2\frac{1-r}{1+r}\right).
\end{equation}
For cold streams feeding galaxies, $r=0$, we get $\Evspat \approx 1.2/\sqrt{\delta}$. A similar value was used by \citet{Dimotakis1987}. We therefore propose using \Cref{eq:entrain-delta} for the entrainment ratio in cold streams feeding galaxies, bearing in mind the risk of a small underestimation.

\begin{figure}
	\centering
	\includegraphics[trim={0.25cm 0.5cm 1cm 1cm}, clip, width=0.475\textwidth]{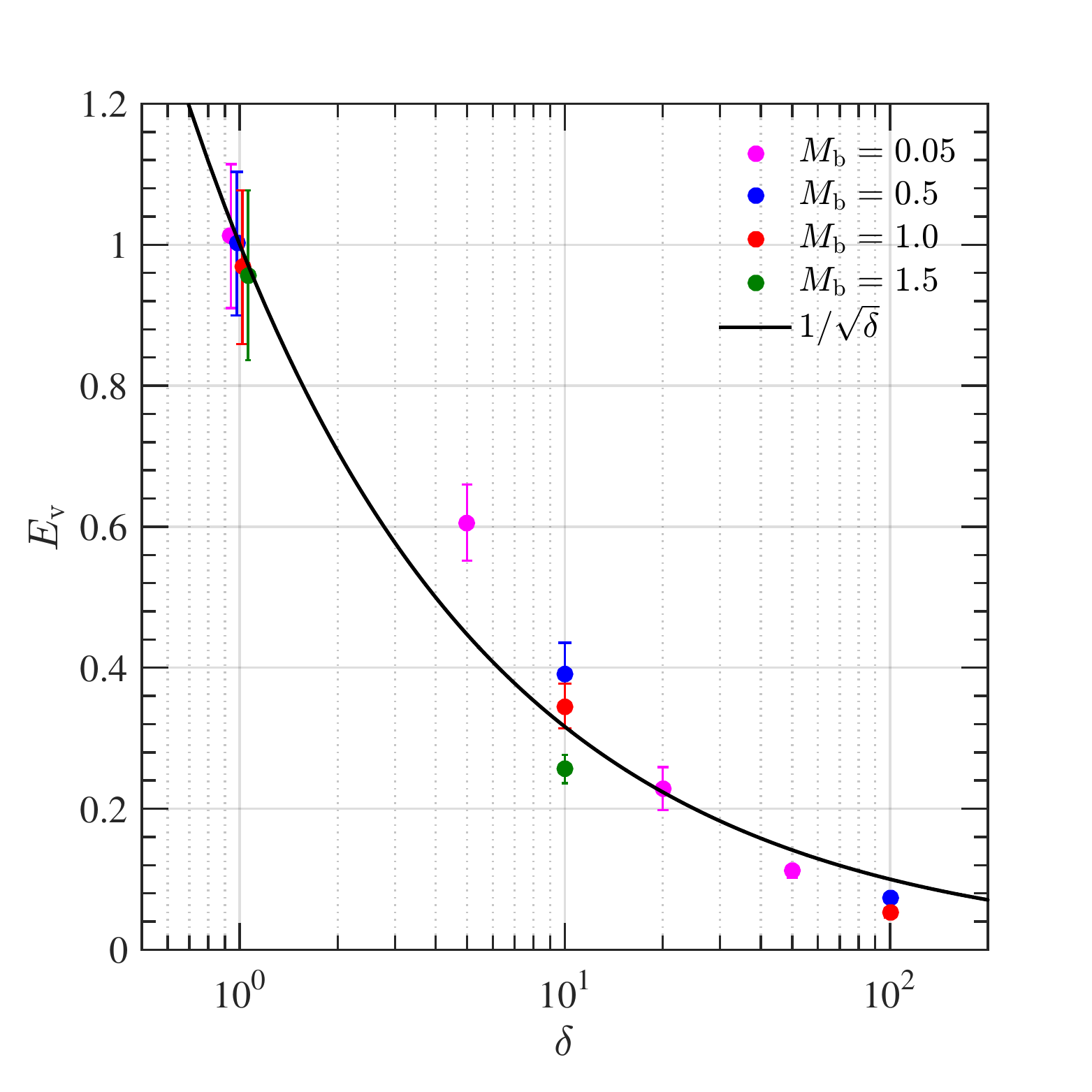}
	\caption{Volume entrainment ratios, $\Ev=\hs/\hb$, based on the one-sided thicknesses \Cref{eq:hb,eq:hs} with $\varepsilon=0.02$, averaged on snapshots in which the shear layer is already well-developed (see \Cref{sec:surface-results-initial-perturbations}). At this stage, $\hbs$  grow linearly with time, so fluctuations around the mean entrainment ratio are small. The colors correspond to different values of $\Mb$. The solid black line corresponds to the analytic prediction, \Cref{eq:entrain-delta}. For each combination $\unpert$ we ran three realizations of broadband white noise interface-only perturbations. Circles mark the mean entrainment ratio. The errorbars are due to the span of $\Ev$ for different realizations. The points at $\delta=1$ were shifted slightly along the horizontal axis to make them distinguishable.}
	\label{fig:entrainment-ratio}
\end{figure}

\Cref{fig:entrainment-ratio} shows the actual volume entrainment ratio in our set of simulations. Overall, the numerical results agree with the predictions of our simple model, \Cref{eq:entrain-delta}. Some scatter about the predicted curve is observed. Around $\delta=10$, the simulated values differ by as much as $30\%$ from the mean curve. At $\delta=100$, the simulations seem to produce systematically lower entrainment ratios than the model predicts, by a factor of $\lsim 2$.

Using \Cref{eq:alpha,eq:entrain-delta,eq:temporal-to-spatial}, we can write separate expressions for $\hbs$ in temporal and spatial growth, 
\begin{align}
\label{eq:hbs-temporal}
&\hs(t) = \frac{1}{1+\sqrt{\delta}} \alpha Vt &&\hb(t) = \frac{\sqrt{\delta}}{1+\sqrt{\delta}} \alpha Vt \\
\label{eq:hbs-spatial}
&\hs(z) = \frac{1}{\sqrt{\delta}} \alpha z &&\hb(z) = \alpha z.
\end{align}

\subsubsection{Planar Slab Growth Rates}
\label{sec:surface-results-compare-slab-sheet}

\begin{figure}
	\centering
	\includegraphics[trim={0.25cm 0.5cm 1cm 1cm}, clip, width=0.475\textwidth]{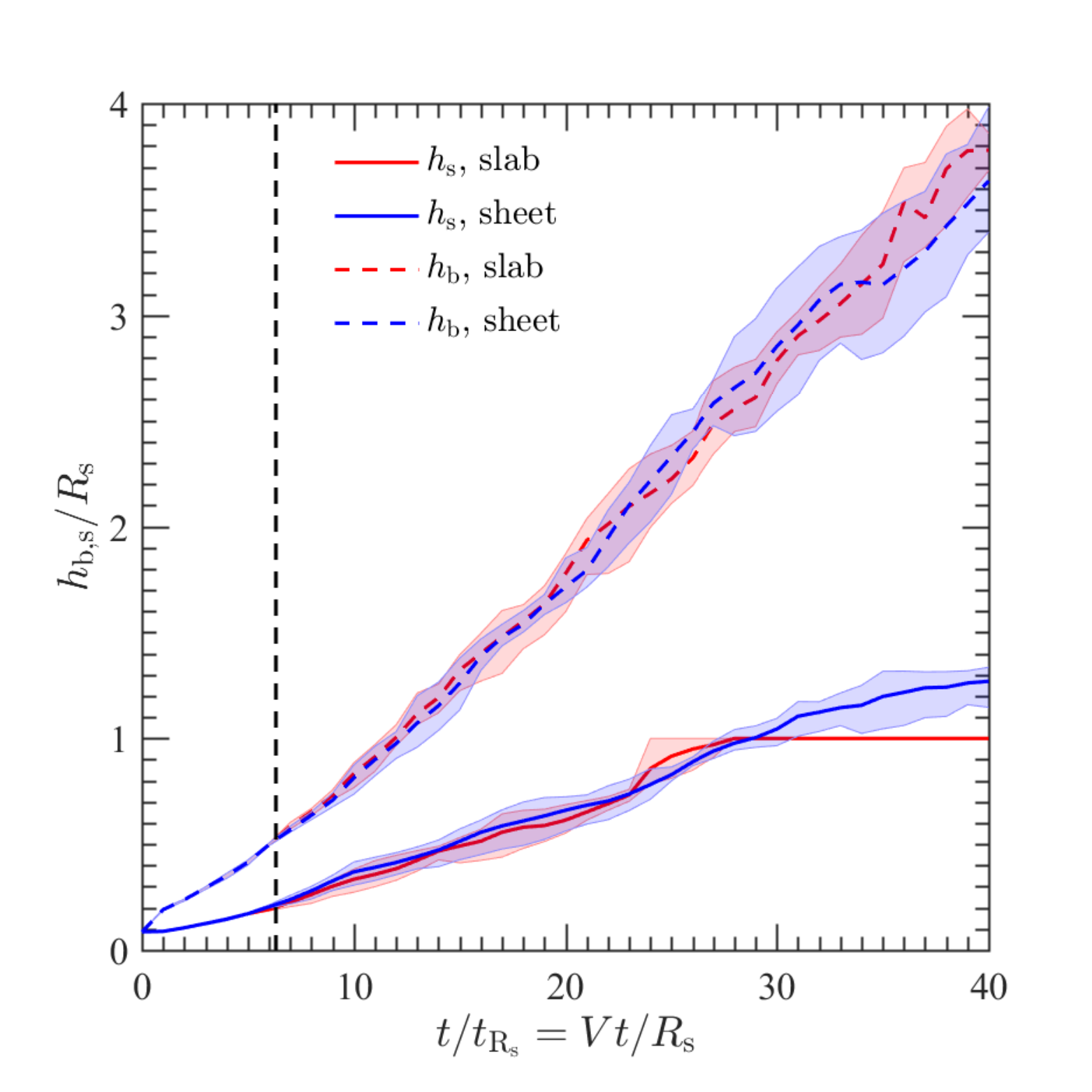}
	\caption{Shear layer growth in planar slab (red) and planar sheet (blue) simulations. The vertical axis shows the stream-side thickness $\hs$ (solid) and background-side thickness $\hb$ (dashed) normalized by the stream radius in the slab simulations $\Rs=1/64$. The unperturbed initial conditions are $(\Mb=1.0,\delta=10)$. Three different realizations of broadband white noise interface-only perturbations were used. For each realization, the $x=-\Rs$ interface in the slab simulation was seeded with exactly the same initial perturbations as the $x=0$ interface in the corresponding planar sheet simulation. The plotted values of $\hbs$ refer to this interface. The shaded regions are spanned due to the variability among the different realizations and the lines correspond to the mean values. The dashed black vertical line marks the stream sound crossing time, $t=\tsc$. The shear layer consumes the entire stream at $t\simeq 27\tRs$, after which $\hs(t)=\Rs$ by definition.}
	\label{fig:compare-slab-sheet}
\end{figure}

The growth rates predicted from planar sheet simulations, shown in \Cref{fig:alpha}, apply to planar slab geometry as well, as long as $h\ll\Rs$. Planar slab simulations might be expected to diverge from the planar sheet results when the shear layer thickness is on the order of the slab thickness, $h\approx\Rs$. In \Cref{fig:compare-slab-sheet} we compare shear layer growth in planar sheet and planar slab geometries for $(\Mb=1.0,\delta=10)$. The opposite sides of the slab do not come into causal contact before one stream sound crossing time has passed and are therefore equivalent to two independent planar sheets. This is clearly visible in \Cref{fig:compare-slab-sheet}, where slab and sheet simulations seeded with the same initial perturbations follow exactly the same history for $t<\tsc=2\Rs\slash\cs$, corresponding to $t/\tRs<6$. Although the simulations are no longer identical at later times, the slab growth rates remain indistinguishable from the sheet results as $\hs\to\Rs$. Furthermore, $\hb$ continues to grow at roughly the same rate predicted by \Cref{eq:hbs-temporal} even after the entire slab has been consumed by the shear layer, $\hs=\Rs$. The same behavior was observed for other combinations of $\unpert$. We therefore conclude that the planar sheet growth rates deduced in \Cref{sec:surface-results-density-mach} are applicable to planar slab geometry as well, except that the slab thickness $\hs$ is limited by $\Rs$.

In principle, a different conclusion might be expected for a cylindrical stream. Simulations of 2D axisymmetric cylindrical streams performed using the moving mesh code \texttt{RICH} \citep{Yalinewich2015} show a \emph{decrease} in overall growth rates as $\hs\to\Rs$, possibly by as much as a factor of $\sim 2$ at late times when $\hs \simeq 0.8\Rs$. On the other hand, preliminary \texttt{RAMSES} simulations of cylindrical streams in full 3D show that $\hs$ \emph{increases} compared to the slab result for $\hs \simeq 0.5\Rs$. The growth rate and entrainment ratio in 3D cylindrical streams are studied both analytically and numerically in the next paper of this series (Mandelker et al. in preparation).

\subsubsection{Stream Deceleration}
\label{sec:surface-results-deceleration}

During the process of shear layer growth, the stream shares its momentum with an increasing amount of background fluid, leading to a gradual deceleration of the stream itself. In the context of cosmic cold streams, this deceleration may lead to a slowdown compared to the accelerated free fall in the halo potential and possibly to a reduced inflow rate of cold, dense gas into the galaxy. The deceleration rate can be deduced from simulations by evaluating the center of mass velocity of the stream along the downstream direction,
\begin{equation}
\label{eq:vcm}
\vcm = \int \vz \dms,
\end{equation}
at different times. The integration is done over the entire simulation domain and the integration variable $\ms$ is the mass of the stream fluid only, defined in \Cref{eq:mass-stream-background}. 

\begin{figure*}
	\centering
	\subfloat{
		\includegraphics[trim={0cm 0.25cm 1cm 1.75cm}, clip, height=0.36\textheight]{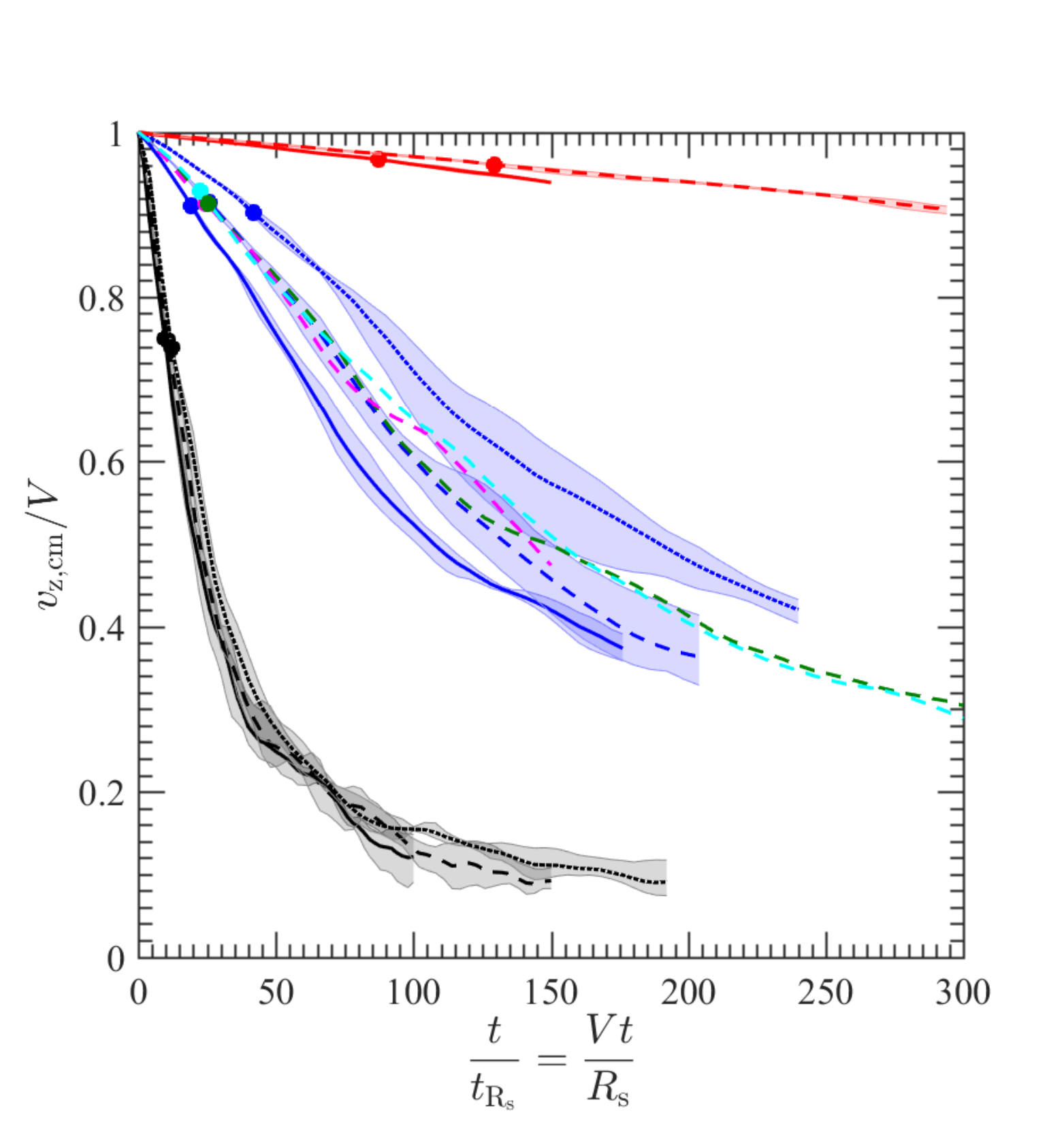}
	}
	\subfloat{
		\includegraphics[trim={1cm 0.25cm 1cm 1.75cm}, clip, height=0.36\textheight]{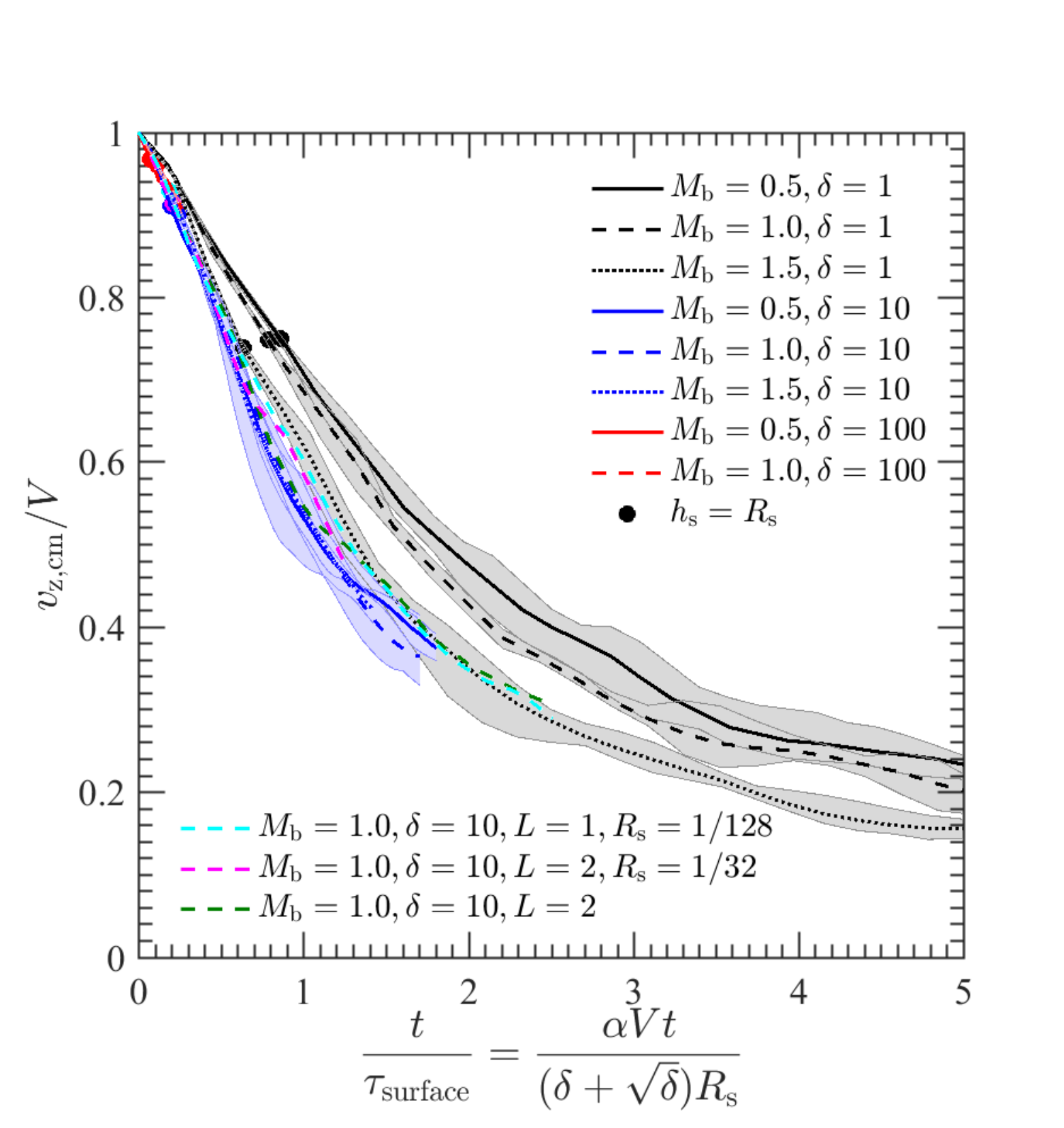}
	}
	\caption{Stream deceleration in planar slab simulations. The vertical axis shows the center of mass velocity of the stream in the downstream direction, $\vcm$, normalized by its initial velocity, $V$. Both panels show the same data, plotted with different scaling for time in the horizontal axis: $t/\tRs$ in the left-hand panel and $t/\tausurf$ in the right-hand panel. The legend in the right-hand panel refers to both panels. The values of $\unpert$ are the same as the eight compressible cases appearing in \Cref{fig:alpha}. For each of these cases we ran three realizations of broadband white noise interface-only initial perturbations, with the fiducial domain size $L=1$ and stream radius $\Rs=1/64$. The different realizations span the black $(\delta=1)$, blue $(\delta=10)$ and red $(\delta=100)$ shaded areas, whereas the mean is marked by solid $(\Mb=0.5)$, dashed ($\Mb=1.0$) and dotted ($\Mb=1.5$) lines. The green, magenta and cyan dashed lines correspond to $(\Mb=1.0,\delta=10)$ with the same realization of initial perturbations but different domain sizes and stream radii. The circles on each of the lines correspond to the time when the stream is entirely consumed by the shear layers growing on either side, i.e. $\hs=\Rs$.}
	\label{fig:deceleration-surface}
\end{figure*}

\Cref{fig:deceleration-surface} shows the evolution of $\vcm$ in various planar slab simulations  (see parameters in \Cref{tab:parameters-surface}). The left-hand panel shows the streams are indeed losing momentum with time, at a rate that depends primarily on the density contrast $\delta$ and only weakly on $\Mb$. For fixed $\delta$, the deceleration rate scales according to 
\begin{equation}
\dot{v}_{\rm z,cm} \sim \frac{V}{\tRs} \sim \frac{V^2}{\Rs},
\end{equation}
as can be predicted from dimensional analysis. The correct scaling with the stream radius is demonstrated by the different simulations with $\Rs=1/128,1/64,1/32$ shown in \Cref{fig:deceleration-surface}. It is interesting to note that only a small amount of deceleration occurs at early times, namely prior to $\hs=\Rs$. The velocities reached at this time are roughly $\vcm=0.75V,0.90V,0.95V$ for $\delta=1,10,100$ respectively. These values are notably higher than the corresponding convection velocities, $\Vc=0.5V,0.76V,0.91V$ respectively (see \Cref{sec:surface-results-convection}). This is to be expected, since the convection velocity is the mean velocity of the entire shear layer, while \Cref{eq:vcm} includes only the stream fluid, which initially carries all of the momentum.

The observed scaling and $\delta$-dependence of the deceleration rate can be explained with the folowing simple argument. We define the deceleration timescale, $\tausurf$, as the time it takes the stream to lose half of its initial velocity. Roughly speaking, this occurs when the stream shares its momentum with a mass of background fluid equal to its own mass. Assuming the stream exchanges momentum efficiently with any background fluid entrained in the shear layer, this translates to
\begin{equation}
\hb L \rhob = \Rs L \rhos,
\end{equation}
for $\hb=\hb\left(t=\tausurf\right)$. Assuming that \Cref{eq:hbs-temporal} holds for $\hb(t)$ at late times, $\hs\gsim\Rs$, as observed in \Cref{sec:surface-results-compare-slab-sheet},  we obtain
\begin{equation}
\label{eq:tausurf}
\tausurf = \frac{\delta+\sqrt{\delta}}{\alpha} \frac{\Rs}{V}.
\end{equation}
The right-hand panel in \Cref{fig:deceleration-surface} shows that this prediction fits the simulations. With the exception of $(\Mb=0.5,\delta=1)$ and $(\Mb=1.0,\delta=1)$, which are the least relevant to cold streams feeding galaxies, our results collapse to a single curve when the  rescaled time $t/\tausurf$ is used for the horizontal axis. Furthermore, we find $\vcm(t=\tausurf) \simeq V/2$, in remarkable agreement with the above derivation.

Simulations with $\delta \gg 1$ must run for a long time, equivalent to several $\Tbox = L/\cb$, before significant deceleration can be observed. In order to test whether the outflow boundary conditions influence the deceleration rate at $t>\Tbox$, we performed a few simulations where the size of the computational domain was made twice as large in both directions, $L=2$. As can be seen in \Cref{fig:deceleration-surface}, these tests produce similar results to those obtained with $L=1$, ruling out any major boundary effects. Furthermore, given sufficient time to run, the stream is expected to distribute its initial momentum over the entire simulation domain and $\vcm(t\to\infty)$ is expected to reach an asymptotic value, determined by the size of the domain. Our simulations were designed not to reach this stage. A comparison of simulations with $L=1$ to those with $L=2$ in \Cref{fig:deceleration-surface} confirms that this is indeed the case, as the final velocities shown for $L=1$ are roughly equal to those in the $L=2$ simulation at the same time.

By varying the stream radius while keeping the size of the grid cells the same, we studied how our conclusions depend on $\Rs\slash\Delta$, the number of cells resolving the thickness of the stream. In these tests, the initial perturbation amplitudes were scaled with the stream radius. As can be seen in \Cref{fig:deceleration-surface}, the change in velocity at $t=\tausurf$ when changing $\Rs\slash\Delta$ by a factor of $4$ is negligible for our purposes. We therefore consider our results to be converged with respect to this numerical metric.

In \Cref{sec:application-inflow}, we use the results of this section to predict the effect of KHI on the inflow rate of cosmic cold streams, taking into account both the deceleration due to KHI and the gravitational acceleration. Comparing these terms, we conclude that the deceleration induced by KHI is too small to account for the constant inflow velocity observed in cosmological simulations, which do not resolve stream instabilities. One important caveat to this conclusion is that cylindrical streams are expected to decelerate more efficiently than planar slabs; preliminary analysis and 2D axisymmetric cylindrical simulations using \texttt{RICH} \citep{Yalinewich2015} suggest that cylinder deceleration timescales may be $\sim 10$ ($\sim3$) times shorter than slab deceleration timescales for $\delta\sim100$ ($\delta\sim10$). This will be presented and compared with full 3D simulations in the next paper of this series (Mandelker et al. in preparation).

\subsubsection{Heating}
\label{sec:surface-results-heating}

The momentum exchange taking place inside the shear layer is accompanied by some conversion of kinetic energy to internal energy, resulting in heating of the stream fluid. This is potentially important in the astrophysical context, since any heat generated in the dense stream will be efficiently radiated away, possibly producing observable emissions, particularly in Lyman-$\alpha$.

\begin{figure*}
	\begin{tabular}{ccc}
		\subfloat{
			\includegraphics[trim={0cm 1.25cm 0.5cm 1.25cm}, clip, height=0.19\textheight]
			{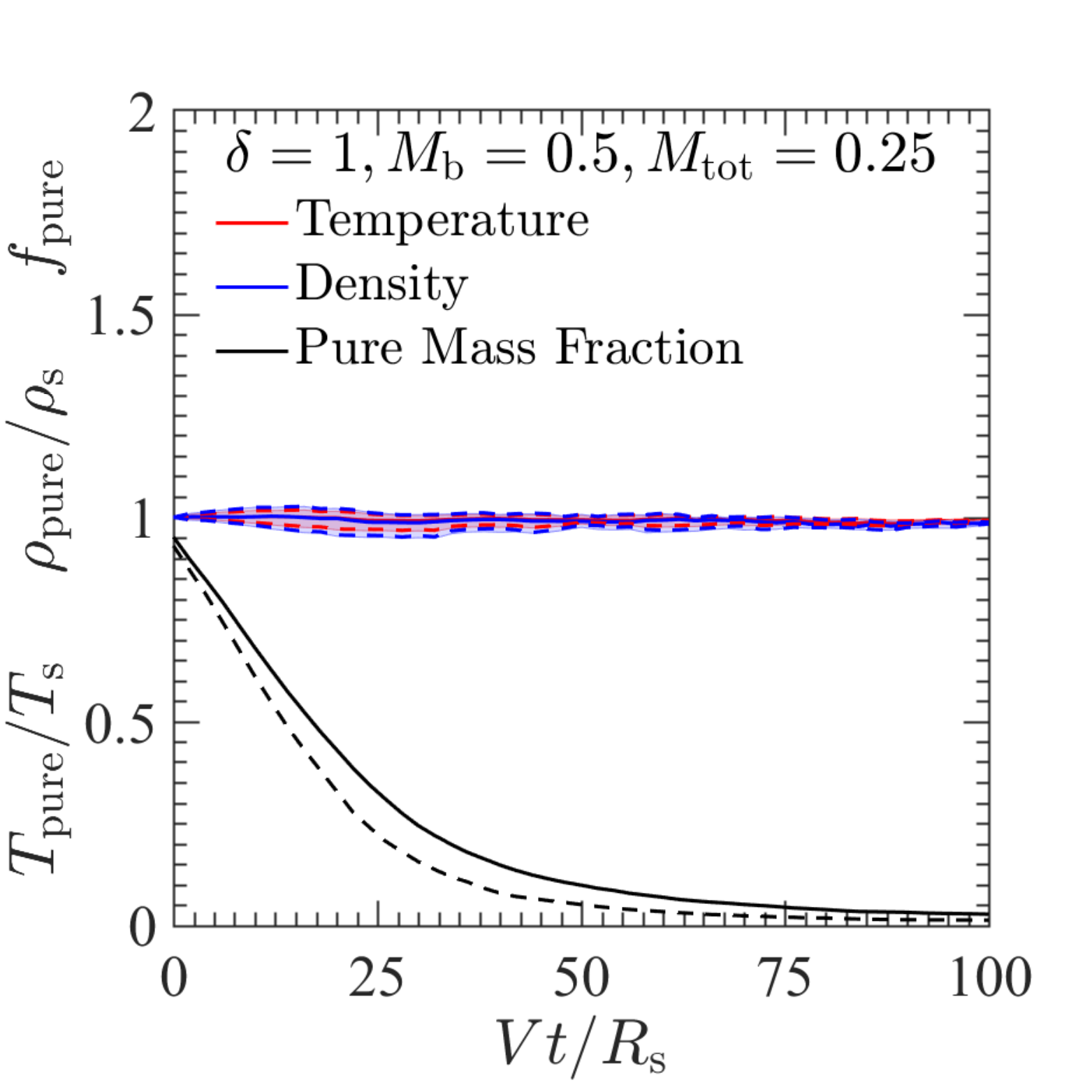}
		}&
		\subfloat{
			\includegraphics[trim={1.25cm 1.25cm 0.5cm 1.25cm}, clip, height=0.19\textheight]
			{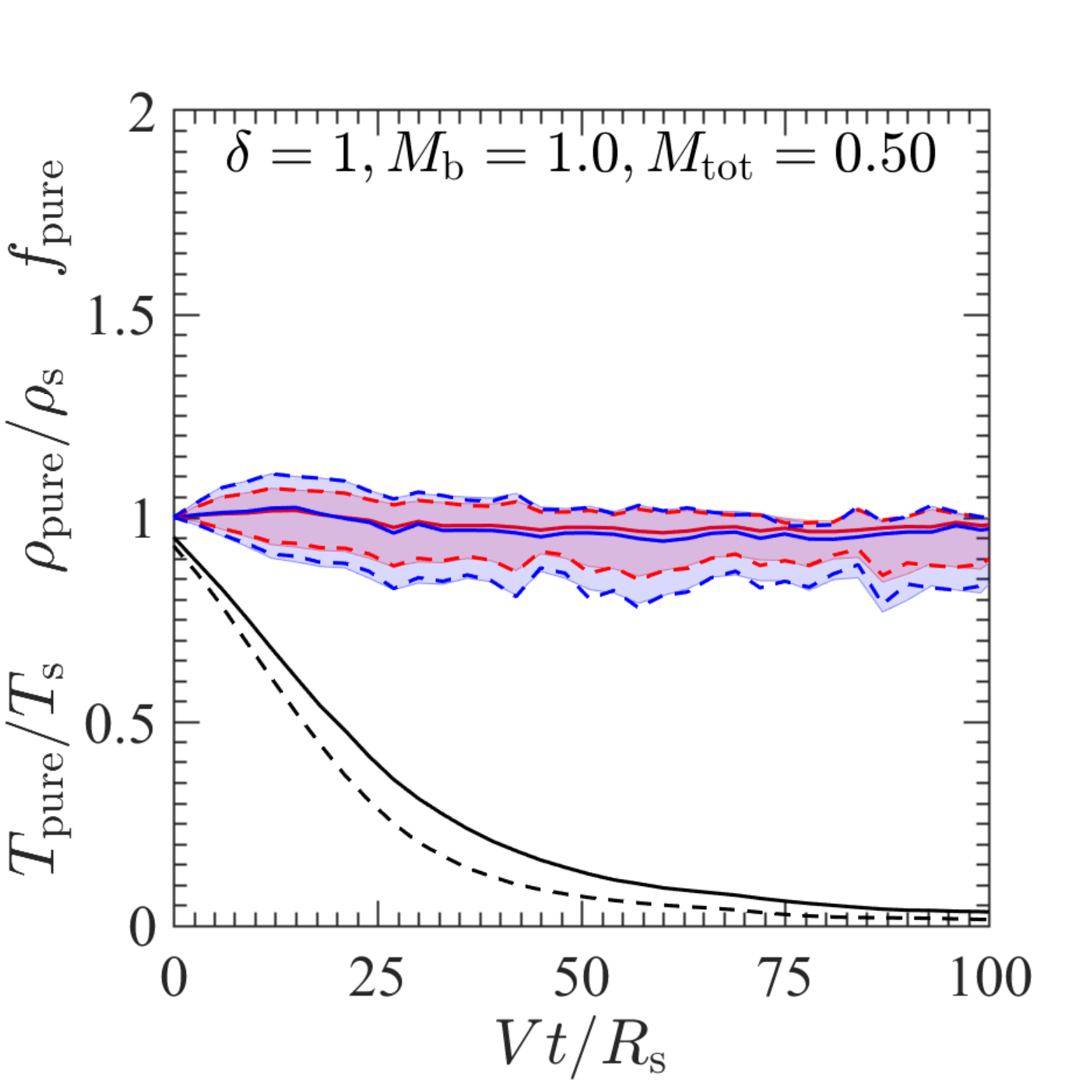}
		}&
		\subfloat{
			\includegraphics[trim={1.25cm 1.25cm 0.5cm 1.25cm}, clip, height=0.19\textheight]
			{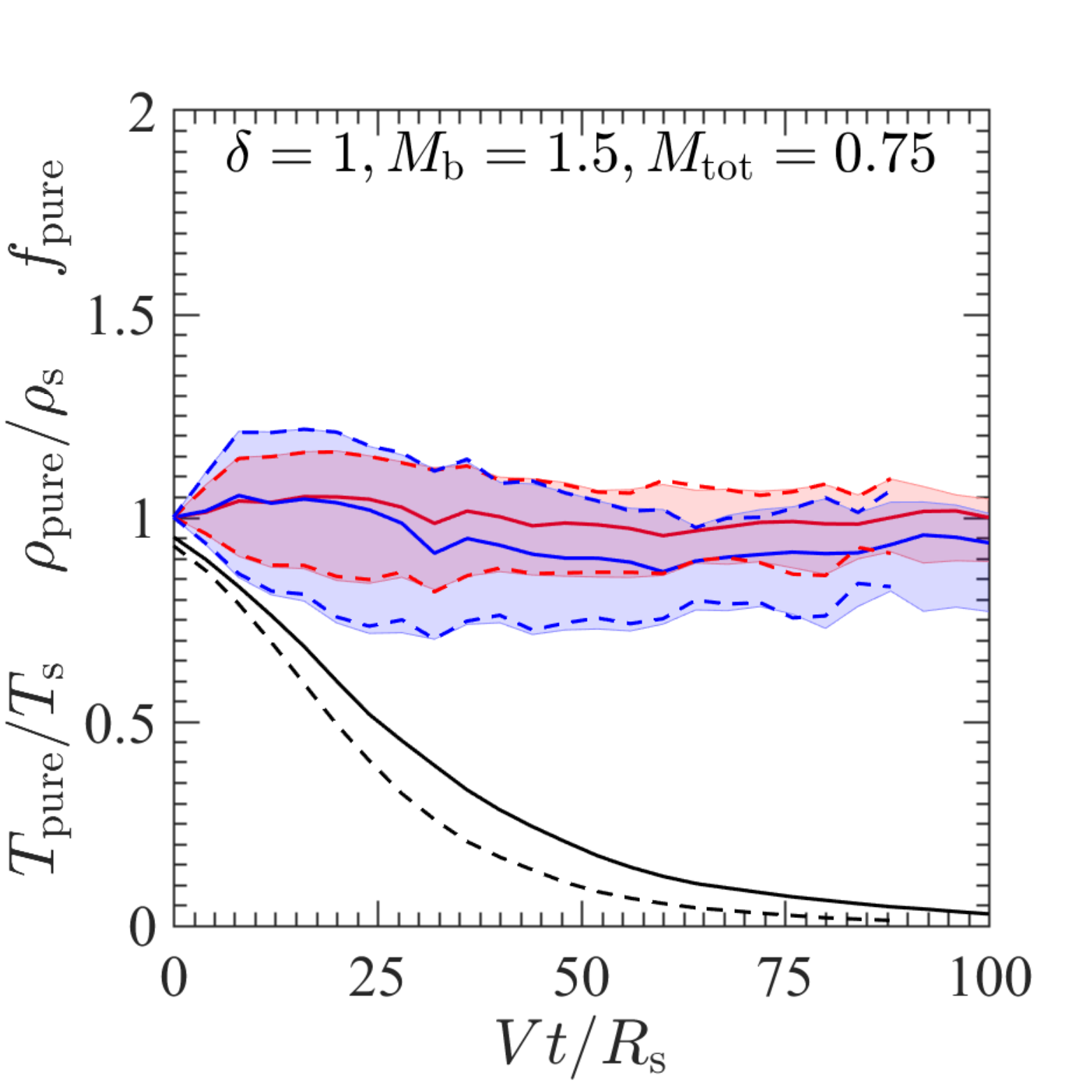}
		}\\
		\subfloat{
			\includegraphics[trim={0cm 1.25cm 0.5cm 1.25cm}, clip, height=0.19\textheight]
			{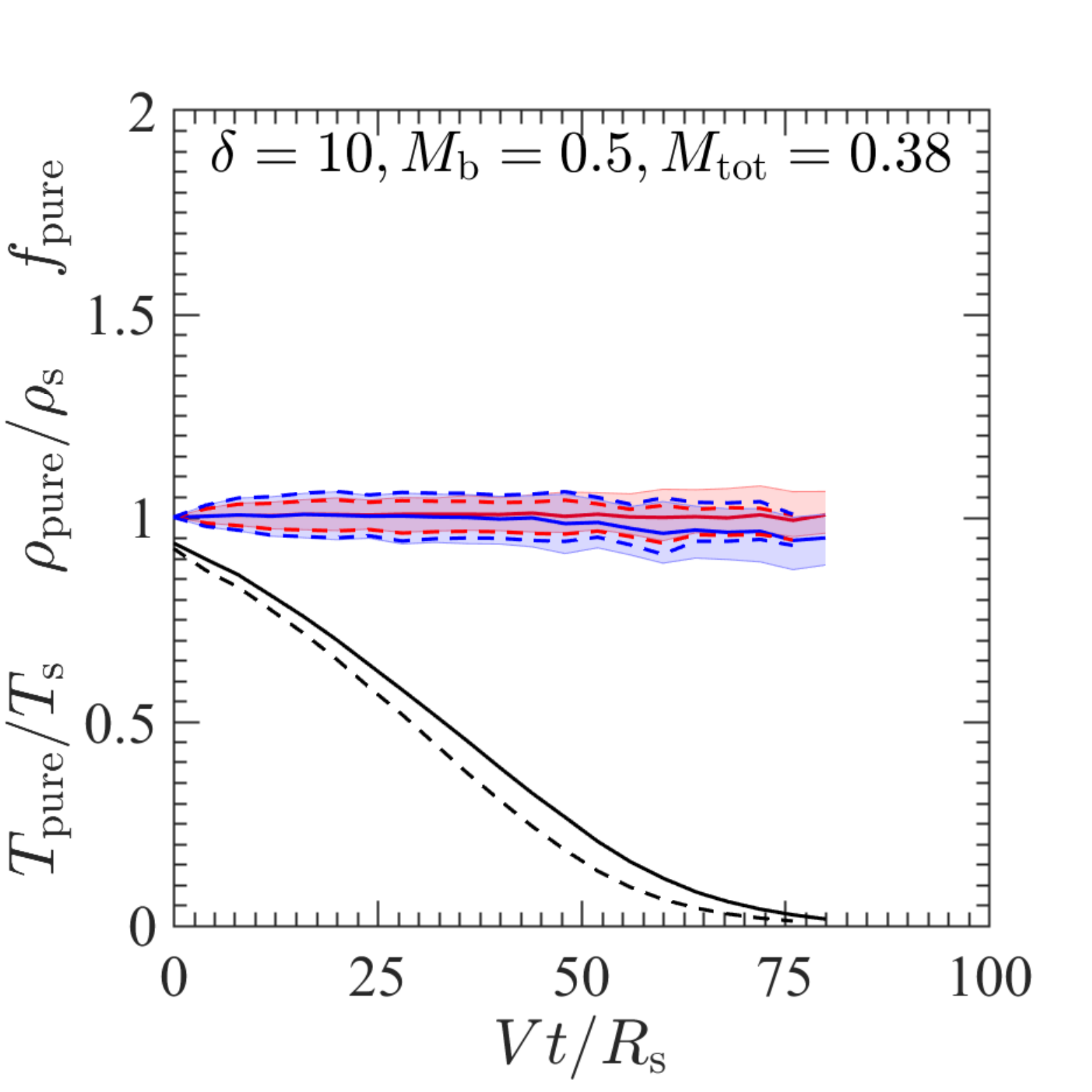}
		}&
		\subfloat{
			\includegraphics[trim={1.25cm 1.25cm 0.5cm 1.25cm}, clip, height=0.19\textheight]
			{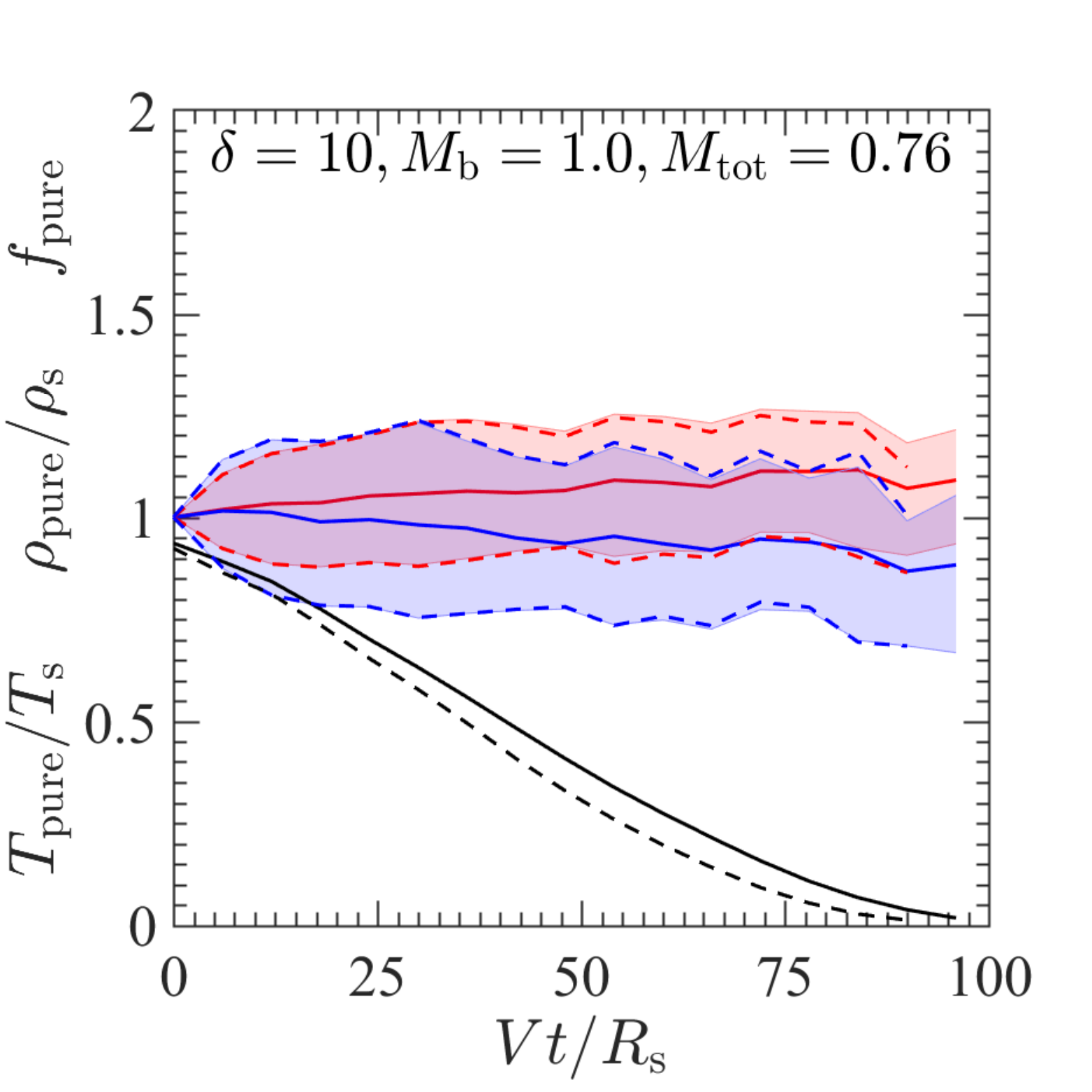}
		}&
		\subfloat{
			\includegraphics[trim={1.25cm 1.25cm 0.5cm 1.25cm}, clip, height=0.19\textheight]
			{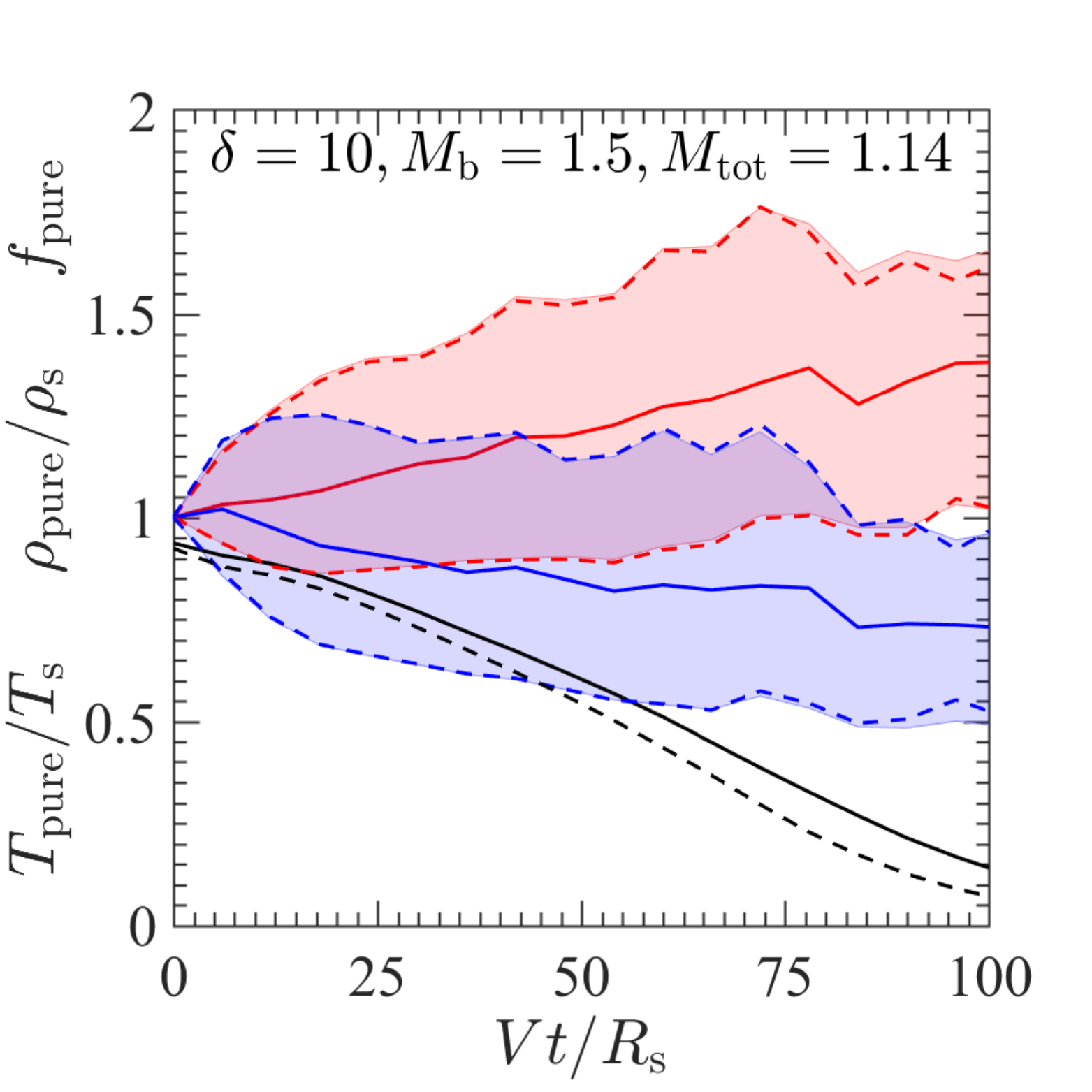}
		}\\
		\subfloat{
			\includegraphics[trim={0cm 0.25cm 0.5cm 1.25cm}, clip, height=0.205\textheight]
			{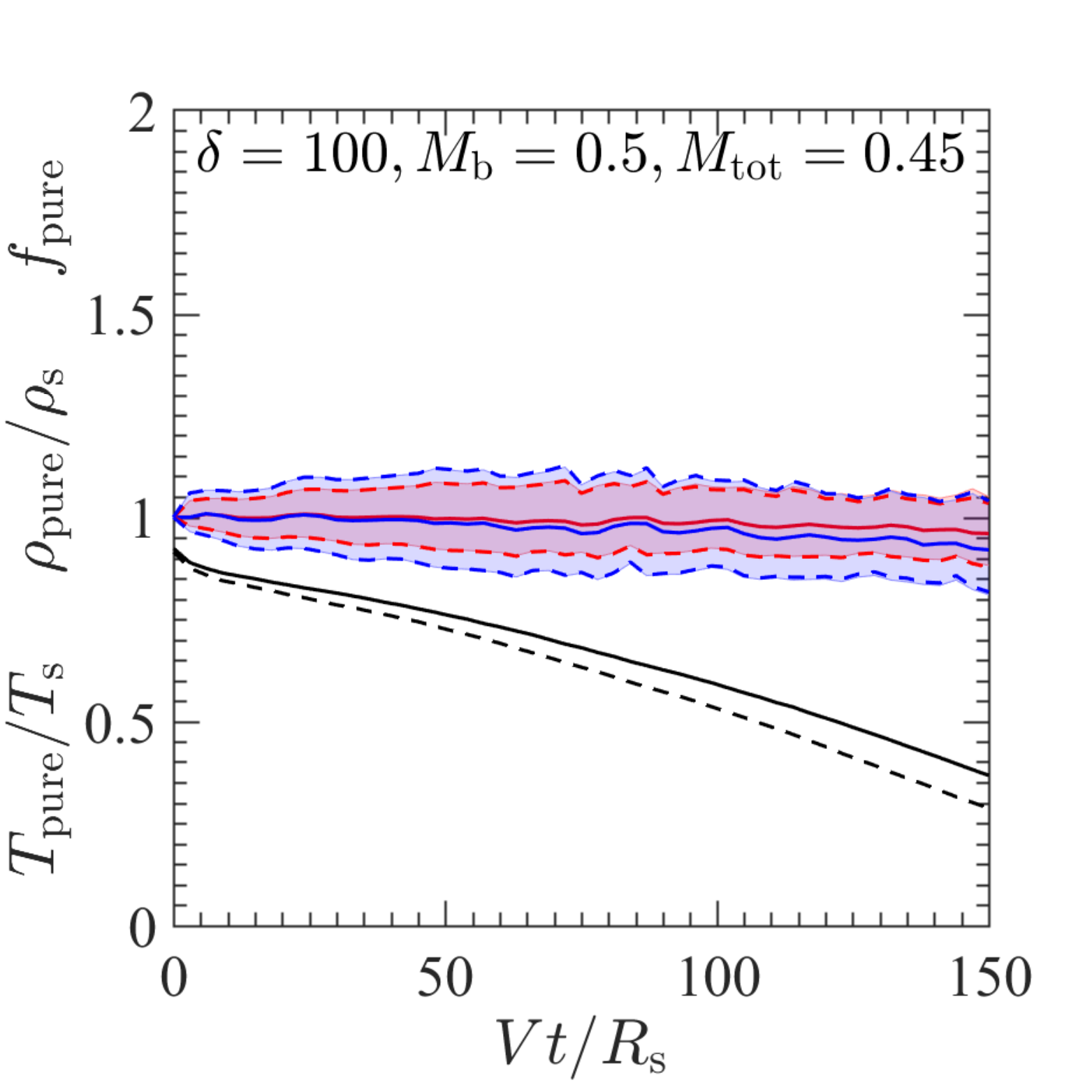}
		}&
		\subfloat{
			\includegraphics[trim={1.25cm 0.25cm 0.5cm 1.25cm}, clip, height=0.205\textheight]
			{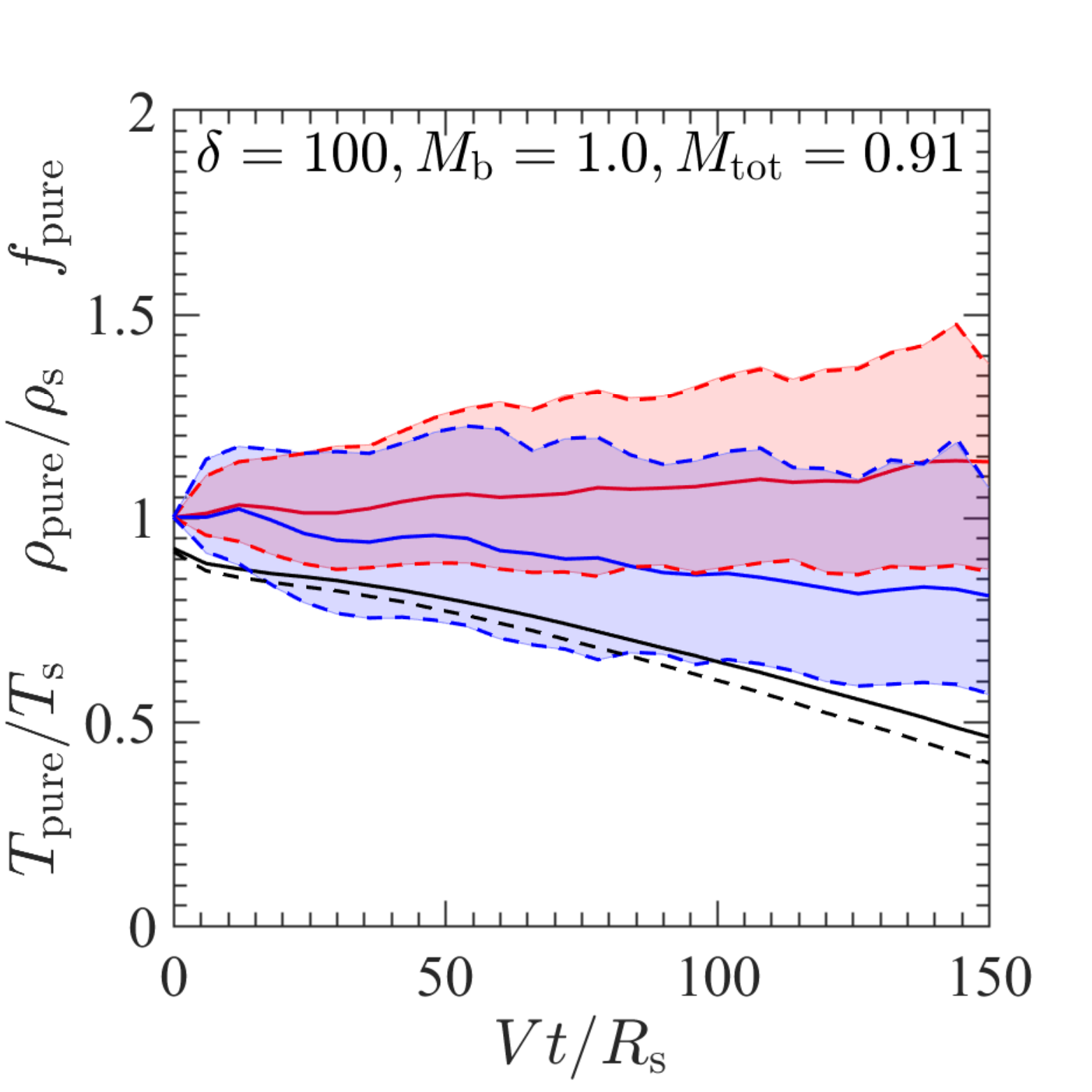}
		}
	\end{tabular}
	\caption{The thermodynamic state of the stream in simulations of shear layer growth in a planar slab. Each panel represents the time evolution of a single run with a different value of $\unpert$. In all cases the initial conditions included the same realization of broadband white noise, interface-only perturbations. Different realizations produce nearly identical results. The vertical axis shows the temperature and density of the unmixed stream fluid, $\Tpure$ and $\rhopure$, normalized by their respective initial values, $\Ts$ and $\rhos$, as well as the mass fraction of unmixed fluid, $\fpure$. The horizontal axis shows the time normalized by $\tRs$. Two different values of the threshold $\epsener$ in  \Cref{eq:pure-stream} are used to define unmixed fluid. For $\epsener=0.1$, the solid lines show $\Tpure$ (red), $\rhopure$ (blue) and $\fpure$ (black), whereas the shaded red/blue areas span the 10th to 90th mass percentiles of temperature/density. For $\epsener=0.01$ the black dashed line shows $\fpure$ and the red/blue dashed lines span the 10th to 90th mass percentiles of temperature/density. The results obtained using both values of $\epsener$ are very similar, indicating that any observed heating is a physical phenomenon affecting the bulk of the stream, rather than a numerical artifact affecting some small boundary layer.}
	\label{fig:thermo-surface}
\end{figure*}

\Cref{fig:thermo-surface} shows how the thermodynamic state of the stream, represented by $\Tpure$ and $\rhopure$, evolves in simulations of shear layer growth in a planar slab. The pure mass fraction, $\fpure$, decreases with time as more fluid is entrained in the shear layers. The time evolution of $\Tpure$ and $\rhopure$ depends on $\unpert$. For $(\Mb=0.5,\delta=1)$, the density and temperature of the stream are completely unaffected, even as the shear layer consumes it entirely. As we increase either $\Mb$ or $\delta$ alone, some scatter around the initial values is observed, but the mean temperature does not indicate any heating. For combinations $(\Mb\gsim 1.0,\delta\gsim 10)$, we find that the temperature of the stream steadily grows, achieving a $20\%-40\%$ increase at $Vt/\Rs\sim 100$. The density decreases by a similar amount. Thus, \Cref{fig:thermo-surface} implies that the amount of heating experienced by the stream depends primarily on the total Mach number, $\Mtot$, defined in \Cref{eq:body-instability-condition}. A non-negligible increase in temperature is observed for $\Mtot\gsim 1$. This is in agreement with the role $\Mtot$ plays in quantifying the effect of compressibility on shear layer growth rates, as discussed in \Cref{sec:surface-results-density-mach} and \Cref{app:compressibility-scaling}. 

Although KHI can cause a non-negligible increase in stream temperature compared to its initial value, the energy deposition is small compared to the total stream energy budget, which is dominated by kinetic and gravitational rather than internal energy. Hence, surface mode instability is a significantly weaker power source for Lyman-$\alpha$ emission from cold streams than the  driving considered in \citet{Dijkstra09} and \citet{Goerdt10}. We elaborate on this point in \Cref{sec:application-thermo}.

The results of this section are based on \emph{adiabatic} simulations, neglecting radiative cooling. Previous studies in other contexts \citep{Hardee1997,Vietri97,Stone97} indicate that radiative cooling can have a significant and non-trivial effect of KHI, which is sensitive to the details of the cooling function. This motivates a study of surface mode instability with the specific cooling function applicable to cosmic cold streams, to be reported in subsequent papers  of this series (Mandelker et al. in preparation).

\section{The Nonlinear Evolution of Body Modes}
\label{sec:body}

This section extends the study of body modes in a planar slab from \PIt~into the nonlinear stage. In \Cref{sec:body-qual} we present a qualitative description. The numerical simulations used for this section are described in \Cref{sec:body-simulations}. The results are presented in \Cref{sec:body-results}.


\subsection{Qualitative Description}
\label{sec:body-qual}

Body modes are fundamentally related to the length scales and dynamical times of the entire stream, represented by its radius $\Rs$ and sound crossing time $\tsc=2\Rs\slash\cs$. This observation is in stark contrast to the scale invariance observed in surface modes. Due to the existence of these preferred physical scales, the evolution and final state of a slab perturbed by body modes will depend on the initial conditions, unlike the results presented in \Cref{sec:surface} for surface modes. Nevertheless, some insight into nonlinear body-mode instability can be gained by considering the results of linear analysis. The arguments presented in this section were first laid out by \cite{Hardee1995} and are reiterated here for completeness. 

According to \Cref{eq:body-tkh}, the linear growth rate diverges as $\lambda\to 0$. At first glance, this suggests that short wavelength perturbations are the first to reach nonlinear amplitudes and therefore dominate the late-time evolution of body modes. However, the opposite conclusion is reached when we consider the maximum amplitude attained by different modes in the linear stage. Linear analysis shows that the transverse displacement of the stream fluid scales as $h\sim u_{\rm x} \slash \omega$, where $u_{\rm x}$ is the transverse velocity perturbation,  (see \Cref{app:displacement}; \citealt{Hardee1995,Hardee1997}). In the linear regime, by definition, the velocity perturbation cannot significantly exceed the speed of sound in the stream, $u_{\rm x}\lsim \cs$. Assuming the transition from linear to nonlinear behavior occurs when this limit is reached, we find $h\propto\omega^{-1}$ \citep{Hardee1995,Hardee1997}. Because $\omega$ decreases as $\lambda$ is increased, we conclude that $h$ must increase with $\lambda$. This implies that long wavelengths dominate the nonlinear evolution of body modes, contrary to the naive assumption; although shorter wavelength modes grow faster, they saturate at smaller amplitudes.

A perturbation is disruptive to the stream when its amplitude satisfies 
\begin{equation}
\label{eq:body-nonlinearity}
h\approx\Rs.
\end{equation}
Consequently, the perturbation expected to ultimately break the stream is the fastest growing among those that satisfy \Cref{eq:body-nonlinearity} in the linear stage. We refer to this as the \emph{critical perturbation mode} or \emph{critical mode}. The precise wavelength $\lambdacrit$ and order $\ncrit$ of the critical mode can be predicted analytically and confirmed in simulations. This was done in previous studies in the context of dilute jets (see references in \Cref{sec:body-simulations}) and is repeated in \Cref{sec:body-results-critical} for the range of parameters applicable to cold streams. In most scenarios, the critical mode is a low-order, long-wavelength mode, roughly in the range $(\ncrit\leq 1,10\Rs \lsim \lambdacrit \lsim 20\Rs)$. Sometimes the disruption of the stream is attributed to a combination of two or three different order modes in this wavelength range, but the fundamental S-mode ($n=0$) is often the dominant among them \citep[e.g.][]{Hardee1995,Bodo1998}.

After the critical mode becomes dominant, it continues to grow, bending the stream and leading to its eventual disintegration. This occurs when the amplitude of the critical mode is comparable to its wavelength, typically $h \approx \lambdacrit/2 \sim 5-10\Rs$. The stream becomes discontinuous and its fragments occupy a broad region, about $\sim 10-20\Rs$ across, perturbed by significant turbulent motion. The evolution beyond this time is driven by turbulence, gradually mixing the stream into the background fluid  \citep[e.g.][]{Bodo1998,Micono2000}.


\subsection{Simulations}
\label{sec:body-simulations}

We performed a comprehensive set of simulations tailored to investigate the nonlinear temporal evolution of body modes in dense streams. The range of parameters studied in our simulations is listed in \Cref{tab:parameters-body}. The chosen values of $\unpert$ represent the values relevant to cosmic cold streams, with a preference for slightly higher Mach numbers in order to steer clear of the region in which body modes and surface modes coexist. This choice makes the simulations less prone to unintended growth of grid-scale surface modes, while preserving the applicability of  our conclusions, which do not depend strongly on the Mach number. The fiducial stream radius was $\Rs=1/128$, half the value used for surface mode simulations (see \Cref{sec:surface-simulations}), in order to compensate for the relatively long dynamical timescales associated with body modes.

\begin{table} 
	\centering
	\caption{Parameters of simulations studying the nonlinear evolution of body modes. The first five entries comprise a survey of the $\unpert$ parameter space with the fiducial stream radius, domain size and smoothing width. For these cases, we performed three simulation runs with different realizations of initial perturbations. The next four entries study the scaling with $\Rs$, the boundary effect and convergence with respect to $\Rs/\Delta$. The last entry tests the effect of doubling the smoothing width, $\sigma$.} \label{tab:parameters-body}
	\begin{tabularx}{\columnwidth}{XXXXXXX}
		\hline
		$\Mb$ &$\delta$ &$\Mtot$ &$\Rs$ &$L$ &$\Rs/\Delta$ &$\Rs/\sigma$ \\
		\hline
		2.5  &5   &1.73 &1/128 &1.0 &64 &32 \\
		2.5  &10  &1.90 &1/128 &1.0 &64 &32 \\
		2.5  &20  &2.04 &1/128 &1.0 &64 &32 \\
		5.0  &1   &2.50 &1/128 &1.0 &64 &32 \\
		5.0  &10  &3.80 &1/128 &1.0 &64 &32 \\
		\hline	 
		2.5  &10  &1.90 &1/64  &1.0 &128 &32 \\
		2.5  &10  &1.90 &1/64  &2.0 &128 &32 \\
		5.0  &1   &2.50 &1/64  &1.0 &128 &32 \\
		5.0  &1   &2.50 &1/64  &2.0 &128 &32 \\
		\hline
		2.5  &10  &1.90 &1/128 &1.0 &64  &16 \\
		\hline
	\end{tabularx}
\end{table}

Each simulation run was initialized with a random realization of interface-only perturbations with the functional form described in \Cref{sec:surface-simulations}. An identical spectrum was used in all cases. The spectrum included harmonics of both the box length and the stream radius, corresponding to resonant wavenumbers of the slab \citepalias[see Appendix H in][]{Mandelker2016b}, specifically $\kj=2\pi\slash\lambdaj$ where $\lambdaj = 2^{j}\Rs$ and $-2 \leq j \leq 4$. This band was selected in order to excite long wavelength modes, known to play an important role in the nonlinear stage, as previously discussed, in addition to a few modes with $\lambda<\Rs$. The random phases assigned to each mode ensure that the initial perturbations represent a mixture of P-modes and S-modes. The initial displacement amplitude was uniform for all wavelengths and scaled with stream radius, $\Hj=2\Delta\times128\Rs$.

The smoothing width parameter in \Cref{eq:smoothing} was also scaled with the stream radius according to $\sigma=2\Delta\times128\Rs$. The resulting transition layer between the slab and the background has a width of $\sim10\%$ the stream radius. In one of the simulations the width parameter was increased by a factor of two in order to test the sensitivity of our results to the ratio $\Rs/\sigma$.

Numerical simulations were previously used to study the nonlinear evolution of body modes, primarily with connection to astrophysical jets. As such, most studies considered hot, dilute jets surrounded by a dense, cold background, i.e. $\delta<1$, the exact inverse of the cosmic cold streams scenario considered here. These include 2D simulations of planar slabs \citep{Hardee_Norman88b,Hardee1989,Bodo1995,Bassett1995}, 2D simulations of axisymmetric cylinders \citep{Bodo94} and full 3D simulations of cylindrical streams \citep{Hardee1995,Bassett1995,Bodo1998}. Two notable exceptions, studying a dense jet confined in a dilute medium, are \citet{Bodo1998}, who studied the case $(\Mb=3.16,\delta=10)$, and \cite{Stone97}, who investigated the spatial evolution of cooling jets with $(\Mb=\{5,20\},\delta=10)$. Our work offers the first comprehensive study of temporal nonlinear growth of body modes in dense slab streams ($\delta>1$) and the first study focusing on  deceleration due to KHI in these systems.


\subsection{Results}
\label{sec:body-results}

Motivated by the ideas presented in \Cref{sec:body-qual} and using the set of simulations described in \Cref{sec:body-simulations}, this section addresses the following quantitative questions:
\begin{itemize}
	\item Does the critical perturbation mode depend on $\unpert$? What are the critical modes for values relevant to cold streams feeding galaxies?
	\item How does the stream width evolve with time, prior to and during the growth of the critical mode? Does this behavior depend on $\unpert$?
	\item What is the rate of deceleration of the stream fluid due to body mode instability? How does this depend on $\unpert$?
	\item How much of the stream kinetic energy is converted into stream internal energy due to  nonlinear growth of body modes (potentially producing Lyman-$\alpha$ emission)? 
\end{itemize}
The answers to these questions, presented below, are used in \Cref{sec:application} to predict the outcome of KHI in cold streams feeding massive SFGs at high redshift.

\subsubsection{Critical Perturbation Modes}
\label{sec:body-results-critical}

Following the qualitative discussion in \Cref{sec:body-qual}, we present more accurate analytic predictions for the critical perturbation modes and compare them to simulations. This is accomplished by carefully defining the transition from linear to nonlinear dynamics, following the procedure first presented by \citet{Hardee1995} and \citet{Hardee1997}.

An eigenmode perturbation displaces each fluid element in the transverse direction by a certain amount $\xix(x,z,t)$, which grows exponentially with time for unstable modes. This quantity is derived in detail in \Cref{app:displacement} based on the linear stability analysis performed in \PIt. At some finite time the displacement $\xix$ causes fluid elements somewhere in the stream to cross. This defines the time of transition to nonlinear evolution, $\tNL$, because crossing streamlines, combined with transonic velocities, imply the formation of shocks. The temporal and spatial dependences of $\xix$ vary with the wavelength $\lambda$ and the order $n$ of the mode, resulting in different $\tNL$ for different eigenmodes. In order to compare the amplitude reached by different eigenmodes in the linear stage, we define $\hNL=\max_{z}{\left|\xix(\pm\Rs,z,\tNL)\right|}$, i.e. the magnitude of displacement to the stream/background interface at time $t=\tNL$. We note that while the amplitude of transition to nonlinearity, $\hNL(\lambda,n)$, does not depend on the initial perturbation amplitude, the transition time, $\tNL(\lambda,n)$, satisfies 
\begin{equation}
\label{eq:tNL}
\tNL = \ln\left(\frac{\hNL}{H}\right)\tkh
\end{equation}
where $H$ is the initial displacement amplitude given to the eigenmode in question and $\tkh$ is its Kelvin-Helmholtz time. The logarithmic dependence of $\tNL$ on $\hNL$ is a result of taking the inverse of the exponential growth in the linear stage.

\begin{figure*}
	\centering
	\subfloat{
		\includegraphics[trim={0cm 0.25cm 1cm 1.0cm}, clip, height=0.35\textheight]{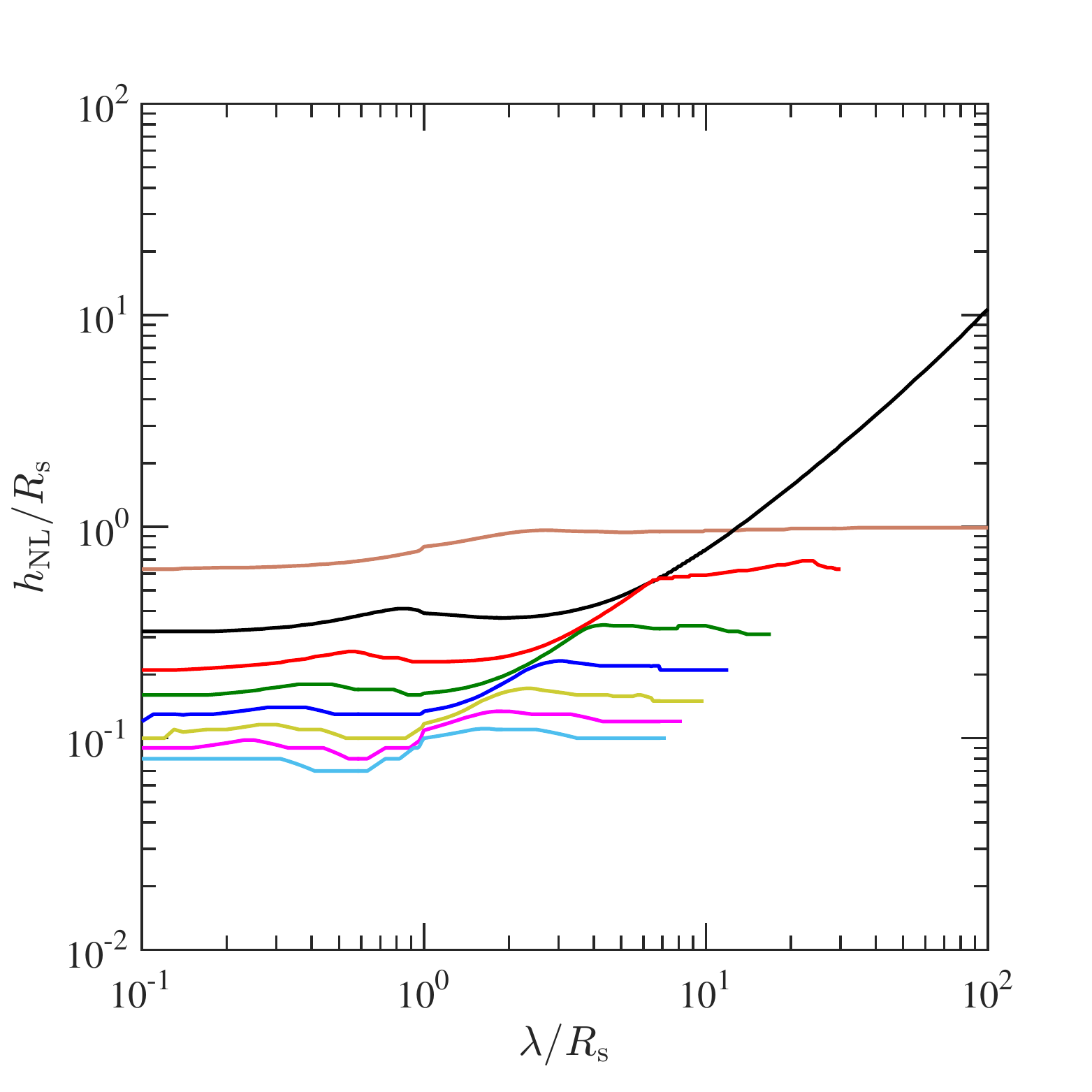}
	}
	\subfloat{
		\includegraphics[trim={0.cm 0.25cm 1cm 1.0cm}, clip, height=0.35\textheight]{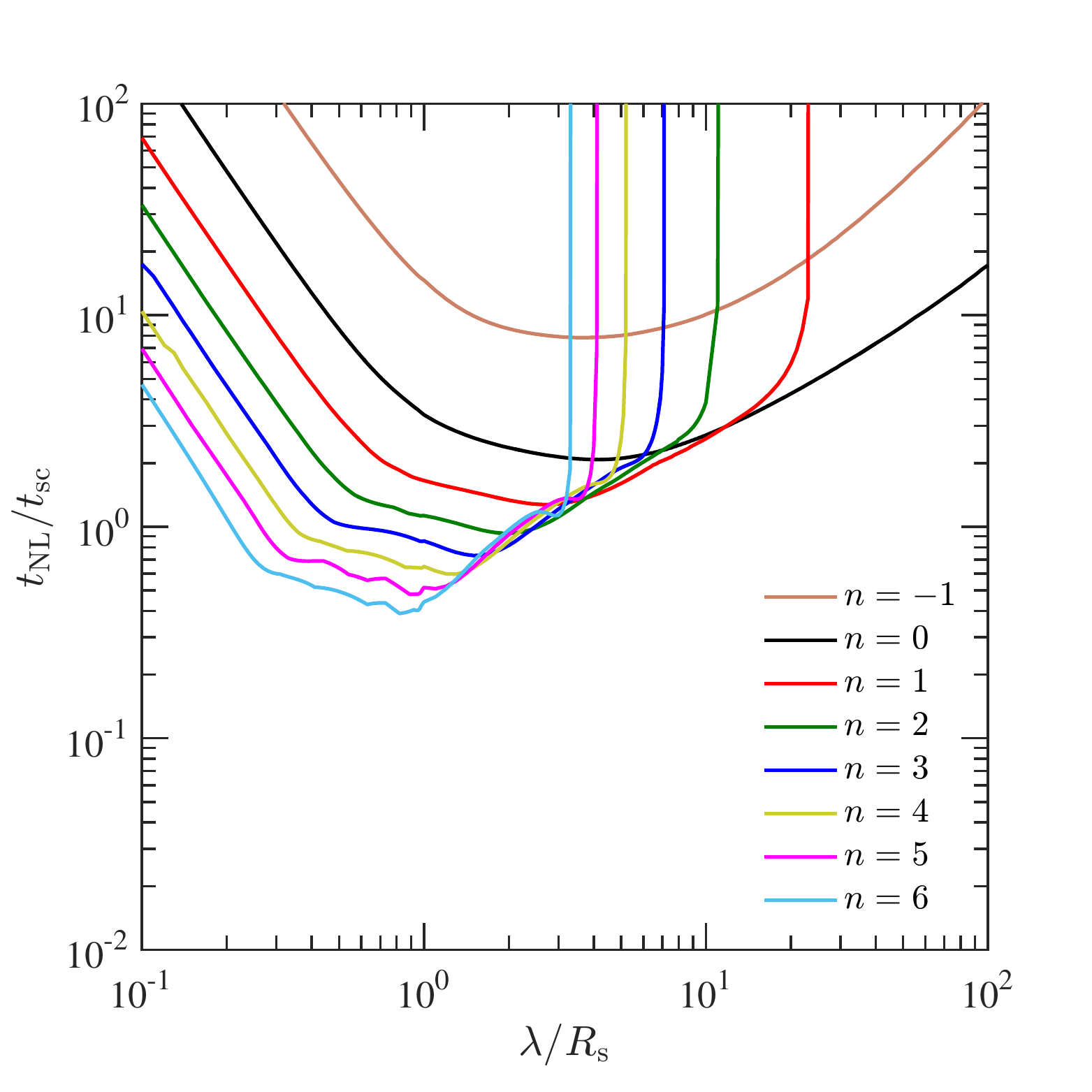}
	}
	\caption{Transition from linear to nonlinear evolution of body modes as predicted by linear analysis for $(\Mb=2.5,\delta=10)$. The first eight body modes are shown in different colors. The legend on the right-hand panel applies to both. The left-hand panel shows the displacement amplitude reached by each mode at the onset of nonlinearity, $\hNL$, as a function of the wavelength, $\lambda$, with both axes normalized by the stream radius $\Rs$.  The right-hand panel shows the time of transition to nonlinearity, $\tNL$, normalized by the stream sound crossing time, $\tsc$. The initial perturbation amplitude is assumed to be $H=0.03\Rs$ for all wavelengths.}
	\label{fig:thNL}
\end{figure*}

The results of this analysis are shown in \Cref{fig:thNL} for a planar slab with $(\Mb=2.5,\delta=10)$. The initial perturbation amplitude is assumed to be $H=0.03\Rs$ for all wavelengths. The left-hand panel shows $\hNL$ as a function of $\lambda$ for different orders, $n$. For all wavelengths, the largest values of $\hNL$ are achieved by the $n=-1$ and $n=0$ modes, corresponding to the fundamental P-mode and the fundamental S-mode respectively. Both of the modes satisfy \Cref{eq:body-nonlinearity}, the P-mode for $\lambda\gsim 1\Rs$ and the S-mode for $\lambda\gsim 10\Rs$, making them the candidate critical modes in these wavelength ranges. The right-hand panel in \Cref{fig:thNL} shows that the S-mode is the faster growing of the two, having smaller $\tNL$ by a factor of $\sim 5$ across all wavelengths. Thus we conclude that in the case $(\Mb=2.5,\delta=10)$, the critical mode ought to be an S-mode ($\ncrit=0$) with wavelength $\lambdacrit\gsim 10\Rs$. For $H=0.03\Rs$, the breakup of the stream is expected to commence at $\tNL \simeq 3.5\tsc$, corresponding to the transition to nonlinearity for the aforementioned $(\lambdacrit,\ncrit)$.

\begin{figure*}
	\centering
	\subfloat{
		\includegraphics[trim={1cm 1.4cm 1.25cm 1.75cm}, clip,width=0.7\textwidth]{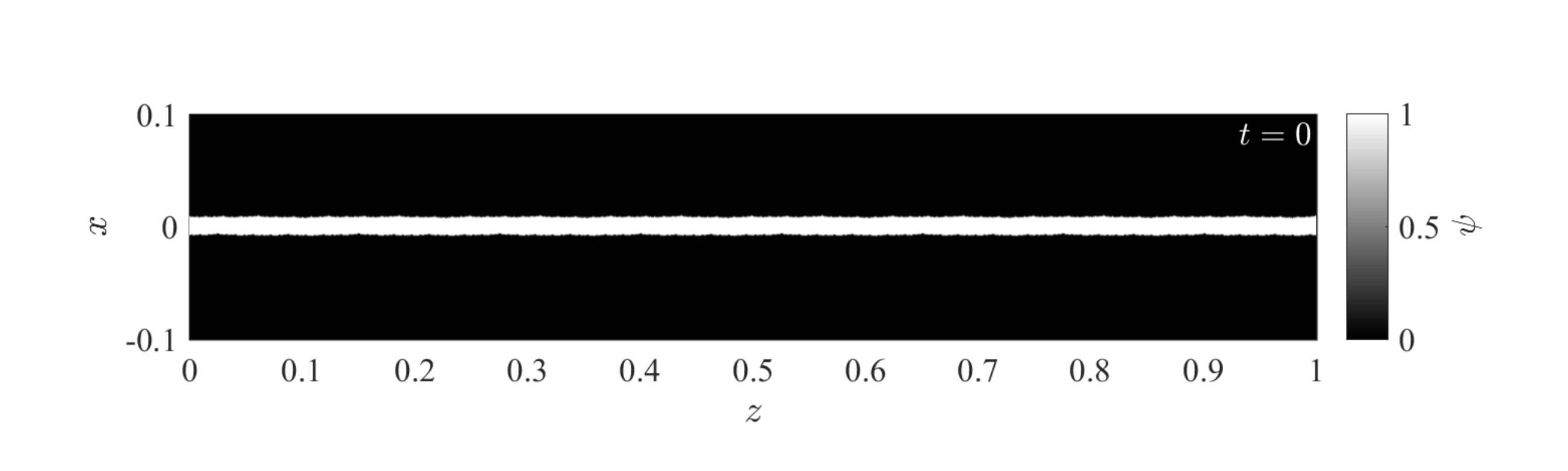}
	}\\
	\subfloat{
		\includegraphics[trim={1cm 1.4cm 1.25cm 1.75cm}, clip,width=0.7\textwidth]{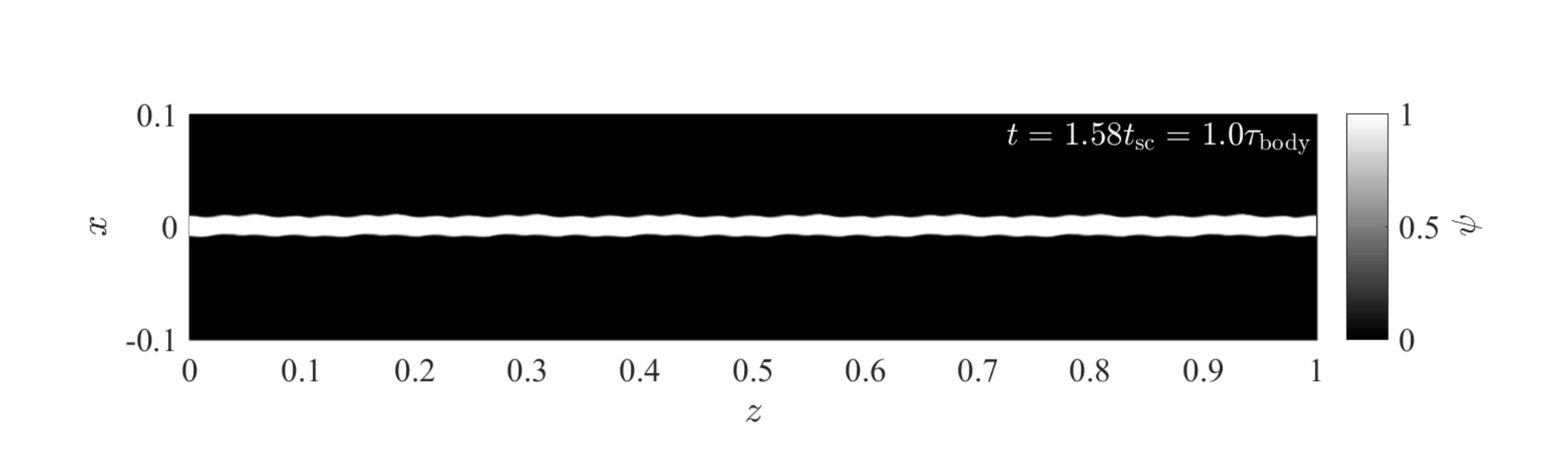}
	}\\
	\subfloat{
		\includegraphics[trim={1cm 1.4cm 1.25cm 1.75cm}, clip,width=0.7\textwidth]{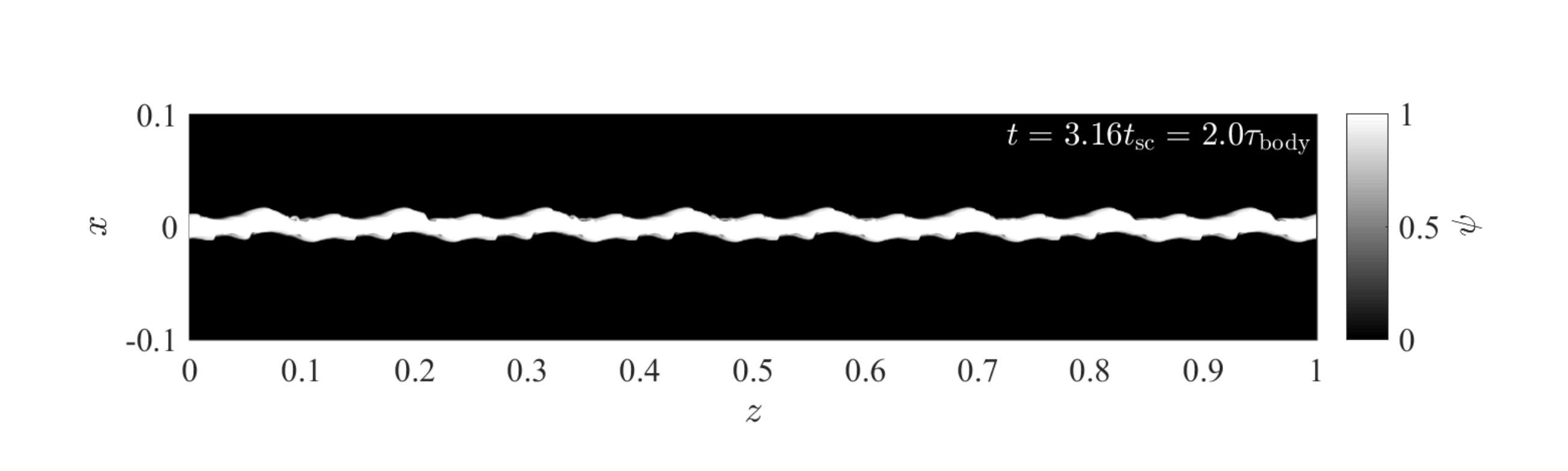}
	}\\
	\subfloat{
		\includegraphics[trim={1cm 1.4cm 1.25cm 1.75cm}, clip,width=0.7\textwidth]{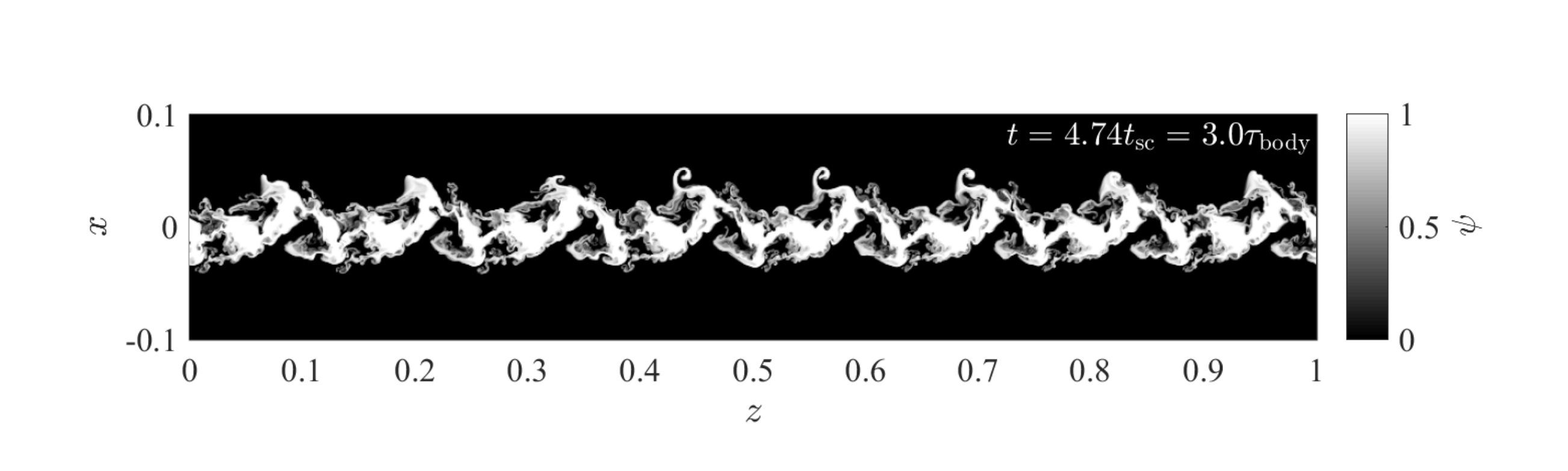}
	}\\
	\subfloat{
		\includegraphics[trim={1cm 0.8cm 1.25cm 1.75cm}, clip,width=0.7\textwidth]{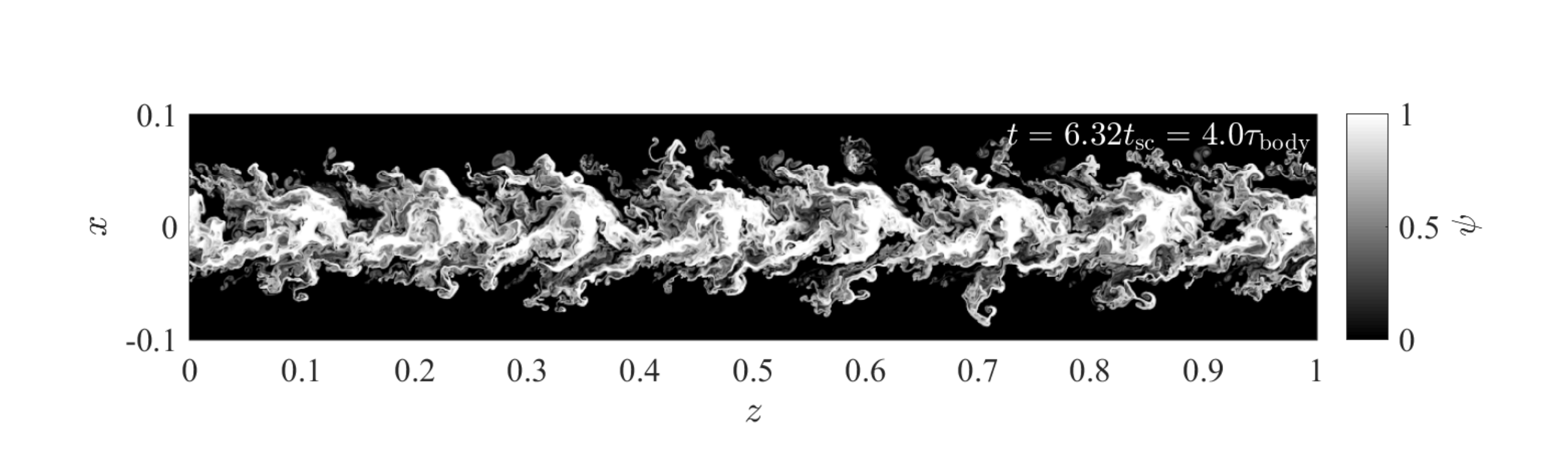}
	}
	\caption{Snapshots of $\psi(x,z)$, a passive scalar field used as a Lagrangian tracer for the stream fluid (see \Cref{sec:methods-analysis}), taken from a planar slab simulation at different times. The times are presented both in terms of the stream sound crossing time, $\tsc$, and the deceleration timescale, $\taubody$ (see \Cref{sec:body-results-deceleration}). The unperturbed initial conditions are $(\Mb=2.5,\delta=10)$, placing the stream in the body-mode-dominated regime. The initial stream radius is $\Rs=1/128$. The initial perturbations are interface-only with a spectrum spanning $0.25\Rs\leq\lambda\leq 16\Rs$ and a fixed initial amplitude $H=2\Delta\simeq0.03\Rs$ (see \Cref{sec:body-simulations}). This setup matches the assumed initial conditions for \Cref{fig:thNL}.}
	\label{fig:demo-body-color}
\end{figure*}

\Cref{fig:demo-body-color} follows a simulation of body modes growing in a planar slab with $(\Mb=2.5,\delta=10)$ and $\Rs=1/128$. The initial perturbations included several wavelengths in the range $0.25\Rs \leq \lambda \leq 16\Rs$ (see \Cref{sec:body-simulations}) with initial amplitude $H=2\Delta\simeq0.03\Rs$, as assumed for the analysis presented in \Cref{fig:thNL}.  The second panel of \Cref{fig:demo-body-color} shows the shape of the stream at $t\simeq 1.6\tsc$. A few of the initial perturbation wavelengths can be identified, most notably an S-mode with $\lambda = 4\Rs$, which is roughly the dominant wavelength predicted by \Cref{fig:thNL} for this time. In the third panel of \Cref{fig:demo-body-color}, corresponding to $t \simeq 3.2\tsc$, the dominant mode is an S-mode with $\lambda = 16\Rs$, although some contribution from $\lambda = 8\Rs$ may be observed. The total displacement at this time is $h \approx \Rs$ relative to the initial stream/background interface. The combination of time, wavelengths and amplitude is in excellent agreement with the analytic prediction for the critical perturbation mode. Accordingly, the subsequent growth of the critical mode leads to the final disintegration of the stream, seen in the bottom two panels in \Cref{fig:demo-body-color}. 

\begin{figure*}
	\centering
	\subfloat{
		\includegraphics[trim={1cm 1.4cm 1.25cm 1.75cm}, clip,width=0.7\textwidth]{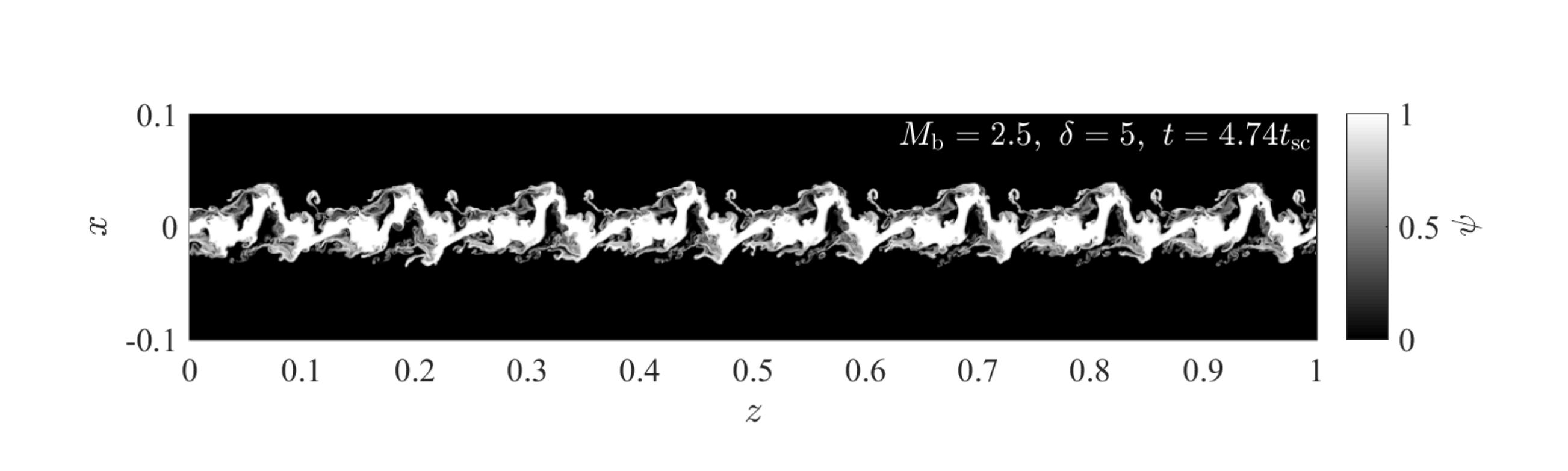}
	}\\
	\subfloat{
		\includegraphics[trim={1cm 1.4cm 1.25cm 1.75cm}, clip,width=0.7\textwidth]{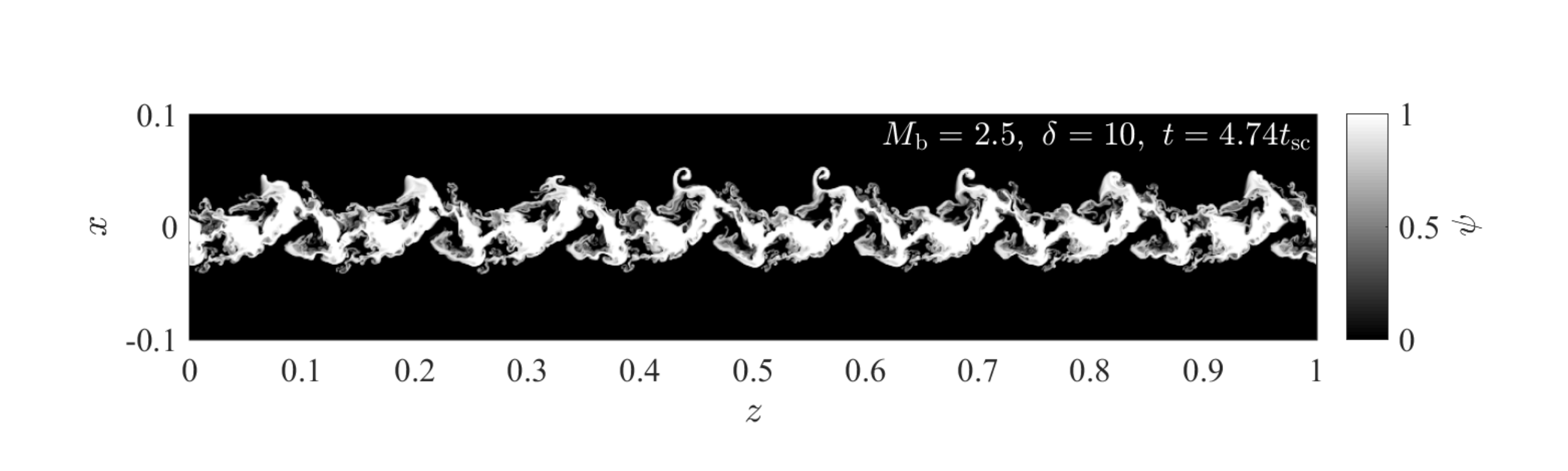}
		\label{fig:demo-body-critical-d10m25}
	}\\
	\subfloat{
		\includegraphics[trim={1cm 1.4cm 1.25cm 1.75cm}, clip,width=0.7\textwidth]{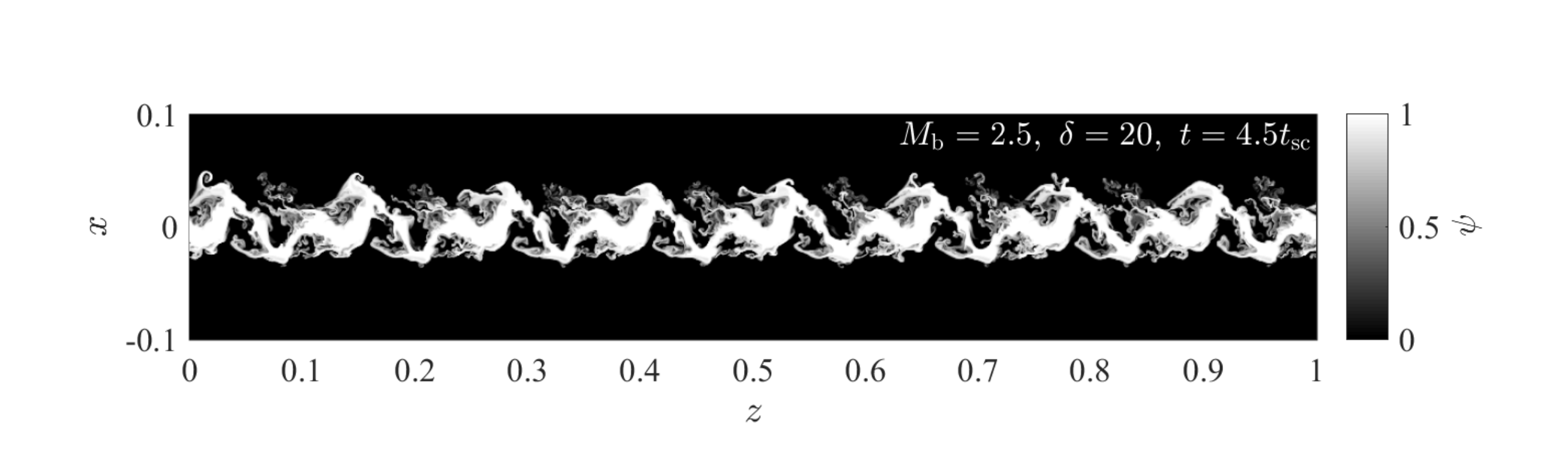}
		\label{fig:demo-body-critical-d20m25}
	}\\
	\subfloat{
		\includegraphics[trim={1cm 1.4cm 1.25cm 1.75cm}, clip,width=0.7\textwidth]{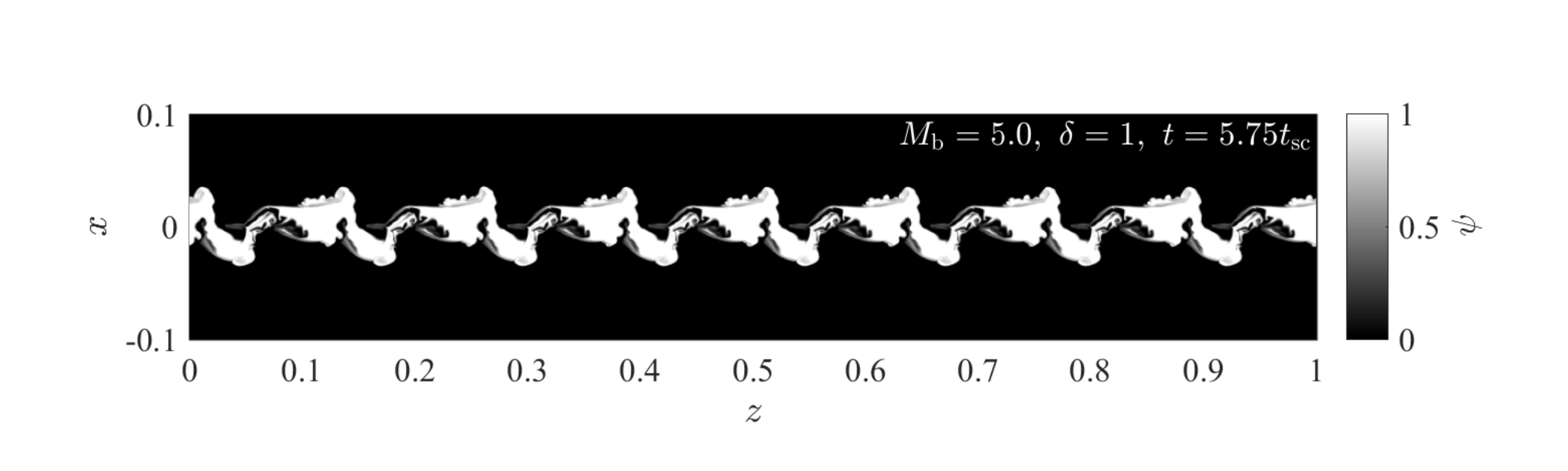}
		\label{fig:demo-body-critical-d1m50}
	}\\	
	\subfloat{
		\includegraphics[trim={1cm 0.8cm 1.25cm 1.75cm}, clip,width=0.7\textwidth]{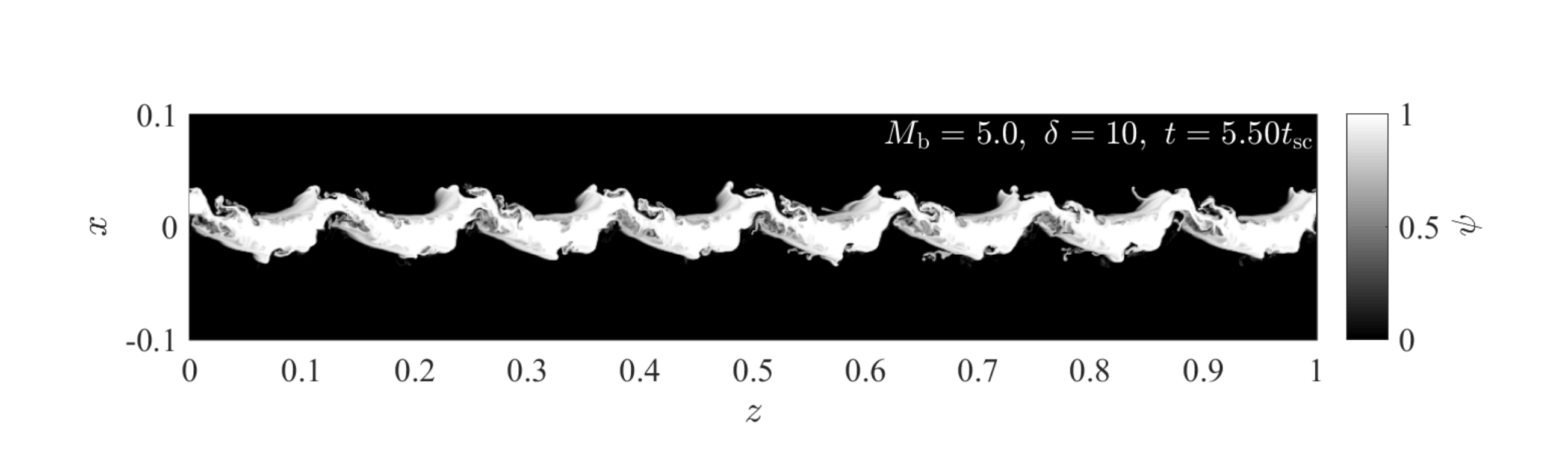}
		\label{fig:demo-body-critical-d10m50}
	}
	\caption{Snapshots of $\psi(x,z)$, a passive scalar field used as a Lagrangian tracer for the stream fluid (see \Cref{sec:methods-analysis}), taken from different body mode simulations a short time after the transition to nonlinear evolution. The simulations have different $\unpert$ but share the same specific realization of interface-only initial perturbations with a spectrum spanning $0.25\Rs\leq\lambda\leq 16\Rs$ and a fixed initial amplitude $H=2\Delta\simeq0.03\Rs$ (see \Cref{sec:body-simulations}). The initial stream radius is $\Rs=1/128$ in all cases. The second panel in this figure is identical to the third panel in \Cref{fig:demo-body-color}.}
	\label{fig:demo-body-critical}
\end{figure*}

The same analytic method is used in \Cref{app:thNL} to produce plots analogous to \Cref{fig:thNL} for other values of Mach number, $\Mb$, and density contrast, $\delta$. These yield nearly identical conclusions, namely $(\lambdacrit\gsim10\Rs,\ncrit=0)$, throughout the range of parameters relevant to cold streams feeding galaxies. \Cref{fig:demo-body-critical} shows a few examples of simulations exhibiting this behavior. Although some differences in morphology are visible, in all cases the stream breaks up due to a sinusoidal perturbation with $\lambdacrit=16\Rs$, following a very similar timetable.

Assuming roughly the same initial amplitude is given to all perturbation modes (white noise), the transition to nonlinearity for a given combination of $\unpert$ is determined by $\tNL$ of the critical perturbation mode,  $\tNLcrit=\tNL(\lambdacrit,\ncrit)$. According to \Cref{eq:tNL}, this depends on the KH time for that mode and on the ratio of the initial amplitude, $H$, to the amplitude at transition, $\hNL$. For the critical perturbation mode, we have $\hNL\simeq\Rs$ by definition. In \Cref{app:tkh-body}, we show that the KH time for the critical perturbation mode in cosmic cold streams is roughly the stream sound crossing time, $\tkh\simeq \tsc = 2\Rs/\cs$, with little dependence on $\unpert$. Therefore, we predict that in cosmic cold streams the transition to nonlinearity will occur at
\begin{equation}
\label{eq:tNLcrit}
\tNLcrit = \ln\left(\frac{\Rs}{H}\right)\tsc = \ln\left(\frac{\Rs}{H}\right) \frac{2\Rs}{\cs}
\end{equation}

\subsubsection{Deformation and Deceleration}
\label{sec:body-results-deceleration}

\begin{figure}
	\includegraphics[trim={0.25cm 0cm 1.0cm 1.25cm}, clip,width=0.475\textwidth]{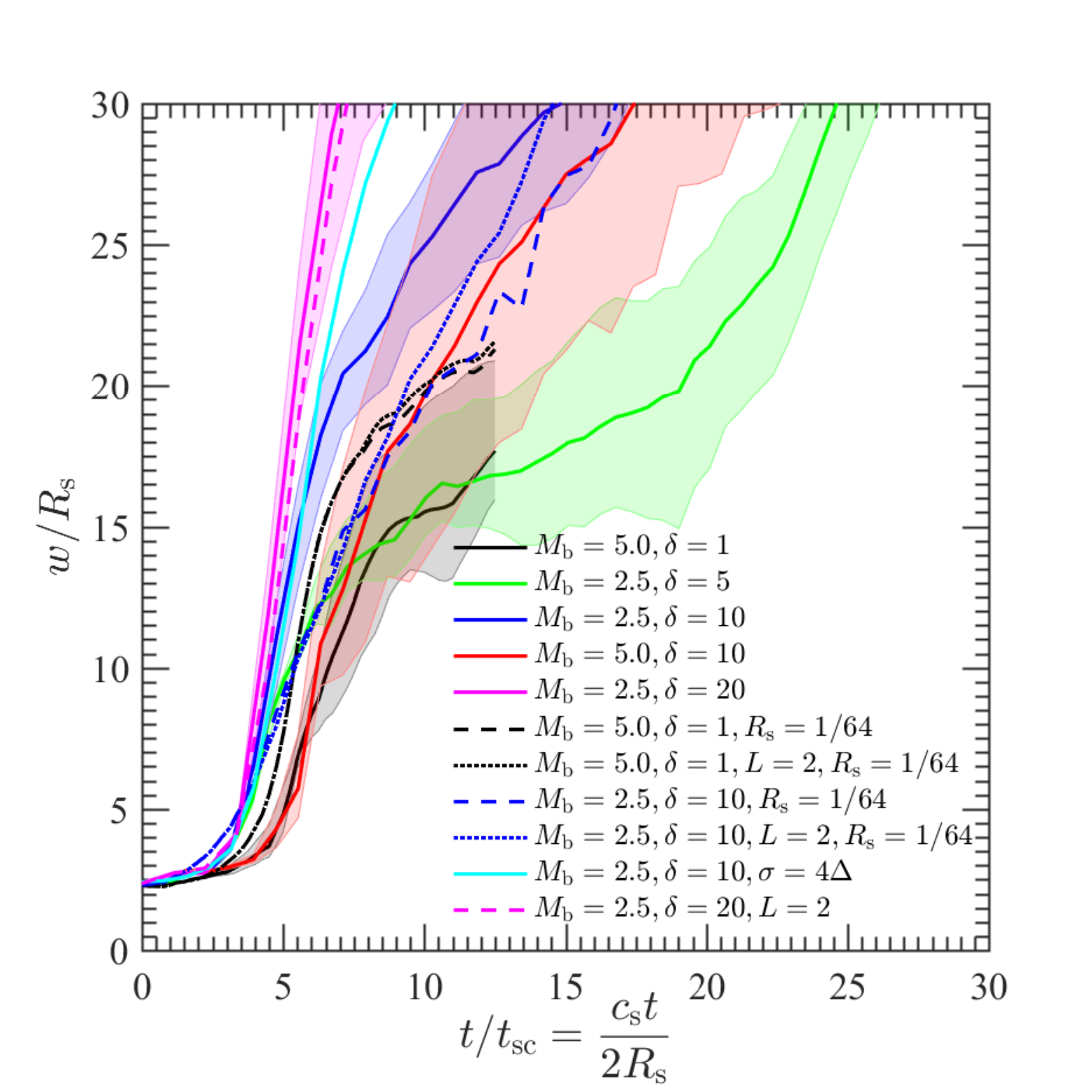}
	\caption{Nonlinear growth of body modes in different planar slab simulations listed in \Cref{tab:parameters-body}. The vertical axis shows the stream width, $w$, normalized by its initial radius, $\Rs$. The initial width is $w(t=0)=2\Rs$. For the fiducial domain size $L=1$ and stream radius $\Rs=1/128$, each combination of $\unpert$ was repeated with three realizations of interface-only initial perturbations, as described in \Cref{sec:body-simulations}. The shaded areas of different colors correspond to the range of results spanned by the different realizations, whereas the solid lines show mean values. The dashed lines show runs with a larger stream radius, $\Rs=1/64$, with one realization for each case. The dotted lines show runs with both a larger stream radius, $\Rs=1/64$, and a larger domain size, $L=2$. The solid cyan line marks a simulation with a larger smoothing width, $\sigma$, in the initial conditions.}
	\label{fig:width-body}
\end{figure}

The nonlinear growth of the critical mode and the subsequent disruption of the stream can be observed by following the evolution of its width, defined in \Cref{eq:stream-width}.  \Cref{fig:width-body} shows $w(t)$ for the simulations listed in \Cref{tab:parameters-body}. In all cases, the streams experience only modest widening at early times, with no more than a $\sim 20\%$ increase in width during the first $\sim 3\tsc$. Around  this point in time, corresponding to the analytic prediction for $\tNL$ (see \Cref{sec:body-results-critical}), we observe a sharp transition to rapid growth. This is interpreted as the critical perturbation mode taking over and initiating the disintegration of the stream. The width of the stream increases to approximately a factor of $\sim5-10$ of its initial value within $\sim1-2\tsc$, at which point the stream breaks up and the growth rate starts to decrease again. 

At later times, \Cref{fig:width-body} shows significant scatter among different combinations of Mach number and density contrast, with larger values of $\delta$ corresponding to higher late-time growth rates. Nevertheless, both the time of transition to nonlinearity and the maximum growth rate are similar in all cases. All of the cases with $\Mb=2.5$, which are the most relevant to cold streams feeding galaxies, show $\tNLcrit\simeq3.5\tsc$ regardless of $\delta$, in excellent agreement with the analytic predictions. 

\begin{figure*}
	\centering
	\subfloat{
		\includegraphics[trim={1cm 1.4cm 1.25cm 1.75cm}, clip,width=0.7\textwidth]{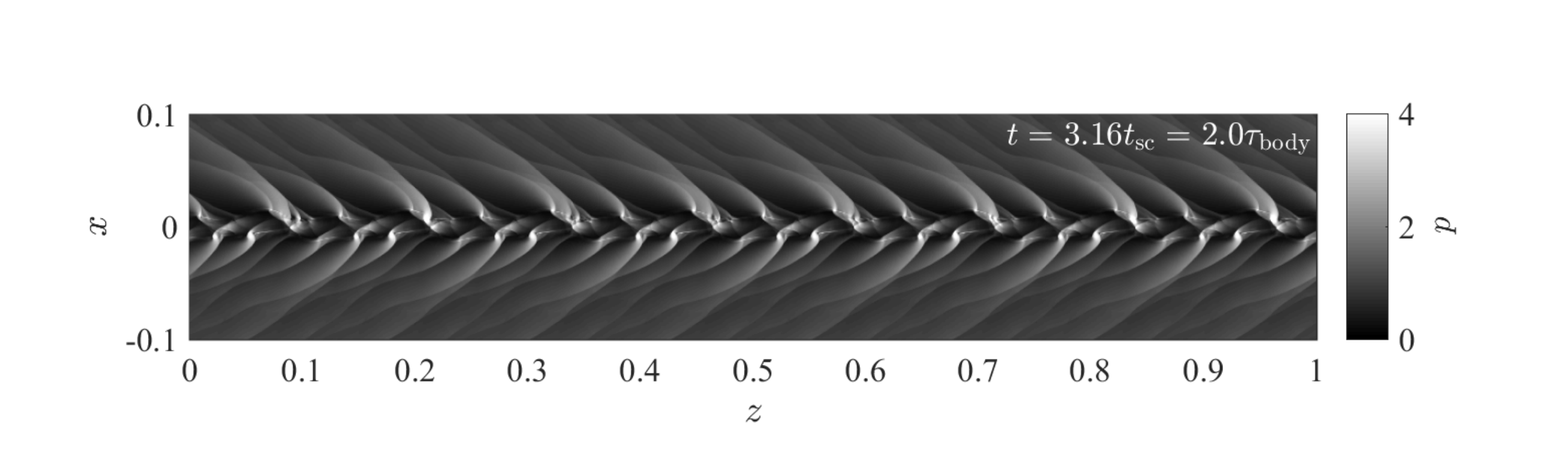}
		\label{fig:demo-body-prs2}
	}\\
	\subfloat{
		\includegraphics[trim={1cm 1.4cm 1.25cm 1.75cm}, clip,width=0.7\textwidth]{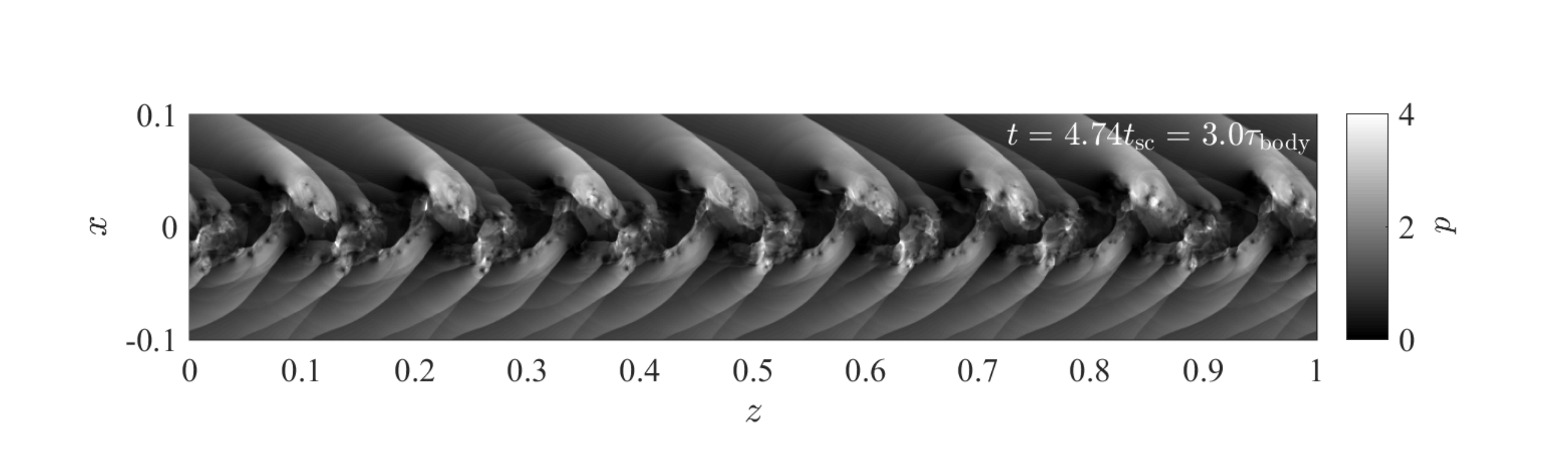}
		\label{fig:demo-body-prs3}
	}\\
	\subfloat{
		\includegraphics[trim={1cm 0.8cm 1.25cm 1.75cm}, clip,width=0.7\textwidth]{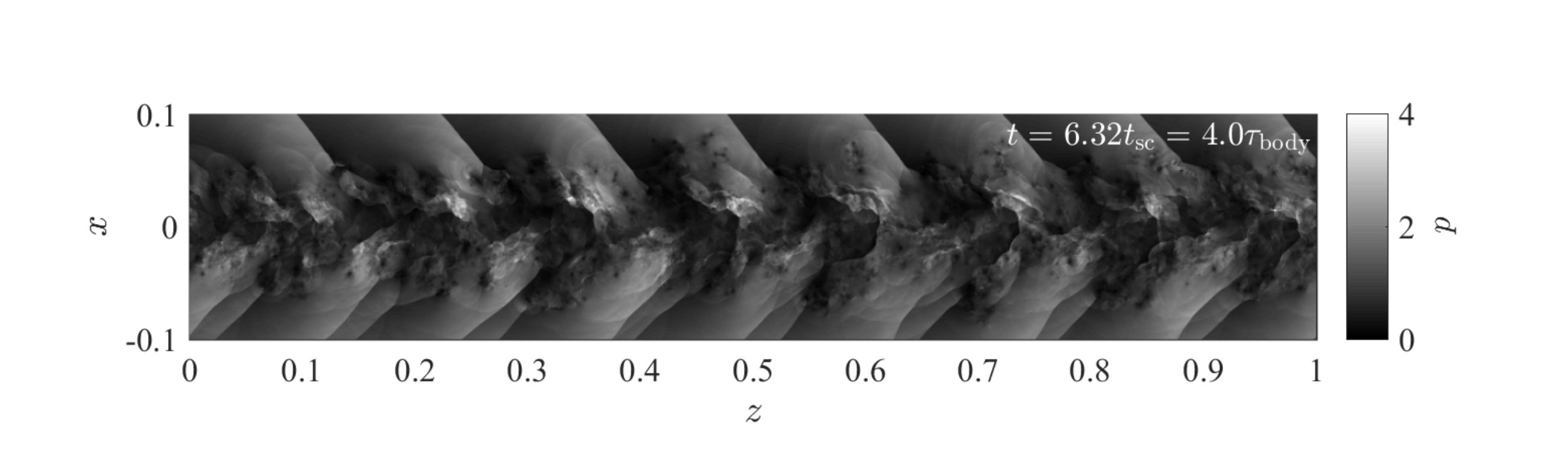}
		\label{fig:demo-body-prs4}
	}
	\caption{Snapshots of pressure, $p(x,z)$, taken from the same run as in \Cref{fig:demo-body-color}. The times are presented both in terms of the stream sound crossing time, $\tsc$, and the deceleration timescale, $\taubody$ (see \ref{sec:body-results-deceleration}). The initial pressure was $p=1$ in simulation units throughout the computational domain.}
	\label{fig:demo-body-prs}
\end{figure*}

The critical perturbation mode bends the stream into a sinusoidal shape, effectively driving a piston through the background medium at every crest of the sinusoid. This produces a periodic pattern of weak shocks, propagating away from the stream at approximately the speed of sound $\cb$, as shown in \Cref{fig:demo-body-prs}. These waves facilitate the transfer of momentum from stream to background fluid. We denote the characteristic time for deceleration due to this interaction by  $\taubody$. The appropriate timescale can be derived by requiring that the mass of background fluid overtaken by the outward-propagating waves be equivalent to the mass of the stream, namely $\cb \taubody L \rhob  \sim \Rs L \rhos $. Thus we obtain the definition
\begin{equation}
\label{eq:taubody}
\taubody \equiv \frac{\delta\Rs}{\cb} = \frac{\sqrt{\delta}}{2}\tsc.
\end{equation}

\begin{figure}
	\includegraphics[trim={0.25cm 0cm 1.0cm 1.25cm}, clip,width=0.475\textwidth]{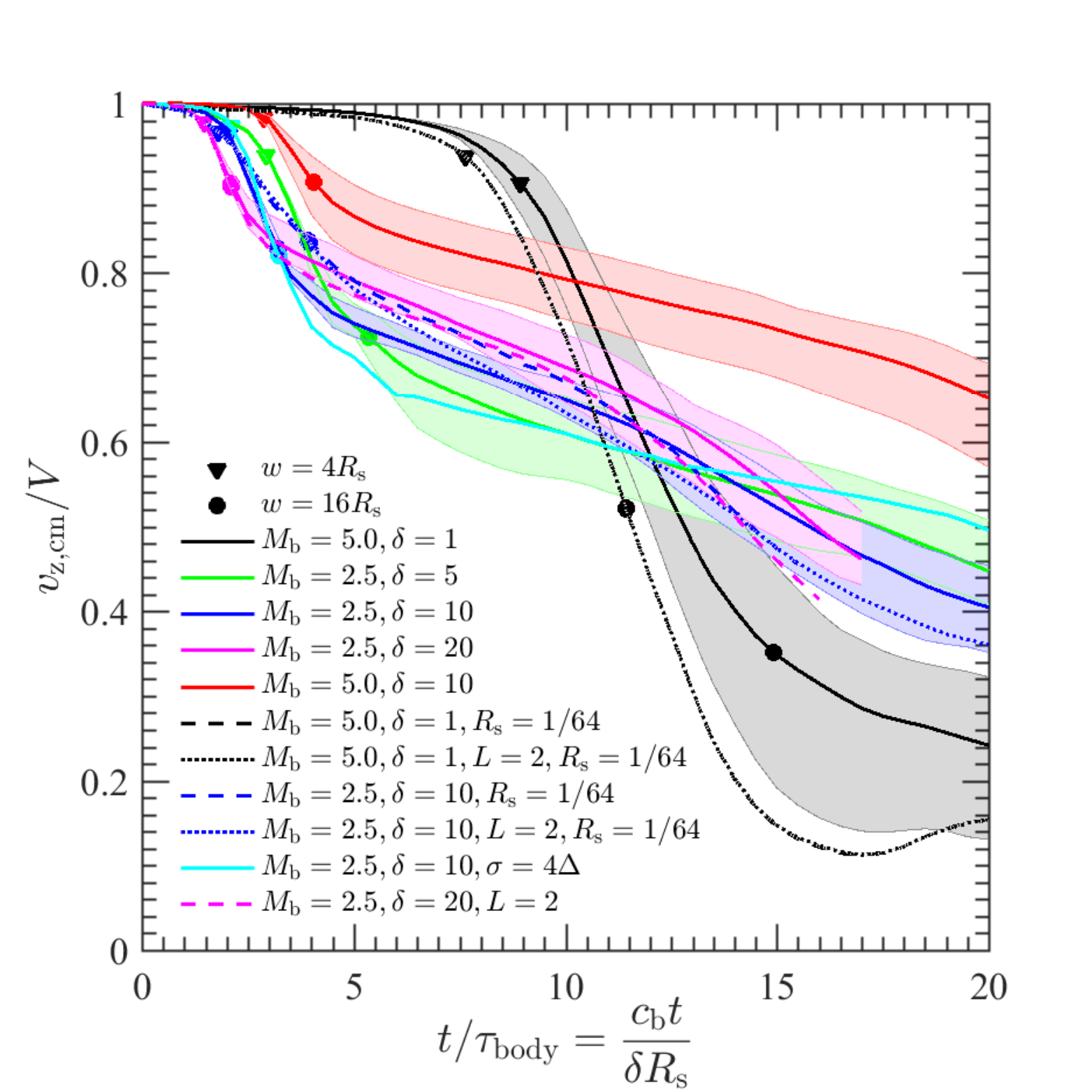}
	\caption{Stream deceleration due to nonlinear growth of body modes in the planar slab simulations listed in \Cref{tab:parameters-body}. The vertical axis shows the center of mass velocity of the stream in the downstream direction, $\vcm$, normalized by its initial velocity, $V$. The legend is identical to \Cref{fig:width-body}. The triangles on each of the lines correspond to $w=4\Rs$, marking the transition to nonlinear evolution, whereas the circles correspond to $w=16\Rs$, marking the disruption of the stream.}
	\label{fig:deceleration-body}
\end{figure}

\Cref{fig:deceleration-body} shows the evolution of $\vcm$, the center of mass velocity of the stream defined in \Cref{eq:vcm}, for the same set of simulations used previously. The results are consistent with those in \Cref{fig:width-body}. During the first $\sim 3\tsc$ the streams maintain their initial velocity, followed by a period of sharp deceleration due to the nonlinear evolution of the critical perturbation mode. This corresponds to the rapid growth in width seen in \Cref{fig:width-body}, as demonstrated by the triangles and circles marking $w=4\Rs$ and $w=16\Rs$ respectively in \Cref{fig:deceleration-body}. The rate of deceleration during this stage fits the predicted timescale, \Cref{eq:taubody}. All cases show roughly the same slope when the velocity is normalized by the initial velocity, $V$, and the time is normalized by the deceleration timescale, $\taubody$, namely
\begin{equation}
\label{eq:deceleration-rate-body-major}
\dot{v}_{\rm z,cm} \simeq -0.12\frac{V}{\taubody}
\end{equation}
regardless of $\unpert$. After the stream breaks, roughly at the time corresponding to $w=16\Rs$ (i.e. $h\approx\lambdacrit/2$), we observe a significant reduction in the deceleration rate, to values in the neighborhood of
\begin{equation}
\label{eq:deceleration-rate-body-minor}
\dot{v}_{\rm z,cm} \simeq -0.016\frac{V}{\taubody},
\end{equation}
with somewhat greater variability with $\unpert$.

\begin{figure}
	\includegraphics[trim={0.25cm 0.5cm 1cm 1cm}, clip, width=0.475\textwidth]{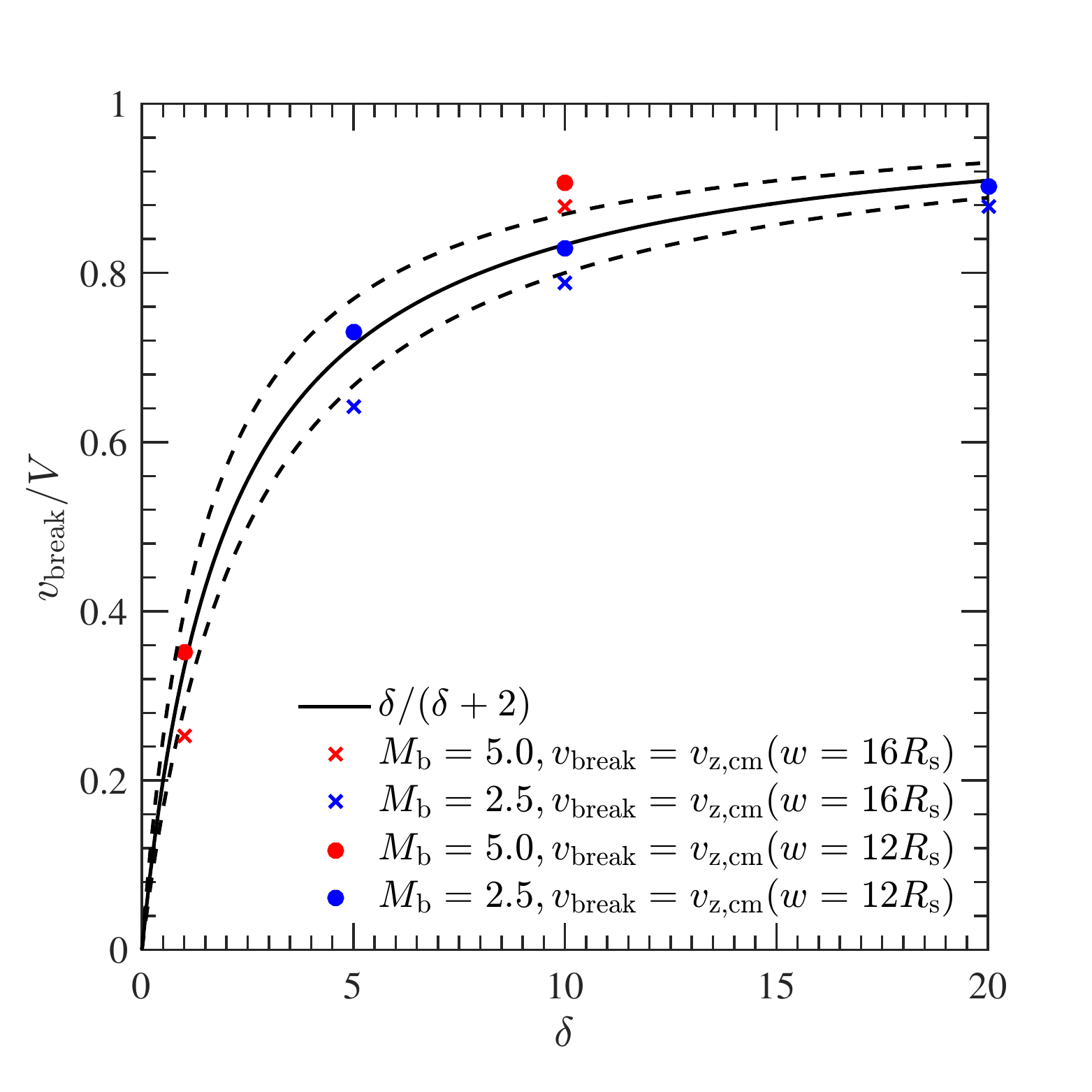}
	\caption{Stream velocity at the time of disintegration. Two definitions are used: $v_{\rm break}=\vcm(w=12\Rs)$ marked by circles and $v_{\rm break}=\vcm(w=16\Rs)$ marked by crosses. The colors represent different Mach numbers: $\Mb=2.5$ (blue) and $\Mb=5.0$ (red). The data is based on the same simulations shown in \Cref{fig:deceleration-body}. The solid black line corresponds to the toy model, \Cref{eq:vbreak}, with the calibrated value $x_{\rm eff}=3\Rs$. The dashed black lines use $x_{\rm eff}=2.5\Rs,3.5\Rs$ as a crude representation of the uncertainty.}
	\label{fig:vbreak-body}
\end{figure}

The velocity of the stream immediately after its disintegration, denoted $\vbreak$, can be estimated by evaluating $\vcm$ at the moment corresponding to a suitable width threshold. \Cref{fig:vbreak-body} shows $\vbreak$ using two different criteria for the moment of breaking, $v_{\rm break}=\vcm(w=12\Rs)$ and $v_{\rm break}=\vcm(w=16\Rs)$, marked by circles and crosses respectively. Intuitively, denser streams end up with higher $\vbreak$. 

The observed dependence of $\vbreak$ on $\delta$ can be interpreted on the basis of the following toy model. By the time it breaks, the stream shares its momentum with some amount of background fluid, presumed to initially occupy the volume $\Rs<|x|<x_{\rm eff}$. Conservation of momentum between the initial state and the final state then indicates $2\Rs L \rhos V = \left[2\Rs L\rhos + 2(x_{\rm eff}-\Rs)L\rhob\right]\vbreak$. Assuming that $x_{\rm eff}$ is independent of $\unpert$ we find
\begin{equation}
\label{eq:vbreak}
\vbreak(\delta) = \frac{\delta}{\delta+\frac{x_{\rm eff}-\Rs}{\Rs}}V.
\end{equation}
\Cref{fig:vbreak-body} shows that our results fit this form\footnotemark~with $x_{\rm eff} \approx 3\Rs$,
\begin{equation}
\label{eq:vbreak-calibrated}
\vbreak(\delta) \simeq \frac{\delta}{\delta+2}V.
\end{equation}
\footnotetext{The value of $x_{\rm eff}$ is notably less than the extent of the perturbed region after the stream has been broken, $w/2 \sim \lambdacrit/2 \sim 8\Rs$, implying that the momentum exchange in this process is relatively inefficient.}

In addition to the dependence on density contrast, \Cref{fig:vbreak-body} shows that $\vbreak$ changes with Mach number. However, when considering the actual range of Mach numbers for which cosmic cold streams are susceptible to body modes, $1.5\lsim \Mb\lsim 2$, this dependence can be largely ignored and \Cref{eq:vbreak-calibrated} can be used to predict $\vbreak$ with reasonable accuracy.

The results presented in this section show little sensitivity to the specific realization of initial perturbations used, the value of the smoothing width, $\sigma$, and the domain size, $L$. The transition time, $\tNL$, the deceleration rate, \Cref{eq:deceleration-rate-body-major}, and the velocity at disintegration, $\vbreak$, are particularly robust. The resolution effect is somewhat more important. In particular, the $(\Mb=2.5,\delta=10)$ case shows some variability both in deceleration rate and in stream width when $\Rs$ is changed at a fixed resolution, changing the number of cells resolving the width of the stream, $\Rs/\Delta$. Nevertheless, these difference can be considered small for the purposes of our analysis of KHI in cold streams feeding galaxies.

The results of this section are used in \Cref{sec:application-inflow} to predict the effect of KHI-induced deceleration on the inflow rate of cosmic cold streams. We find that streams unstable to body modes undergo negligible deceleration, too small to account for the constant inflow velocity observed in (unresolved) cosmological simulations. Whether this result extends to cylindrical streams in 3D remains to be seen (Mandelker et al. in preparation).

\subsubsection{Heating}
\label{sec:body-results-heating}

The nonlinear growth of the critical perturbation mode drives weak, oblique shock waves through the stream fluid itself, seen in \Cref{fig:demo-body-prs}, in addition to the outward-propagating waves mentioned in \Cref{sec:body-results-deceleration}. These waves reverberate back and forth inside the stream, causing it to heat up. As in the case of surface modes (see \Cref{sec:surface-results-heating}), this heating is interesting in the astrophysical context primarily due to its potential to produce observable Lyman-$\alpha$ or other emissions.

\begin{figure*}
	\begin{tabular}{ccc}
		\subfloat{
			\includegraphics[trim={0cm 0.2cm 1.0cm 1.5cm}, clip, height=0.25\textheight]
			{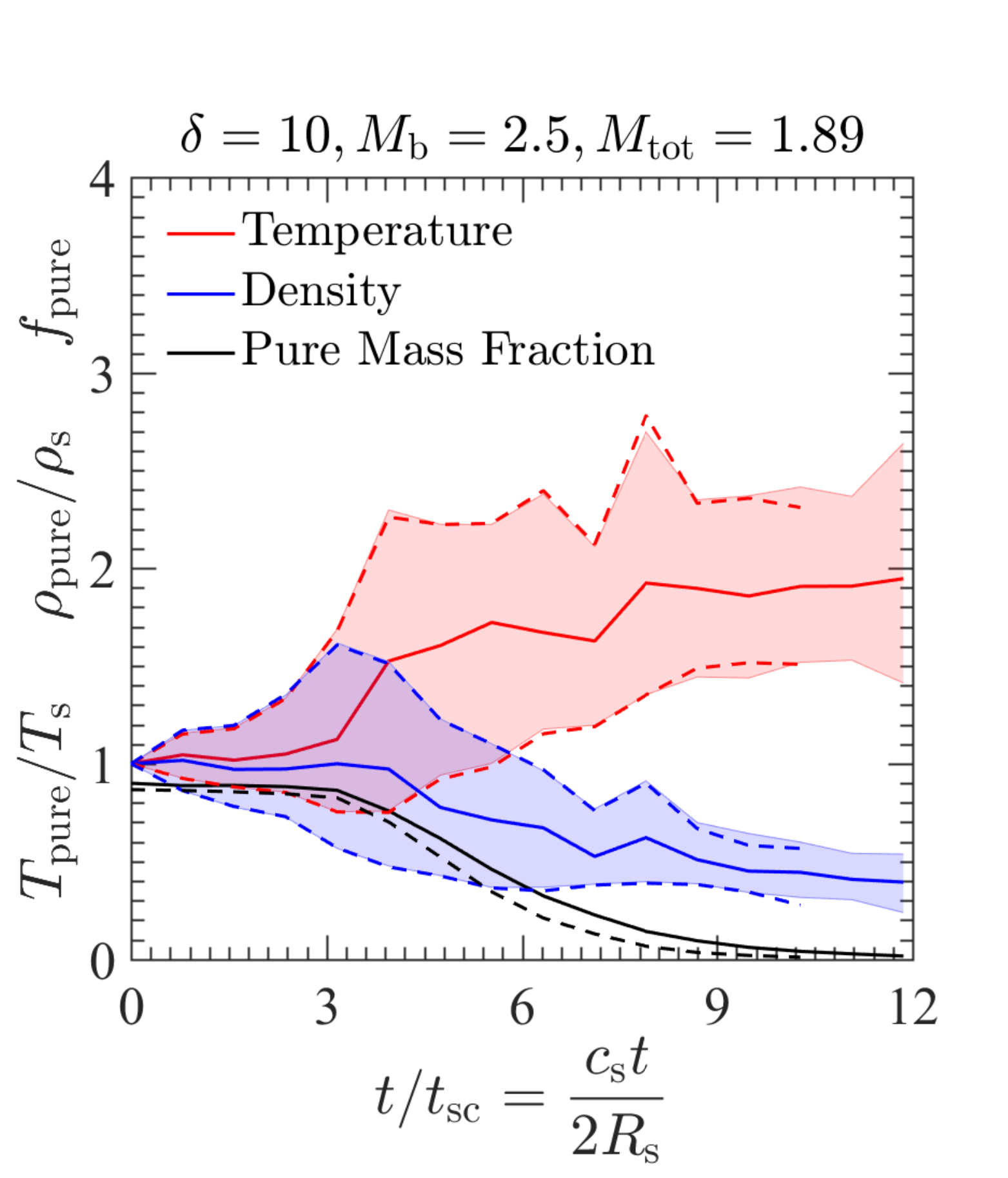}
		}&
		\subfloat{
			\includegraphics[trim={1.25cm 0.2cm 1.0cm 1.5cm}, clip, height=0.25\textheight]
			{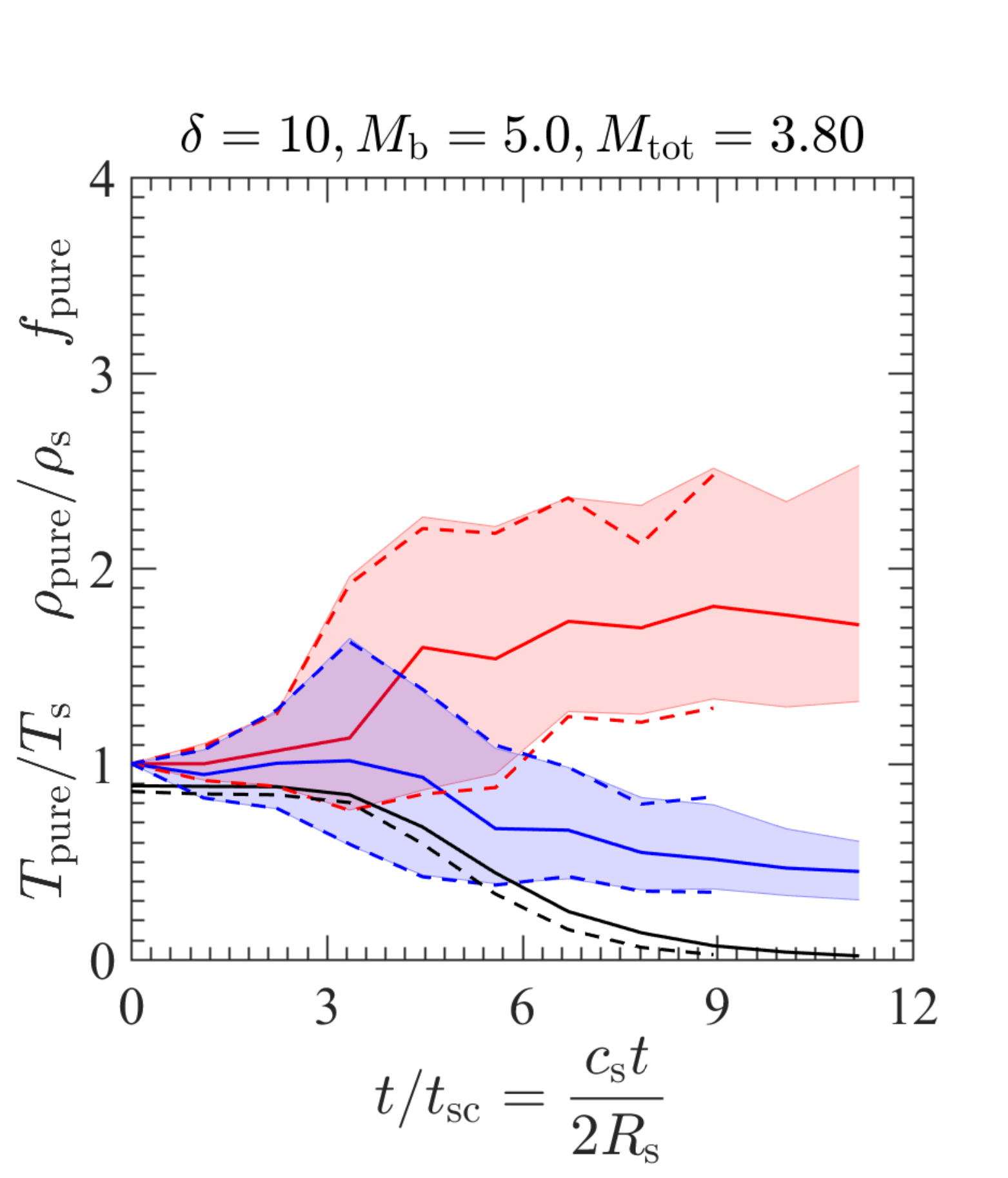}
		}&
		\subfloat{
			\includegraphics[trim={1.25cm 0.2cm 1.0cm 1.5cm}, clip, height=0.25\textheight]
			{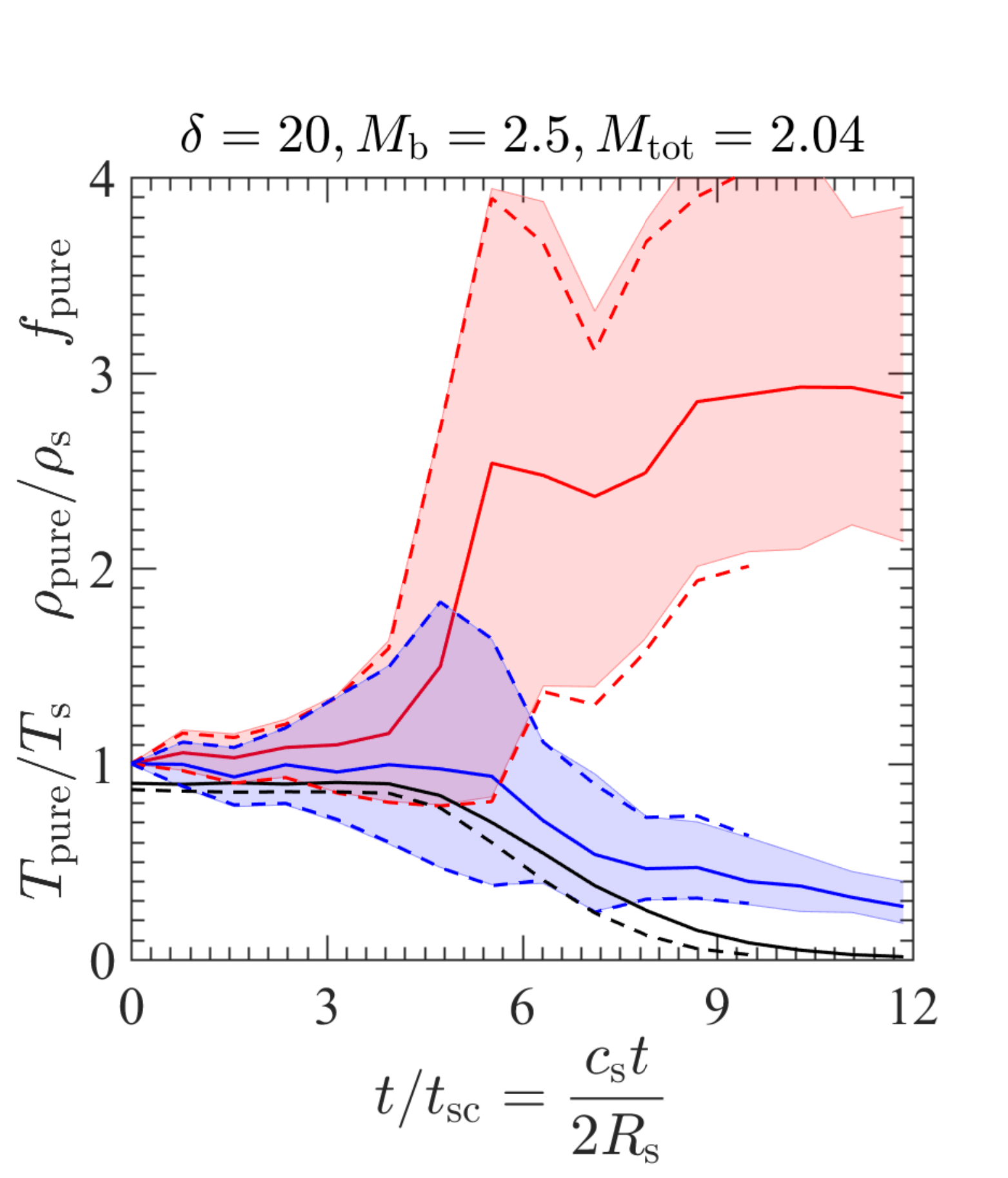}
		}
	\end{tabular}
	\caption{The thermodynamic state of the stream in simulations of nonlinear growth of body modes. Each panel represents the time evolution of a single run with a different value of $\unpert$. In all cases the initial conditions included the same realization of interface-only perturbations. Different realizations produce nearly identical results. The vertical axis shows the temperature and density of the unmixed stream fluid, $\Tpure$ and $\rhopure$, normalized by their respective initial values, $\Ts$ and $\rhos$, as well as the mass fraction of unmixed fluid, $\fpure$. The horizontal axis shows the time normalized by the sound crossing time, $\tsc$. Two different values of the threshold $\epsener$ in \Cref{eq:pure-stream} are used to define unmixed fluid. For $\epsener=0.1$, the solid lines show $\Tpure$ (red), $\rhopure$ (blue) and $\fpure$ (black), whereas the shaded red/blue areas span the 10th to 90th mass percentiles of temperature/density. For $\epsener=0.01$ the black dashed line shows $\fpure$ and the red/blue dashed lines span the 10th to 90th mass percentiles of temperature/density. The results obtained using both values of $\epsener$ are very similar, indicating that any observed heating is a physical phenomenon affecting the bulk of the stream, rather than a numerical artifact affecting some small boundary layer.}
	\label{fig:thermo-body}
\end{figure*}

\Cref{fig:thermo-body} shows how the temperature and the density of the stream evolve with time in three simulations with different $\unpert$. All cases show a mild increase in temperature at early times, followed by a short period of rapid heating around $t\simeq \tNL\simeq 3.5\tsc$, corresponding to stream breakup, and a plateau at later times. For $\Mb=2.5$, the mean temperature of the stream increases by a factor of $\sim1.5-2$, with a weak dependence on $\delta$. For $(\Mb=5.0,\delta=10)$, the shock waves are stronger, resulting in a factor of $2.5-3$ increase in temperature. The densities decrease by a similar factor.

The increase in stream temperature seen in \Cref{fig:thermo-body} for body modes is greater, in relative terms, than the increase seen in \Cref{sec:surface-results-heating} for surface modes. Nevertheless, this still represents a small fraction of the overall energy budget of the stream. As was the case for surface modes, this means that body mode instability is a significantly weaker power source for Lyman-$\alpha$ emission than those appearing in previous works \citep{Goerdt10,Dijkstra09}. We discuss this point in detail in \Cref{sec:application-thermo}.

The results of this section are based on \emph{adiabatic} simulations and therefore do not take into account the effect of radiative cooling on KHI, which are potentially significant. In the same vein as \Cref{sec:surface-results-heating}, a careful study of body mode instability with a realistic cooling function is necessary. This will be reported in a forthcoming paper (Mandelker et al. in preparation).

\section{Application to Cold Streams Feeding Galaxies}
\label{sec:application}

In this section, we use the results obtained in \Cref{sec:surface,sec:body} to predict the outcome of KHI in cold streams feeding galaxies. In \Cref{sec:application-parameters}, we estimate the values of the parameters controlling KHI in cold streams. In \Cref{sec:application-breakup} we predict the potential for breakup of streams. \Cref{sec:application-inflow} addresses the effect on their inflow rate. Finally, in \Cref{sec:application-thermo} we comment on heating of the streams, in particular with regard to the potential for Lyman-$\alpha$ emission.

\subsection{Estimating $\Mb$, $\delta$ and $\Rs/\Rv$ for Cold Streams}
\label{sec:application-parameters}

Our results depend primarily on three dimensionless parameters: the Mach number, $\Mb$, the density contrast, $\delta$, and ratio of stream radius to virial radius, $\Rs/\Rv$. In order to estimate these three parameters, we assume that cosmological simulations can be used to infer the large-scale properties of cold streams and the hot halo gas, reduced to a number of more fundamental parameters listed in \Cref{tab:stream-parameters}. The values therein are meant as a rough guide, to be refined in future work. We use the resulting expressions for $(\Mb,\delta,\Rs/\Rv)$ to assess, based on our numerical and analytic models of KHI, the effects of KHI on cold streams.

Cosmological simulations indicate that the stream velocity is approximately constant during infall \citep{Dekel09,Goerdt15a,Goerdt10} and comparable to the halo virial velocity,
\begin{equation}
\label{eq:V-to-Vvir}
V = \eta \Vv,
\end{equation}
with $0.5\lsim \eta \lsim 1$ \citep{Goerdt15a}. Using the simplifying assumption that the halo CGM is isothermal, with a temperature proportional to the virial temperature, $\Tb=\Theta_{\rm b}\Tv$, the sound speed in the halo is 
\begin{equation}
\label{eq:cb}
\cb = \sqrt {\frac{\gamma K_{\rm B}\Tb}{\mu m_{\rm p}}} = \sqrt{\frac{\gamma K_{\rm B}\Theta_{\rm b}\Tv}{\mu m_{\rm p}}},
\end{equation}
where $K_{\rm B}$ is the Boltzmann constant, $\Tv$ is the virial temperature, $\mu m_{\rm p}$ is the mean particle mass, $\gamma=5/3$ is the adiabatic index of the gas and $\Theta_{\rm b} \gsim 3/8$ \citep{db06}. The temperature, $\Tv$, can be related to the velocity, $\Vv$, using virial equilibrium, 
\begin{equation}
\label{eq:virial-equilibrium}
\frac{3}{2}K_{\rm B}\Tv \simeq \frac{1}{2}\frac{G\Mv \mu m_{\rm p}}{\Rv} = \frac{1}{2}\mu m_{\rm p} \Vv^2,
\end{equation}
where $\Mv$ is the virial mass of the halo. Combining \Cref{eq:virial-equilibrium,eq:cb} we obtain $\Vv/\cb = \sqrt{3/\gamma\Theta_{\rm b}} \lsim 2.2$, and plugging this into \Cref{eq:V-to-Vvir} we find
\begin{equation}
\label{eq:Mb}
\Mb = \frac{V}{\cb} = \eta \frac{\Vv}{\cb} \lsim 2.2 \eta
\end{equation}
Taking the uncertainties in $\Theta_{\rm b}$ and $\eta$ into account, we assume values in the range $\Mb\sim 0.75-2.25$, as was done in \PIt.

To estimate $\delta$, we assume that the halo CGM and the stream are in pressure equilibrium. Hence, as in the previous sections, the density contrast is the inverse of the temperature ratio, $\delta = \rhos/\rhob = \Tb / \Ts$. Assuming that both the halo CGM and the stream are isothermal, this ratio is constant at any halo-centric radius. Due to efficient cooling in the dense streams, they do not support a stable shock at the virial radius, and their temperature is set by the cooling curve, $\Ts = \Theta_{\rm s} 10^4\K$, with typical values of $1\lsim\Theta_{\rm s}\lsim3$ \citep{db06}. Therefore we write
\begin{equation}
\label{eq:delta-Tv}
\delta = \frac{\Theta_{\rm b}}{\Theta_{\rm s}} \frac{\Tv}{10^4\K} = \Theta \frac{\Tv}{10^4\K},
\end{equation}
where $\Theta\equiv\Theta_{\rm b}/\Theta_{\rm s}$ is roughly in the range $0.3\lsim\Theta\lsim1.1$.

The virial temperature can be related to the redshift and the virial mass \citep[see appendix A in][]{db06}
\begin{equation}
\label{eq:Tvir}
\Tv = 1.5\times 10^6\K \times M_{12}^{2/3} (1+z)_{3}^{-1},
\end{equation}
where $(1+z)_{3} = (1+z)/3$ and $M_{12} = \Mv/10^{12}\Msun$. Substituting \Cref{eq:Tvir} for $\Tv$ in \Cref{eq:delta-Tv} we get
\begin{equation}
\label{eq:delta}
\delta = 150\times \Theta M_{12}^{2/3} (1+z)_{3}^{-1}.
\end{equation}
Using the aforementioned range of $\Theta$, this results in $40\lsim\delta\lsim 160$ for $\Mv=10^{12}\Msun$ halos at redshift $z=2$. This is somewhat higher than the range we assumed in \PIt, which we adopt here as well for consistency, $\delta\sim 10-100$. The latter is also consistent with cosmological simulations \citep{Ocvirk08,Dekel09,Goerdt10}.

In \PIt~we assumed $\Rs/\Rv\sim 0.005-0.05$, based on cosmological simulations. This can be improved by relating $\Rs/\Rv$ to the density contrast and Mach number, using the cosmological accretion rate as a constraint, as shown below. A similar idea is used to estimate dark matter filament properties in \citet{Birnboim2016}.

On the one hand, the accretion rate of gas with density $\rhos$ flowing with velocity $V$ along a stream with radius $\Rs$ is
\begin{equation}
\label{eq:Msdot-flow}
\dot{\Ms} = \pi \Rs^2 \rhos V.
\end{equation}
On the other hand, the overall accretion rate of matter into the virial radius can be expressed as \citep{Dekel2013},
\begin{equation}
\label{eq:Mvdot}
\dot{\Mv} = s \Mv (1+z)^{5/2},
\end{equation}
where $s = \dot{\Mv}/\Mv$ is the specific accretion rate. The accretion rate of gas along the stream is then
\begin{equation}
\label{eq:Msdot-cosmo}
\dot{\Ms} = f_{\rm s} f_{\rm g} s \Mv (1+z)^{5/2},
\end{equation}
where $f_{\rm s}$ is the fractional contribution of the stream to the overall accretion rate and $f_{\rm g}$ is the gas fraction of the accreting matter. A typical major stream is assumed to carry roughly half of the overall accretion rate, whereas the second and third streams typically carry a quarter or less \citep{Danovich12}, so $0.2\lsim f_{\rm s} \lsim0.5$. The gas fraction $f_{\rm g}$ will cancel out in the final expression for $\Rs/\Rv$. 

The stream gas density in \Cref{eq:Msdot-flow} can be expressed in terms of the density contrast and the mean virial density, 
\begin{equation}
\label{eq:rhos}
\rhos = \delta \rhob = \delta f_{\rm h} f_{\rm g} \tilde{\Delta}(r) \rho_{\rm V}
\end{equation}
where $f_{\rm h}$ is the hot-to-total gass mass fraction in the halo and $\tilde{\Delta}(r)$ is the local overdensity factor at a halo centric radius $r$ relative to the mean virial density. In cosmological simulations we find $0.3\lsim f_{\rm h}\lsim 0.4$ (Roca-Fabrega et al. in preparation). For a standard NFW profile, the overdensity at the virial radius is $\tilde{\Delta}(\Rv)=1/6$, reaching unity at $\sim 0.5\Rv$. The mean virial density is simply 
\begin{equation}
\label{eq:rhovir}
\rho_{\rm V} = \frac{3}{4\pi}\frac{\Mv}{\Rv^3}.
\end{equation}

Using \Cref{eq:V-to-Vvir,eq:virial-equilibrium,eq:rhovir,eq:delta}, we express the velocity in  \Cref{eq:Msdot-flow} in terms of the virial density and radius, 
\begin{equation}
\label{eq:V-to-GrhoR}
V = \eta \sqrt{\frac{G\Mv}{\Rv}} \simeq \left(\frac{\pi}{3}\right)^{1/2} \Mb \left(G\rho_{\rm V}\right)^{1/2} \Rv.
\end{equation}

Plugging \Cref{eq:rhos,eq:V-to-GrhoR} into \Cref{eq:Msdot-flow} we get
\begin{equation}
\label{eq:Msdot-flow-end}
\dot{\Ms} \simeq \left(\frac{\pi^3}{3}\right)^{1/2} f_{\rm h} f_{\rm g} \tilde{\Delta} \delta \Mb \Rs^2 \Rv G^{1/2} \rho_{\rm V}^{3/2}.
\end{equation}
Using \Cref{eq:rhovir}, we rewrite \Cref{eq:Msdot-cosmo} as
\begin{equation}
\label{eq:Msdot-cosmo-end}
\dot{\Ms} = \left(\frac{4\pi}{3}\right) f_{\rm s} f_{\rm g} s \Rv^3 \rho_{\rm V} (1+z)^{5/2}
\end{equation}

We equate \Cref{eq:Msdot-flow-end,eq:Msdot-cosmo-end} to obtain
\begin{equation}
\label{eq:RsRv}
\frac{\Rs}{\Rv} \simeq 
\left( \frac{16}{3\pi} \right)^{1/4} 
\left( \frac{f_{\rm s}} {f_{\rm h} \tilde{\Delta} \delta \Mb} \right)^{1/2} 
\left( s \tv \right)^{1/2}
(1+z)^{5/4}
\end{equation}
where $\tv=\Rv/\Vv=(G\rho_{\rm V})^{-1/2}$ is the virial crossing time. In the Einstein-deSitter (EdS) regime, which is a good approximation for $z>1$, the virial crossing time is given by $\tv \simeq 0.49 \Gyr \times (1+z)_3^{-3/2}$ \citep[e.g.][]{Dekel2013}. Plugging this into \Cref{eq:RsRv} we find
\begin{equation}
\label{eq:RsRv-fiducial}
\frac{\Rs}{\Rv} \simeq 0.06 \times 
\left( \frac{f_{\rm s,0.5} s_{0.03} } {f_{\rm h,0.4} \tilde{\Delta}_{0.17} } \right)^{1/2} \left(\frac{1}{\delta_{75}M_{\rm b,1.5}}\right)^{1/2}
(1+z)_3^{1/2},
\end{equation}
where all the parameters were set to their fiducial values. For redshift $z=2$, taking all parameters to their extreme values (see \Cref{tab:stream-parameters}) results in $\Rs/\Rv \sim 0.01-0.10 \times \delta_{75}^{-1/2} M_{\rm b,1.5}^{-1/2}$. This is similar to the range used in \PIt, but includes the dependence of $\Rs/\Rv$ on $\unpert$, as a consequence of constraining \Cref{eq:Msdot-flow} to the cosmological inflow rate \Cref{eq:Msdot-cosmo}.  

\begin{table} 
	\centering
	\caption{Different quantities used to evaluate the Mach number $\Mb$, the density contrast  $\delta$ and the ratio of stream radius to virial radius, $\Rs/\Rv$.}
	\label{tab:stream-parameters}
	\begin{tabularx}{\columnwidth}{lXc}
		\hline
 		                  &description                                         &values \\
		\hline
		$\eta$            &ratio of stream velocity to virial velocity         &0.5-1.0 \\ 	  	
		$\Theta_{\rm b}$  &ratio of halo CGM temperature to virial temperature &$\lsim 3/8$ \\
		$\Theta_{\rm s}$  &ratio of stream temperature to $10^{4}K$            &$1-3$ \\
		$\Theta$          &$\Theta_{\rm b}/\Theta_{\rm s}$                     &$0.3-1.1$ \\
		$f_{\rm s}$       &fractional contribution of the stream to the overall mass accretion rate     &$0.2-0.5$  \\     
		$f_{\rm h}$       &hot-to-total mass gas fraction in the halo CGM      &$0.3-0.4$ \\
		$s$               &specific cosmological accretion rate of total matter &$0.015-0.060\Gyr^{-1}$ \\
		$\tilde{\Delta}$  &local overdensity factor relative to the mean virial density &$0.17-1$ \\			
		\hline
	\end{tabularx}
\end{table}

\subsection{Breakup}
\label{sec:application-breakup}

The terms ``breakup'' or ``disintegration'' are to be interpreted slightly differently in the surface-mode-dominated regime and the body-mode-dominated regime, defined by \Cref{eq:sheet-instability-condition,eq:body-instability-condition} respectively.

In the surface-mode-dominated regime, we define breakup as the entrainment of the entire stream in the shear layer, $\hs=\Rs$. We now estimate the critical stream radius, $\Rsbsurface$, to satisfy this condition just as the stream reaches the central galaxy, namely
\begin{equation}
\label{eq:Rsb-surface-def}
\hs(z=\Rv) = \Rsbsurface.
\end{equation}
Substituting \Cref{eq:hbs-spatial} for $\hs(z)$ we find
\begin{equation}
\label{eq:Rsbsurface}
\frac{\Rsbsurface}{\Rv} = \frac{\alpha(\Mb,\delta)}{\sqrt{\delta}},
\end{equation}
where $\alpha$ can be reasonably approximated by using \Cref{eq:alpha-dimotakis}. Streams with $\Rs<\Rsbsurface\unpert$ are expected to disintegrate due to surface mode instability while traversing the hot CGM.

In the body-mode-dominated regime, the stream becomes discontinuous approximately $1-2$ stream sound crossing times after the critical perturbation mode begins nonlinear evolution (see \Cref{fig:demo-body-color} and \Cref{fig:width-body}). This provides a natural definition for the breakup time,
\begin{equation}
\label{eq:tbreakbody}
\tbreakbody \simeq \tNLcrit + \tsc
\end{equation}
where $\tNLcrit$ is the time of transition to nonlinearity for the critical perturbation mode. Substituting \Cref{eq:tNLcrit} for $\tNLcrit$ results in
\begin{equation}
\tbreakbody \simeq \left[1+\ln{\left(\frac{\Rs}{H}\right)}\right]\tsc =  \left[1+\ln{\left(\frac{\Rs}{H}\right)}\right] \frac{2\Rs}{\cs}.
\end{equation}
Since body modes cause negligible deceleration prior to breakup, this can be easily translated into a breakup distance, $V\tbreakbody$. Requiring $V\tbreakbody=\Rv$ yields the critical stream radius for breakup to occur just as the stream reaches the central galaxy, 
\begin{equation}
\label{eq:Rsbbody}
\frac{\Rsbbody}{\Rv} = \left[ 2\sqrt{\delta}\Mb \left(1+\ln{\frac{\Rs}{H}}\right)\right]^{-1}.
\end{equation}
Streams with $\Rs<\Rsbbody\unpert$ are expected to disintegrate due to body mode instability while traversing the hot CGM.

\begin{figure*}
	\centering
	\subfloat{
		\includegraphics[trim={0cm 1.0cm 3cm 2.0cm}, clip, height=0.34\textheight]{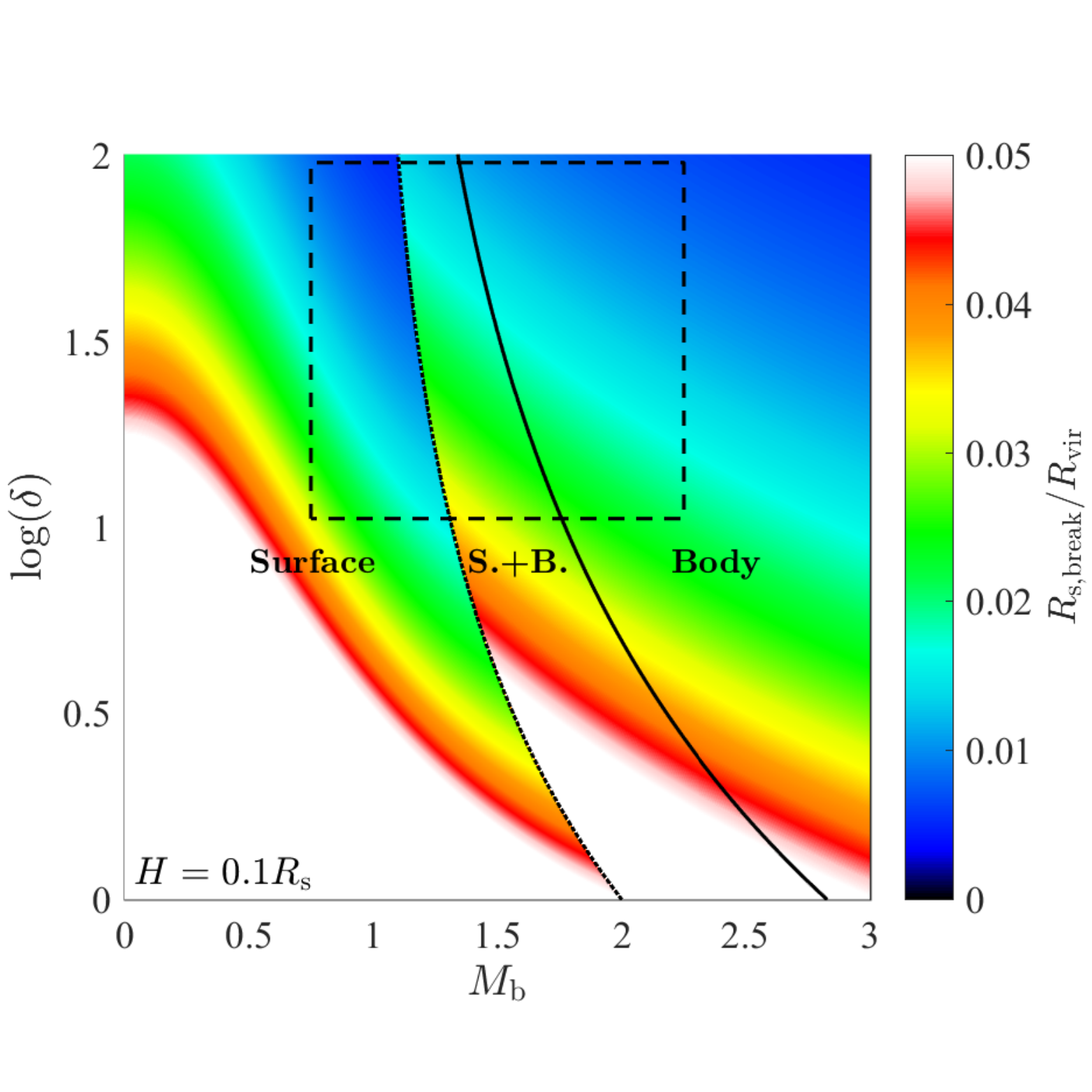}
	}
	\subfloat{
		\includegraphics[trim={0.75cm 1.0cm 0cm 2.0cm}, clip, height=0.34\textheight]{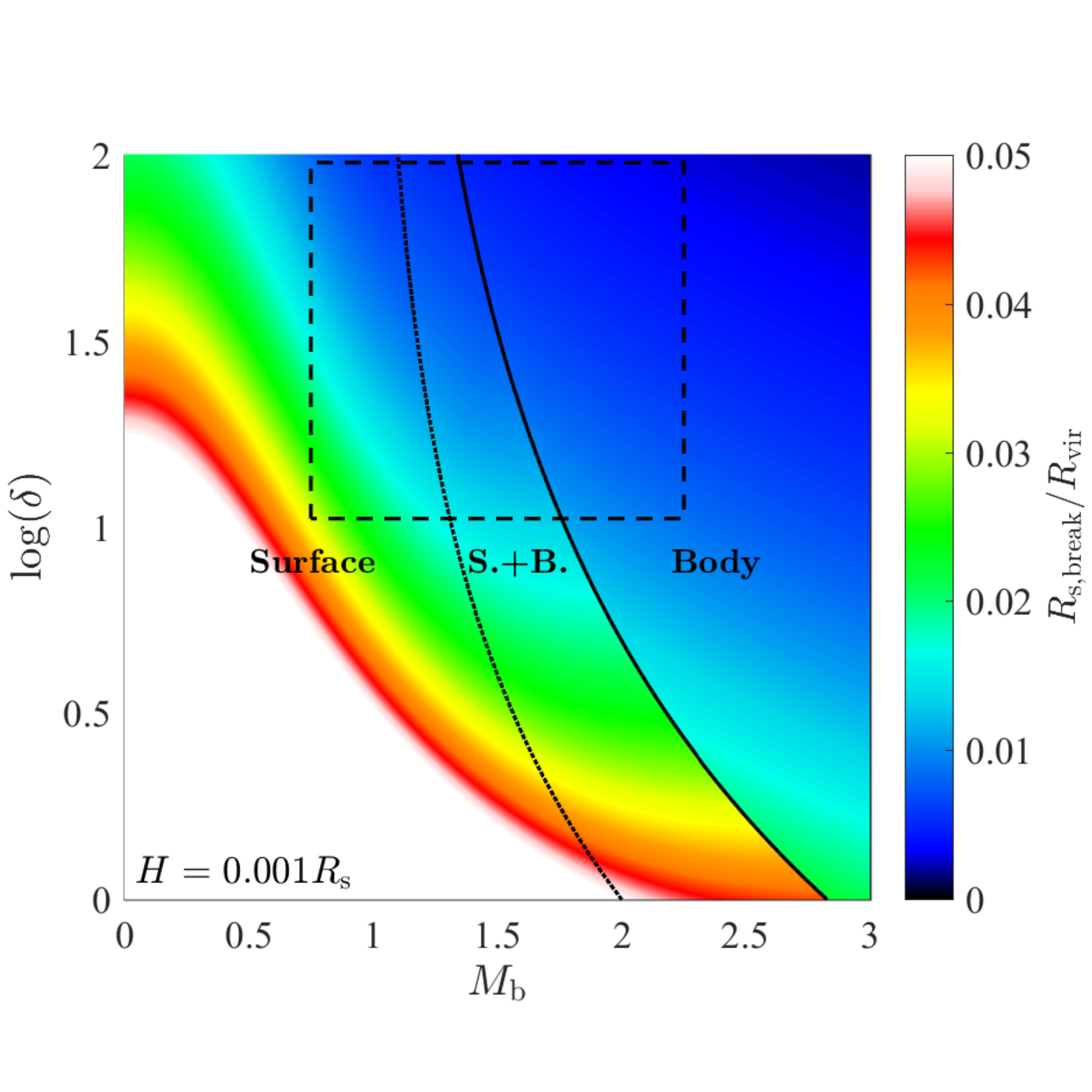}
	}
	\caption{Critical stream radius for breakup due to nonlinear KHI, \Cref{eq:Rsbsurface,eq:Rsbbody,eq:Rsb}, as a function of the density contrast, $\delta$, and Mach number, $\Mb$. Realistic values for cold streams feeding galaxies are $\Mb\sim0.75-2.25$ and $\delta\sim10-100$, shown in the dashed square. Streams with $\Rs<\Rsb$ are expected to disintegrate before they reach the central galaxy. The colormap is identical for both panels, with realistic values for the stream radius being $\Rs/\Rv\sim 0.005-0.05$. These can be related to $\unpert$ using \Cref{eq:RsRv-fiducial}. The condition for breakup ranges from $\Rs<0.003\Rv$ to $\Rs<0.03\Rv$ for realistic values of $\unpert$. The temporal shear layer growth rate, $\alpha$, appearing in \Cref{eq:Rsbsurface}, is estimated using \Cref{eq:alpha-dimotakis}. The solid black line shows the \emph{maximal} Mach number allowing surface mode instability, \Cref{eq:sheet-instability-condition}. The dotted black line shows the \emph{minimal} Mach number allowing body mode instability, \Cref{eq:body-instability-condition}. In the body-mode-dominated regime, $\Rsb$ depends on the initial displacement amplitude of the critical perturbation mode, $H$. The different panels show different values of $H$. The results in the surface-mode-dominated regime are independent of $H$. The discontinuity in $\Rsb$ across the dotted line in the left panel and across the solid line in the right panel corresponds to the transition from surface modes to body modes in \Cref{eq:Rsb}.}
	\label{fig:Rsb}
\end{figure*}

We combine \Cref{eq:Rsbsurface,eq:Rsbbody} into a critical $\Rs$ for breakup by either surface modes or body modes,
\begin{equation}
\label{eq:Rsb}
\Rsb = \max\left\{\Rsbsurface,\Rsbbody\right\},
\end{equation}
where we define $\Rsbsurface(\Mb>\Mcrit)=0$ and $\Rsbbody(\Mtot<1)=0$ in order to comply with \Cref{eq:sheet-instability-condition,eq:body-instability-condition}. \Cref{fig:Rsb} shows $\Rsb$ as a function of $\unpert$. For the parameter range relevant to cold streams ($0.75<\Mb<2.25,10<\delta<100$), we find that streams up to a radius of $\approx 0.03\Rv$ may disintegrate prior to reaching the galaxy, depending on the exact values of $\unpert$. Overall, the critical stream radius decreases with both density contrast and Mach number, decreasing from roughly $0.03\Rv$ for $(\Mb=0.75,\delta=10)$ to $0.003\Rv$ for $(\Mb=2.25,\delta=100)$. The critical stream radius at the highest relevant density contrast and Mach number ($\Mb\simeq2.25$ and $\delta=100$) weakly depends on the initial amplitude of the critical mode, varying in the range $0.003\Rv<\Rsb<0.007\Rv$ for $0.001\Rs<H<0.1\Rs$. 

It is interesting to note that the dominant type of modes in the coexistence region ($\Mb<\Mcrit$ and $\Mtot>1$) is determined by $H$. If $H/\Rs$ is sufficiently small, less than a few percent, then the coexistence region is dominated by surface modes and the transition to body modes occurs only when surface modes stabilize completely, as shown in the right-hand panel of \Cref{fig:Rsb}. If the opposite is true, then the coexistence region is dominated by body modes and the transition occurs close to $\Mtot=1$, as in the left-hand panel of \Cref{fig:Rsb}. The overall conclusions remain the same for both cases.

\subsection{Inflow Rate}
\label{sec:application-inflow}

The deceleration of stream fluid due to KHI is interpreted in the cosmological context as a reduction in cold mass inflow rate into the central galaxy. As a crude estimation of the the magnitude of this effect, we evaluate the critical stream radius for a factor of $2$ reduction in inflow rate to occur due to KHI before the stream reaches the central galaxy, which we denote by $\Rsi$. This does not take into account gravitational acceleration compensating for some or all of this reduction in inflow rate.

In the surface-mode-dominated regime, deceleration from the initial velocity $\vcm=V$ to $\vcm=V/2$ occurs at an approximately constant deceleration rate (see \Cref{sec:surface-results-deceleration}, in particular \Cref{fig:deceleration-surface}),
\begin{equation}
\label{eq:vdot-surface}
\dot{v}_{\rm z,cm} \simeq -\frac{1}{2} \frac{V}{\tausurf}
\end{equation}
By requiring $\vcm(\tv)=V/2$ we find the critical stream radius,
\begin{equation}
\label{eq:Rsisurface}
\frac{\Rsisurface}{\Rv} = \frac{\alpha\unpert}{\delta+\sqrt{\delta}},
\end{equation}
Streams with $\Rs<\Rsisurface\unpert$ are expected to lose more than $50\%$ of the inflow rate they had at the virial radius while traversing the CGM. 

In the body-mode-dominated regime, we use the approximate deceleration profile obtained in \Cref{sec:body-results-deceleration},
\begin{equation}
\label{eq:vdot-body}
\dot{v}_{\rm z,cm}=
\begin{cases}
0 &t<\tNL \\
-0.12 V/\taubody &\tNL<t<t(\vcm=\vbreak) \\
-0.016 V/\taubody &t(\vcm=\vbreak)<t
\end{cases},
\end{equation}
where $\vbreak/V \simeq \delta/(\delta+2)$. Using \Cref{eq:vdot-body} and requiring $\vcm(\tv)=V/2$, we find the critical stream radius,
\begin{equation}
\label{eq:Rsibody}
\frac{\Rsibody}{\Rv} = 
\begin{cases}
\left[2\sqrt{\delta}\Mb\left(\ln{\dfrac{\Rs}{H}}+2.1\sqrt{\delta}\right)\right]^{-1} 
& \delta<2 \\
\left[2\sqrt{\delta}\Mb\left(\ln{\dfrac{\Rs}{H}}+\dfrac{15.6\,\delta-22.8}{\delta+2}\sqrt{\delta}\right) \right]^{-1} 
& \delta>2
\end{cases}.
\end{equation}
Streams with $\Rs<\Rsibody\unpert$ are expected to lose more than $50\%$ of the inflow rate they had at the virial radius while traversing the CGM. 

\begin{figure*}
	\centering
	\subfloat{
		\includegraphics[trim={0cm 1.0cm 3cm 2.0cm}, clip, height=0.34\textheight]{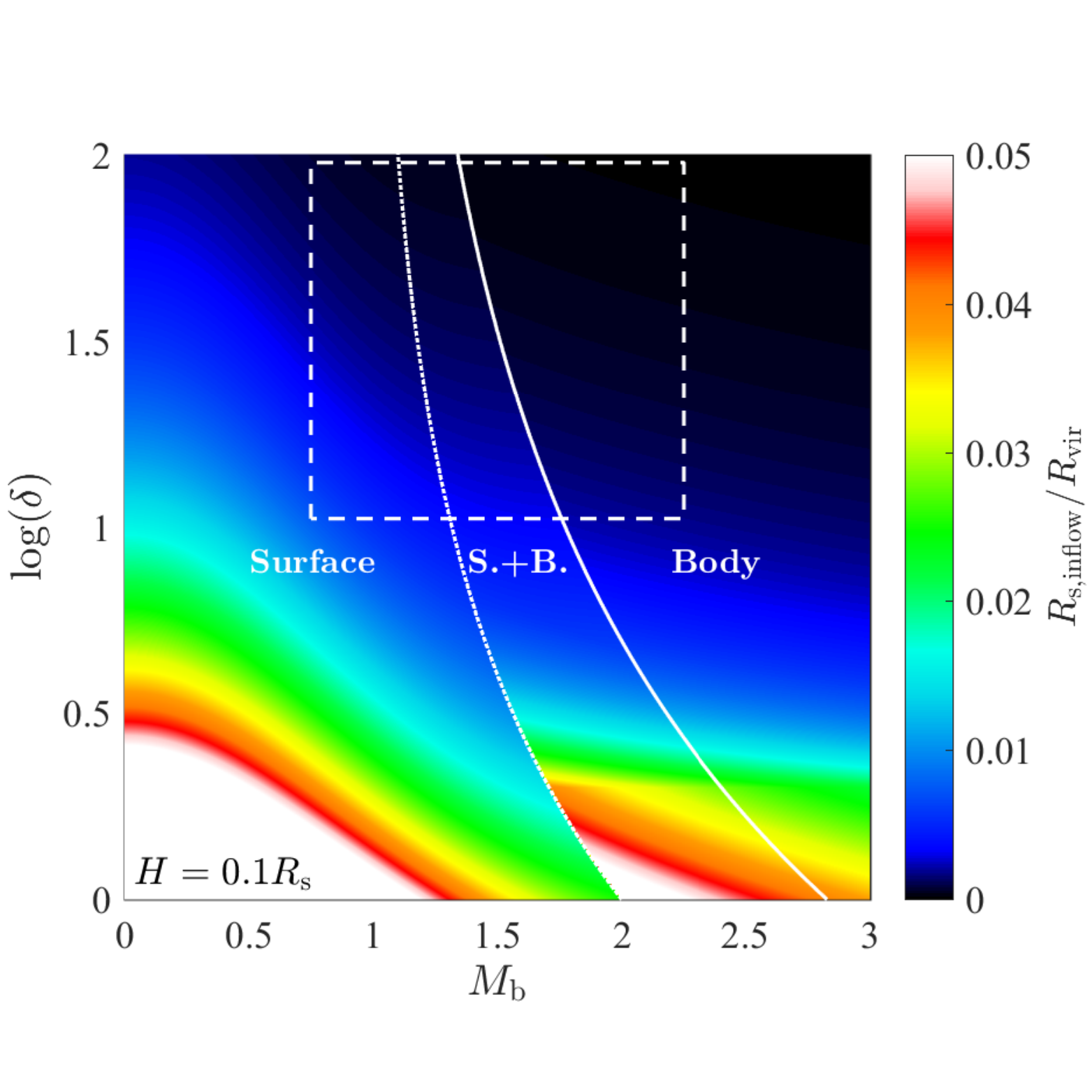}
	}
	\subfloat{
		\includegraphics[trim={0.75cm 1.0cm 0cm 2.0cm}, clip, height=0.34\textheight]{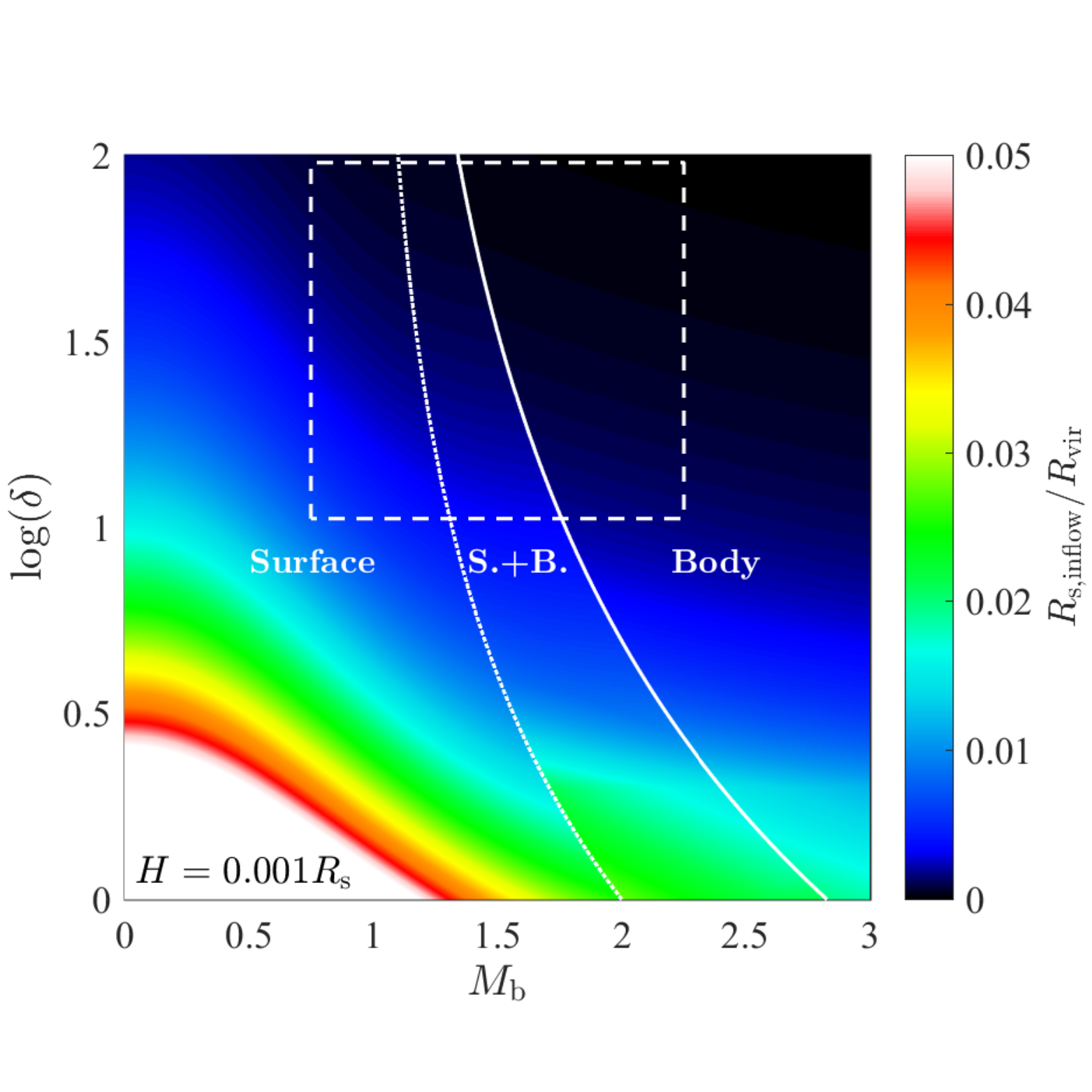}
	}
	\caption{Critical stream radius for a factor of $2$ reduction in inflow velocity due to nonlinear KHI, \Cref{eq:Rsisurface,eq:Rsibody,eq:Rsi}. The axis, dashed square, lines and colormap are the same as in \Cref{fig:Rsb}. Streams with $\Rs<\Rsi$ lose at least half of their initial mass inflow rate by the time they reach the central galaxy. In the relevant range of Mach number, $\Mb$, density contrast, $\delta$, and stream radius, $\Rs$, deceleration due to KHI is found to be weak compared to the gravitational acceleration, which is absent from this figure altogether; gravity is supposed to increase the inflow velocity by a factor of $\sim 2$ or more (see \Cref{sec:application-inflow,app:inflow-velocity-toy-model}). Significant deceleration is expected only for very narrow streams. We note that a 3D analysis of cylindrical streams, to be reported in the next paper of this series (Mandelker et al.  in preparation), is likely to raise the estimated critical stream radius.}
	\label{fig:Rsi}
\end{figure*}

As in the previous section, we combine \Cref{eq:Rsisurface,eq:Rsibody} by using the expression
\begin{equation}
\label{eq:Rsi}
\Rsi = \max\left\{\Rsisurface,\Rsibody\right\},
\end{equation}
where we define $\Rsisurface(\Mb>\Mcrit)=0$ and $\Rsibody(\Mtot<1)=0$ in order to comply with \Cref{eq:sheet-instability-condition,eq:body-instability-condition}. \Cref{fig:Rsi} shows $\Rsi$ as a function of $\unpert$. We find that most streams in the relevant parameter range experience very little deceleration. Even when streams are expected to disintegrate, the high density contrast means that their inertia is too large for significant deceleration to occur in a virial time. Only dilute ($\delta \simeq 10$), slow ($\Mb \simeq 1$) and narrow ($\Rs \leq 0.01\Rv$) streams are expected to lose $50\%$ or more of their inflow rate before they reach the central galaxy. By virtue of \Cref{eq:RsRv-fiducial}, streams at the dilute, slow and narrow extreme of the allowed parameter space should be a rare occurrence, since such cases will deviate considerably from the cosmological accretion rate.

Our findings are robust with regard to the assumed value of the initial amplitude of the critical mode, $H$, as demonstrated in \Cref{fig:Rsi}. For initial amplitudes in the range $0.001\Rs\leq H\leq0.1\Rs$,  shown in the figure, the values of $\Rsi$ vary by less than $\sim25\%$ for realistic values of $\unpert$. Even if the critical mode is seeded with nonlinear initial amplitude, $H/\Rs\sim 1$, we find that $\Rsi$ remains largely below the lower bound for $\Rs$ in  cold streams feeding galaxies. 

As mentioned previously, \Cref{fig:Rsi} neglects the gravitational acceleration acting on the inflowing stream. \Cref{app:inflow-velocity-toy-model} presents a simple toy model for the evolution of the inflow velocity, including an acceleration term due to gravity and a deceleration term due to KHI, for the specific case of surface mode instability. The results  therein show that, in most cases, the contribution of KHI to the stream inflow rate is negligible compared to the gravitational acceleration. Except for dilute and narrow streams ($\delta \simeq 10, \Rs \simeq 0.01\Rv$), the inflow velocity in the toy model approximately follows the free fall solution. 

In cosmological simulations, stream velocity is nearly constant during the infall \citep{Dekel09,Goerdt15a}, indicating that the gravitational acceleration is counteracted by a rapid dissipation process. This dissipation process has yet to be studied in detail and its physical origins are unclear. The extent to which insufficient resolution contributes to the high dissipation rate observed in cosmological simulations is also uncertain. Based on the discussion in this section, we conclude that KHI is \emph{not} a potential source for efficient dissipation in cold streams feeding massive SFGs.

An important caveat to the conclusions above arises when considering potential differences between deceleration in 2D slabs, studied here, and 3D cylinders, which resemble the shape of cosmic cold streams more closely. In the surface-mode-dominated region, analysis and preliminary simulations suggest that stream deceleration timescales are $\sim10$($\sim3$) times shorter than slab deceleration timescales for $\delta\sim100$($\delta\sim10$). The differences between body modes in slabs and cylinder are expected to be similar or smaller. In terms of the toy model studied in \Cref{app:inflow-velocity-toy-model} this corresponds to multiplying the parameter $B$ by the same factors. Performing this manipulation one sees that the KHI-induced deceleration still amounts to a small effect on the stream infall velocity profile. These issues will be addressed in the following paper of this series (Mandelker et al. in preparation).

\subsection{Lyman-$\alpha$ Emission}
\label{sec:application-thermo}

\citet{Goerdt10} predicted the Lyman-$\alpha$ luminosity from cold streams based on cosmological simulations where the stream velocity is roughly constant, implying any gravitational energy released during the infall is completely dissipated into internal energy and subsequently radiated away. The resulting Lyman-$\alpha$ sources have properties consistent with observed Lyman-$\alpha$ ``blobs'' (LABs), with a luminosity of $10^{42}$ to $10^{44}\rm{erg~s^{-1}}$ for halos in the mass range $10^{11}-10^{13}\Msun$ at $z=3$. Similarly, \cite{Dijkstra09} worked out an analytic model for Lyman-$\alpha$ luminosity from cold streams based on the general properties of the streams reported in simulations, where the fraction of gravitational energy radiated away, $f_{\rm grav}$, is a free parameter. This work concludes that emission from cold streams is comparable to observed LABs if $f_{\rm grav}\gsim 0.2$. The cosmological simulations considered by \citet{Goerdt10} effectively have $f_{\rm grav}\simeq1$.

Neither \citet{Goerdt10} nor \citet{Dijkstra09} investigate the dissipative process powering the  radiation. As can be expected from the discussion in  \Cref{sec:application-inflow}, we find that KHI cannot provide the necessary high dissipation rate. This is explained in detail below.

The \emph{adiabatic} simulations presented in \Cref{sec:surface-results-heating,sec:body-results-heating} demonstrate that non-negligible heating of the stream fluid due to KHI occurs if $\Mtot = \Mb \sqrt{\delta}/(1+\sqrt{\delta})\gsim1$. In the surface-mode-dominated regime, this heating takes place at a roughly constant rate as the shear layer grows, raising the temperature of the stream fluid by $\lsim 40\%$ until it reaches the halo center. In the body-mode-dominated regime, the heating occurs abruptly when the stream breaks up, raising its temperature by a factor $\sim2-3$.

The overall energy budget of a cold stream can be broken into three components, 
\begin{equation}
\label{eq:energy-budget}
e = e_{\rm grav} + e_{\rm kin} + e_{\rm int},
\end{equation}
where $e$ is the specific energy and the different terms correspond to gravitational, kinetic and internal energy respectively. As the stream flows into the halo, gravitational energy is converted into kinetic energy, which can then be dissipated into internal energy. At the initial time, when the stream fluid enters the halo at radius $\sim \Rv$ with a velocity $\sim \Vv$, we have
\begin{align}
\label{eq:egrav}
e_{\rm grav}(t=0) &= \frac{G\Mv}{\Rv} = \Vv^2, \\
\label{eq:ekin}
e_{\rm kin}(t=0) &= \frac{1}{2}V^2, \\
\label{eq:eint}
e_{\rm int}(t=0) &= \frac{3}{2}\frac{K_{\rm B}\Ts}{\mu m_{\rm p}}.
\end{align}
Combining \Cref{eq:energy-budget,eq:egrav,eq:ekin,eq:eint} we obtain
\begin{equation}
\label{eq:initial-energy}
\frac{e(t=0)}{e_{\rm grav}(t=0)}
=
1 + \frac{\eta^2}{2} + \frac{3\eta^2}{2\gamma}\frac{1}{\Ms^2}
=
1 + \frac{\eta^2}{2} + \frac{9\eta^2}{10}\frac{1}{\Ms^2},
\end{equation}
where $\gamma=5/3$ is used in the last equality and $0.5\lsim\eta=V/\Vv\lsim1$. 

\Cref{eq:initial-energy} shows that the energy budget of the stream is initially dominated by the gravitational and kinetic terms, with the internal energy contributing only $\sim \Ms^{-2} = \delta^{-1} \Mb^{-2}$ of the total energy. For the relevant values of $\unpert$, this factor is in the range $\sim 0.001-0.1$. Hence, although the aforementioned gain in stream temperature due to KHI is non-negligible compared to the initial state, it corresponds to the dissipation of only a small fraction of the gravitational energy released during the infall. This is true for both surface modes and body modes\footnotemark. In conclusion, we expect KHI to add only $\sim0.1\%-10\%$ to the Lyman-$\alpha$ luminosity estimated by \cite{Goerdt10} from cosmological simulations, depending on $\unpert$.
\footnotetext{Streams disrupted by body mode instability heat up more than those susceptible to surface mode instability (compare \Cref{fig:thermo-surface,fig:thermo-body}). This is compensated by the fact that $\Ms$ is typically higher in the body-mode-dominated regime than in the surface-mode-dominated regime, so the amount of energy dissipated remains small compared to the total stream energy.}

Our adiabatic simulations do not account for the effects of radiative cooling on KHI. Previous investigations done in other contexts \citep{Hardee1997,Vietri97,Stone97} found that the evolution KHI in a cooling fluid can deviate considerably from the adiabatic case, both in the linear and in the nonlinear phase. Whether this strengthens or weakens the instability depends strongly on the details of the cooling function. Cold streams feeding massive SFGs at $z\sim2$ are at a temperature of a few $10^4\K$, slightly below the peak associated with Lyman-$\alpha$ emission from atomic hydrogen \citep{bd03,db06}. The cooling function rises steeply in this region \citep{Sutherland1993}, increasing by more than an order of magnitude for a factor of $\sim2$ increase in temperature. Linear analysis for a similarly steep power-law cooling function shows reduced KHI growth rates \citep{Hardee1997}. On the other hand, nonlinear simulations with a full cooling function, qualitatively similar to the \citet{Sutherland1993} function, exhibit more violent disintegration of the stream than identical  adiabatic simulations \citep{Stone97}. In a forthcoming paper (Mandelker et al. in preparation), we perform 3D simulations of nonlinear KHI and study the effect of realistic radiative cooling on both surface modes and body modes.

The arguments above offers a strong indication that the heat deposited in cold streams due to KHI, $q$, is small compared to the gravitational energy scale, $e_{\rm grav}$. Although this does not imply that the instantaneous power must obey the same hierarchy at any instant, i.e. ${\rm d}q/{\rm d}t<{\rm d}e_{grav}/{\rm d}t$, this is the most likely conclusion (see \Cref{app:inflow-velocity-toy-model}). A more careful study of the power balance in cold streams feeding galaxies awaits later stages in our long-term campaign, when we will turn to examine the combined effects of cooling and gravity.

Throughout this work, we referred primarily to Lyman-$\alpha$ emission, because it is expected to be a dominant coolant in cold streams feeding massive galaxies. However, our analysis is not limited to any single cooling mechanism. The conclusions above rely on estimations of the overall energy deposition due to KHI, without considering the specific mechanisms expected to radiate the deposited energy away. The relative importance of different cooling channels will be assessed as part of our future work on simulations including realistic cooling functions (Mandelker et al. in preparation).


\section{Summary and Conclusions}
\label{sec:conclusions}

Motivated by the conclusions of the linear analysis of \PIt, we presented a detailed study of the nonlinear stage of  purely-hydrodynamic, adiabatic Kelvin-Helmholtz Instability (KHI) in 2D planar sheet and slab geometries, using a combination of analytic models and \texttt{RAMSES} numerical simulations. We then applied our results to the problem of cold streams feeding massive star-forming galaxies (SFGs) at high redshift. Our main results can be summarized as follows.

\begin{enumerate}
	\item For surface modes, which dominate at lower Mach numbers, the nonlinear evolution is driven by vortex mergers resulting in self similar shear layer growth. The late-time evolution is independent of the exact properties of the initial perturbations. The temporal growth rate in this regime depends primarily on the total Mach number, $\Mtot = \Mb \sqrt{\delta}/(1+\sqrt{\delta})$, consistent with previously reported simulations and terrestrial experiments performed in other contexts with significantly smaller density contrasts. We find good agreement between our analytic models and our  simulation results in terms of the convection (drift) velocity of the vortices, the entrainment ratio (stream/background asymmetry) and the stream deceleration rates. Overall, as either the density contrast, $\delta$, or the Mach number, $\Mb$, are increased, the shear layer growth rates decrease, the entrainment ratio diverges from unity and the deceleration rates decrease.
	\item For body modes, which are slower to grow but dominate at higher Mach numbers, the nonlinear evolution is driven by long-wavelength sinusoidal perturbations. We found good agreement between the analytical predictions for the critical perturbation mode that is expected to break the stream, and the dominant mode observed at late times in numerical simulations. We proposed simple scaling relations for the deceleration rate, which fit the simulation results. The disruption and deceleration timescales are largely independent of $\Mb$ and decrease with $\delta$.
	\item The range of parameters relevant to cold streams feeding massive galaxies at $z\sim2$ is estimated, refining the arguments made in \PIt~and obtaining $0.75\lsim\Mb\lsim2.25$, $10 \lsim\delta\lsim 100$ and $0.005\lsim\Rs/\Rv\lsim0.05$. In addition, we find a mutual constraint on these three parameters based on the cosmological accretion rate, expressed by  $\Rs/\Rv\propto\delta^{-1/2}\Mb^{-1/2}$.
	\item Using these estimates, we predict the following outcomes of KHI in a virial time. In part of the allowed parameter range, streams are expected to disintegrate prior to reaching the central galaxy. The upper limit for the stream radius that results in disintegration is between $0.003\Rv$ for the fastest ($\Mb\sim2.5$), densest ($\delta\sim100$) streams and $0.03\Rv$ for the slowest ($\Mv\sim0.75$), most dilute ($\delta\sim10$) streams. 
	\item \label{conclusion:deceleration} In most cases, only a small effect on the inflow rate is expected in a virial time, except for very narrow streams. This is due to the significant inertia of the stream compared to the CGM.
	\item \label{conclusion:ly-alpha} For streams with $\Mtot\gsim1$, some heating of the stream fluid is observed. For surface modes this heating is gradual, whereas for body modes it occurs abruptly when the stream disintegrates. Compared to the overall energy budget of cosmic cold streams, the observed dissipation due to KHI is negligible. Therefore, KHI is not a viable power source for the efficient Lyman-$\alpha$ emission considered in \citet{Goerdt10} and \citet{Dijkstra09}.
\end{enumerate}

Two caveats to \cref{conclusion:deceleration,conclusion:ly-alpha} are worth noting. First, preliminary estimates of the deceleration rates in cylinders suggest that they may be significantly higher than the results cited above for slab geometry. Nevertheless, when compared with the gravitational acceleration experienced by streams during infall into the halo potential, KHI-induced deceleration is likely to have a minor effect on the infall velocity, even for a cylindrical stream. Second, although the dissipation rates are small compared to the overall energy budget, they can result in a significant increase of the radiative cooling rate, due to the steepness of the cooling curve in the relevant temperature range. This can have a non-trivial effect on the evolution of KHI, either amplifying or damping the instability depending on the exact shape of the cooling function. These effects are absent from our adiabatic simulations. 

Both of the caveats above will be addressed in the next two papers of this series (Mandelker et al. in preparation), where we will present simulations and analysis of 3D cylinderical streams first without and then with radiative cooling. In future work, we will introduce additional physical processes, including the self-gravity of the stream, the gravitational potential of the underlying dark-matter halo, and eventually magnetic fields and thermal conduction. In parallel, we make progress on cosmological simulations with adaptive refinement in the streams to resolve the desired instabilities (Roca-Fabrega et al. in preparation). 


\section*{Acknowledgments}

We acknowledge stimulating discussions with Frederic Bournaud, Andi Burkert,  Ami Glasner, Daisuke Nagai and John Forbes. We thank the anonymous referee for offering insightful comments that contributed to the quality of this work.

YB and DP acknowledge support from  ISF grant 1059/14. AD acknowledges support by the grants ISF 124/12, I-CORE Program of the PBC/ISF 1829/12, BSF 2014-273, PICS 2015-18, GIF I-1341-303.7/2016, DIP STE1869/2-1 GE625/17-1, and NSF AST-1405962. NM acknowledges support from the Klaus Tschira Foundation, through the HITS-Yale Program in Astrophysics (HYPA). MRK acknowledges support from the Australian Research Council's Discovery Projects, grant DP160100695.




\bibliographystyle{mnras}
\bibliography{biblio} 



\appendix

\section{Convergence Tests in Planar Sheet Geometry}
\label{app:convergence}

\begin{figure}
	\centering
	\includegraphics[trim={0cm 0.5cm 1.0cm 1.0cm}, clip, width=0.475\textwidth]{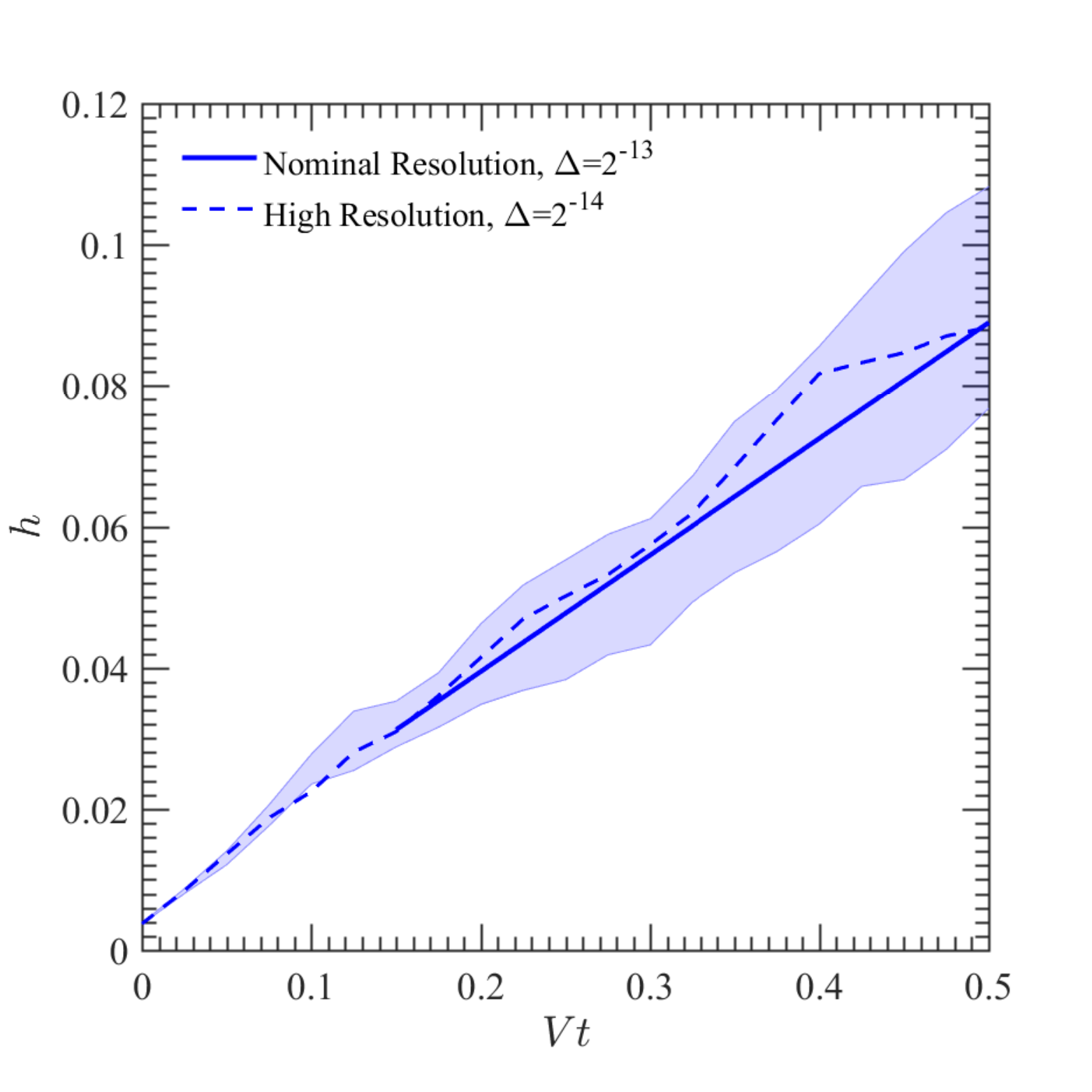}
	\caption{Shear layer thickness (see \Cref{sec:methods-analysis}), $h$, for $\Mb=0.5$ and $\delta=1$. The shaded area is spanned by three runs with different realizations of sparse white noise full eigenmode perturbations (see \Cref{sec:surface-simulations}), all with the nominal resolution described in \Cref{sec:methods-grid}, reaching a minimal cell size of $\Delta=2^{-13}$. The solid line is a linear fit to the mean thickness at times $Vt \geq 0.15$. The dashed line corresponds to a single realization of the same kind of initial conditions, with the resolution doubled throughout the computational grid, reaching a minimal cell size of $\Delta=2^{-14}$.}
	\label{fig:resolution-thickness}
\end{figure}

In this section we demonstrate the resolution effect on the results presented in \Cref{sec:surface-results}, by comparing simulations carried out with the nominal resolution, described in \Cref{sec:methods-grid}, to a simulation with double resolution throughout the computational grid.  \Cref{fig:resolution-thickness} compares the two in terms of shear layer growth. The high resolution run is confined within the span of runs with nominal resolution, and the slopes (growth rates) are consistent. We therefore conclude that our simulations are sufficiently converged for the purposes of evaluating large scale properties, such as the shear layer growth rate and the convection velocity. These conclusions come as no surprise. Although the \emph{linear} growth rates of the shortest wavelength perturbations included in our simulations are somewhat reduced due to the finite resolution, this only slightly delays the onset of \emph{nonlinear} evolution. At later times the extent of the shear layer spans $\approx 1000$ cells at the nominal resolution, and the results are converged.

\section{Compressibility Scaling of Shear Layer Growth Rate and Comparison with Experiments}
\label{app:compressibility-scaling}

There has been considerable debate as to which parameter should be used to scale the compressibility dependence of shear layer growth rates. \citet{Bogdanoff1983} and \citet{Papamoschou1988} suggested that compressibility effects in experiments can be scaled by the \emph{convective} Mach numbers, 
\begin{align}
\label{eq:convective-mach-numbers}
M_{\rm cb} \equiv \frac{\Vc-\Vb}{\cb} \\
M_{\rm cs} \equiv \frac{\Vs-\Vc}{\cs}
\end{align}
where $\Vc$ is the velocity at which large structures are convected downstream inside the shear layer, see \Cref{sec:surface-results-convection}, and $V_{\rm b}<V_{\rm c}<V_{\rm s}$ is assumed. By manipulating \Cref{eq:Vc_general}, one finds that for isentropic flow and identical adiabatic indices in both fluids the Mach numbers should be equal, $M_{\rm cb} \simeq M_{\rm cs}$.

\citet{Papamoschou1989} pointed to the significance of the \emph{total Mach number}, defined in  \Cref{eq:body-instability-condition}. This parameter has been used to scale compressibility effects in most subsequent work\footnotemark. Note that the total Mach number is the sound speed weighted average of the individual convective Mach numbers, $\Mtot = (\cb M_{\rm cb} + \cs M_{\rm cs})/(\cb+\cs)$. Since it is Galilean invariant, $\Mtot$ is viable as a scaling parameter for temporal growth simulations as well as spatial experiments. 

\footnotetext{This parameter is commonly referred to as the total convective Mach number or simply the convective Mach number. It is denoted by  $M_{\rm c}$ in most publications. In the interest of consistency, we retain the notation $\Mtot$ used in \PIt.}

\citet{Slessor2000} proposed an alternative parameter, $\Pic$, motivated by the scaling of kinetic-to-thermal energy conversion in the dimensionless energy equation,
\begin{equation}
\label{eq:Pic}
\Pic \equiv \max_{\rm i} \left[ \sqrt{\gamma_{\rm i}-1}\frac{V}{c_{\rm i}} \right].
\end{equation}
where the index $\rm i$ denotes either of the two fluids. In our set of runs $\gammab=\gammas=5/3$ and $\Ms \geq \Mb$, yielding $\Pic \simeq 0.82\Ms$. For $\gammab=\gammas$ and $\delta \simeq 1$, the new scaling parameter is practically equivalent to the total Mach number. However, if the density contrast is far from unity, $\Pic$ and $\Mtot$ can differ significantly. \citet{Slessor2000} reported improved collapse of experimental growth rates to a single curve, when plotted against $\Pic$ rather than $\Mtot$. Despite this, the new compressibility parameter hasn't gained much popularity in later publications. \citet{Freeman2014} showed that when taking into account a larger database of experimental results, the differences in scatter between both scaling parameters reported in \citet{Slessor2000} become hardly noticeable.

To evaluate the relative merits of both parameters, we follow the normalization procedure \citet{Slessor2000} employed to extract the compressibility dependence of spatial growth rates. Denote the incompressible spatial growth rate as $h'_0$. It can be expressed as $h'_0(r, \delta) = C \Phi(r,\delta)$, where $\Phi$ is a known function for the incompressible $(r,\delta)$ dependence with $\Phi(r=0,\delta=1)=1$. \citet{Slessor2000} allowed individual experimental apparatuses to differ in the value of $C$ and inferred these values directly from the available measurements in each experiment. The measured growth rates, $h'(r,\delta,\Mtot)$, were then normalized by the incompressible growth rates  $h'_0(r,\delta)$, with the appropriate value of $C$ for each dataset. To apply the analogous procedure for our temporal results, we assume that the temporal growth rate depends only on the compressibility parameter, whatever it may be (see discussion in \Cref{sec:surface-results-density-mach}). We then use \Cref{eq:spatial-growth-rate} to write $h'(r,\delta,\Mtot) = 2\alpha(\Mtot) \Phi(r,\delta)$, where $\Phi(r,\delta) = V/(2\Vc)$, and $V/(2\Vc)$ can be substituted for \Cref{eq:Vc_general}. The incompressible growth rate is thus $h'_0(r,\delta) = 2\alpha(\Mtot\to 0) \Phi(r,\delta)$ and the normalized growth rates will read
\begin{equation}
\label{eq:normalized-growth-rate}
\frac{h'(r,\delta,\Mtot)}{h'_0(r,\delta)} = \frac{\alpha(\Mtot)}{\alpha(\Mtot \to 0)}.
\end{equation}
Unsurprisingly, this is simply the ratio of temporal growth rates. 

\begin{figure*}
	\begin{tabular}{cc}
		\subfloat{
			\includegraphics[trim={0.25cm 0.5cm 1cm 1.25cm}, clip, height=0.35\textheight]{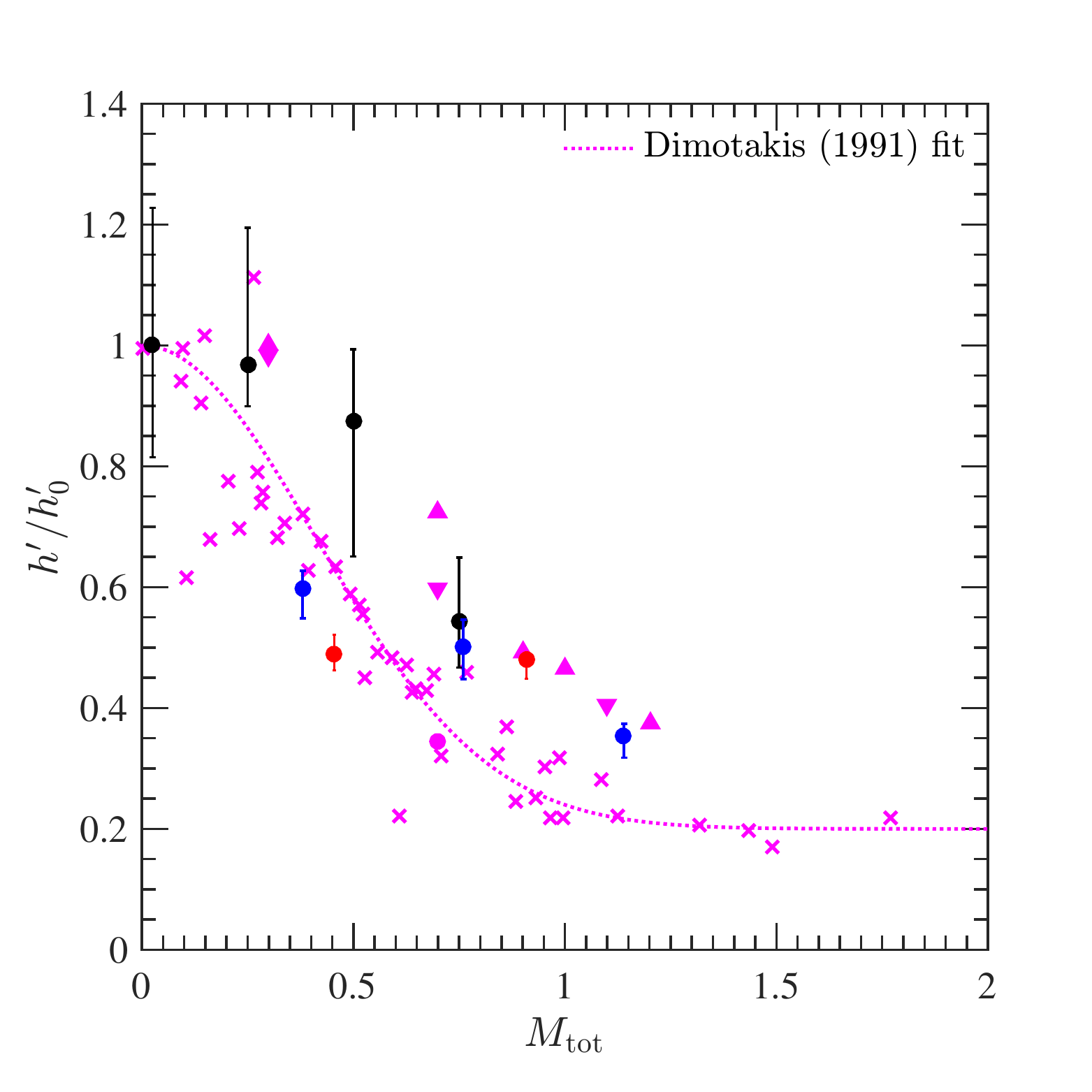}
		}&
		\subfloat{
			\includegraphics[trim={1.0cm 0.5cm 1cm 1.25cm}, clip, height=0.35\textheight]{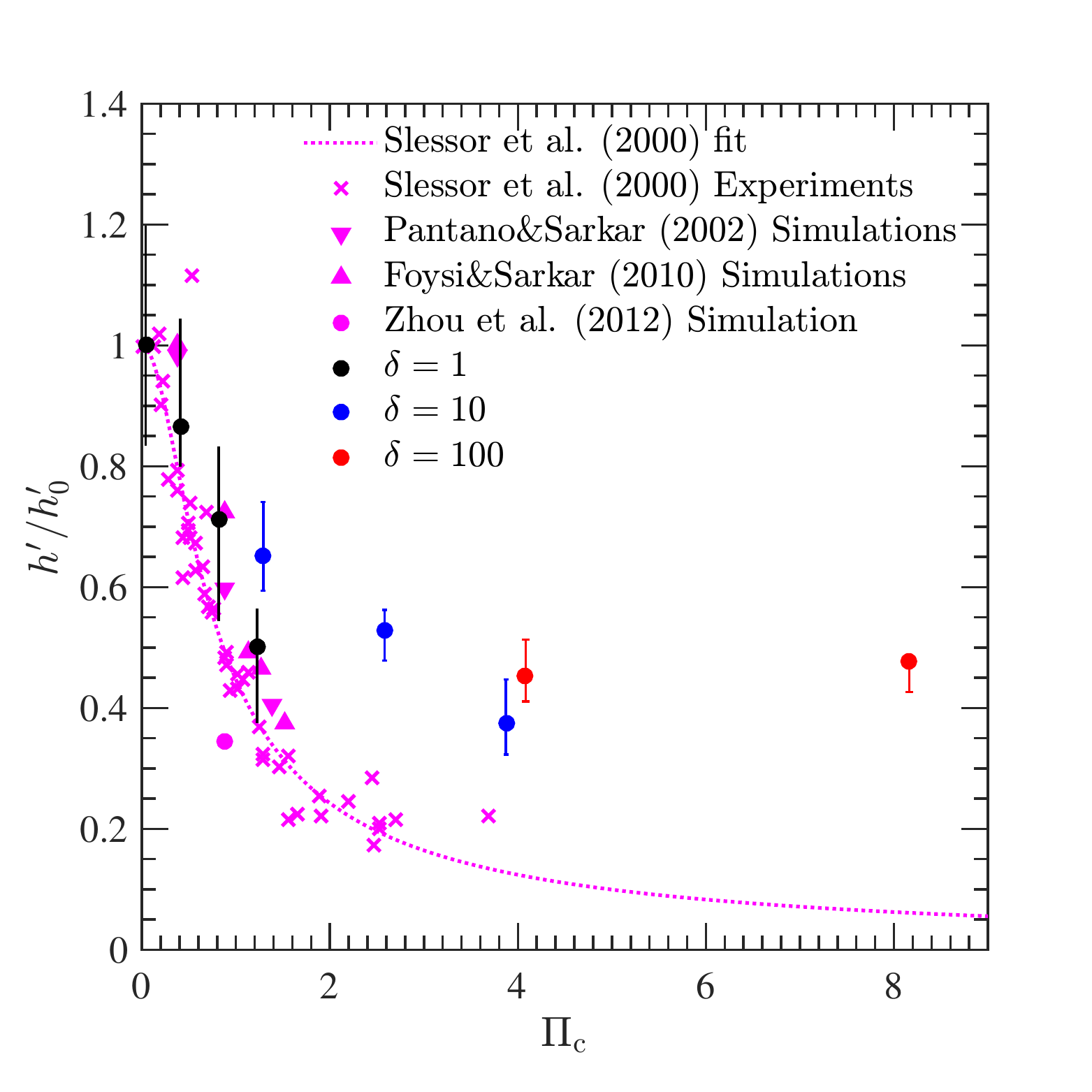}
		}
	\end{tabular}
	\caption{Shear layer growth rate compressibility scaling. The left-hand panel uses $\Mtot$, defined in \Cref{eq:body-instability-condition} as the scaling parameter, while the right-hand panel uses $\Pic$, defined in \Cref{eq:Pic}. The legend on the right-hand panel applies to both panels, except for the dotted Magenta line, which corresponds to the empirical fit $h'/h'_0 = 0.8\exp{(-3\Mtot^2)}+0.2$ in the left-hand panel \citep{Dimotakis1991} and $h'/h'_0=(1+4\Pi_c^2)^{-0.5}$ in the right-hand panel \citep{Slessor2000}. Black, blue and red circles correspond to the results of this work for $\delta=1,10,100$ respectively. These growth rates are obtained by plugging the values of $\alpha(\Mtot)$ from \Cref{fig:alpha} into \Cref{eq:normalized-growth-rate}. The errorbars bound the values consistent with different realizations of initial perturbations. The magenta symbols are experimental and numerical results digitized from \citet{Slessor2000}, \citet{Pantano2002} and \citet{Foysi2010}.}
	\label{fig:compressibility-scaling}
\end{figure*}

Our results of \Cref{sec:surface-results} are plotted against $\Mtot$ and $\Pic$ in \Cref{fig:compressibility-scaling}, along with a compilation of previously published experimental and numerical growth rates. When plotted against either parameter, the normalized growth rates obtained by  \Cref{eq:normalized-growth-rate} are clearly reduced with compressibility. In accordance with \citet{Slessor2000}, our results exhibit less scatter when using $\Pic$ rather than $\Mtot$. However, this is possibly a result of our specific choices of $\unpert$, with rather large strides in $\delta$. More importantly, when plotted against $\Pic$, our results for $\delta\geq10$ are at odds with existing literature. In particular, for $\Pic \gsim 1$, the growth rates reported in this work seem to be much larger than both experimental and numerical studies. 

On the other hand, when plotted against $\Mtot$, the results of this work are in agreement with existing literature throughout the range of $\Mtot$. Although our values typically still lie somewhat higher than most experimental data suggests, they agree very well with previously reported numerical simulations. We therefor adopt $\Mtot$ as the compressibility scaling parameter for \Cref{sec:surface-results}, in line with common practice.

It should be mentioned that $\Pic \gsim 1$ corresponds to cases with $\delta=10$ and $\delta=100$ in this work, while none of the existing experiments or simulations reaches such extreme density contrast, with values typically well under $10$. This is of potential importance because $\Pic$ is expected to differ considerably from $\Mtot$ as $\delta$ is increased. Thus, the discrepancy between our estimated growth rates and the results of previous studies, when presented as a function of $\Pic$, may stem from incorrect $\delta$ dependence introduced by this scaling.

Regardless of the scaling parameter chosen, the agreement between this work and previous publications is far from perfect, although it is greatly improved when comparing only with previous numerical studies. Since different thickness definitions are known to exhibit different scaling with compressibility \citep{Freeman2014}, the cause for the apparent discrepancies between numerical and experimental growth rates may lie in the analysis methods rather than the physics itself. A thorough study of this question is deferred to future work.


\section{Fluid Displacement in Planar Slab Body Modes}
\label{app:displacement}

In \Cref{sec:body-results-critical} we defined the fluid displacement, $\xix(x,z,t)$. As this quantity was not explicitly addressed in \PIt~we define it here using the formalism developed in Sections 2.1-2.3 of \PIt. The basic idea is that velocity perturbations perpendicular to the flow, $u_{\rm x}$, induce spatial displacements of the fluid elements relative to their unperturbed positions, $\xix$. Expanding this in the same Fourier modes used for the other perturbed quantities, we write 
\begin{equation}
\label{eq:Fourier_xi}
\xix(x,z,t) = \xix(x)\exp{\left[i(kz-\omega t)\right]},
\end{equation}
where without loss of generality we have assumed the wave vector to be in the $\hat{z}$ direction, parallel to the bulk flow in the slab. Otherwise, in the analysis below we must change the bulk velocity $V$ to the component of the velocity parallel to the wave vector, $V_k={\vec {V}}\cdot {\hat{k}}$ (see section 2.1 in \PIt). 

To first-order, the displacement $\xix$ is related to the transverse velocity perturbation $\ux$ by 
\begin{equation}
\label{eq:u_to_xi}
\ux = \frac{\partial \xix}{\partial t} + \left(\vec{v}\cdot\vec{\nabla}\right)\xix = i\left(k v_k-\omega\right)\xix.
\end{equation}
Note that \Cref{eq:u_to_xi} is a generalization of Equation 13 from \PIt, which was specific to the displacement of the interface between the fluids, defined therein as $h$. 

Since by definition $\ux$ and $\xix$ have the same dependence on $z$ and $t$ given by the Fourier expansion above, \Cref{eq:u_to_xi} relates the $x$-dependent amplitudes of the two perturbed quantities. Equation 9 of \PIt~relates $\ux$ to the unperturbed density and velocity, $\rhobs$ and $\Vbs$, the Fourier components $k$ and $\omega$, and the $x$-dependent amplitudes of the pressure perturbation $P_1(x)$. Inserting this into \Cref{eq:u_to_xi} yields for the fluid displacement 
\begin{equation}
\label{eq:xi_to_P}
\xix = \frac{1}{\rhobs(k \Vbs - \omega)^2} \frac{\partial P_1}{\partial x}.
\end{equation}
The expression for $P_1(x)$ inside the slab is given by equation 24 in \PIt, repeated here for convenience, 
\begin{equation}
\label{eq:P1_slab}
P_1(x)=
\begin{cases}
A~\dfrac{\sinh(\qs x)}{\sinh(\qs \Rs)}&\quad\text{S-modes}\\
\\
A~\dfrac{\cosh(\qs x)}{\cosh(\qs \Rs)}&\quad\text{P-modes}
\end{cases}\quad,
\end{equation}
where $A$ is an integration constant corresponding to the initial amplitude of the pressure perturbation at the interface between the two fluids at $t=0$, and $q_{\rm s}$ is a modified wavenumber defined in equation 12 of \PIt, 
\begin{equation}
\label{eq:qs}
q_{\rm s} = k\left[1-\left(\frac{\omega-k\Vs}{kc_{\rm s}}\right)^2\right]^{1/2}.
\end{equation}
Note that since in general $\omega$ is complex, $q_{\rm s}$ is complex as well.

Inserting \Cref{eq:P1_slab} into \Cref{eq:xi_to_P} yields the displacement inside the stream
\begin{equation}
\label{eq:xi_slab}
\xix(x)=
\begin{cases}
\dfrac{A\coth(\qs \Rs)}{\rhos(k \Vs - \omega)^2}~\dfrac{\cosh(\qs x)}{\cosh(\qs \Rs)}&\quad\text{S-modes}\\
\\
\dfrac{A\tanh(\qs \Rs)}{\rhos(k \Vs - \omega)^2}~\dfrac{\sinh(\qs x)}{\sinh(\qs \Rs)}&\quad\text{P-modes}
\end{cases}\quad,
\end{equation}
\Cref{eq:xi_slab} can be further simplified by recognizing that the initial amplitude of the fluid displacement at the interface is $H=\xi(\pm\Rs)$. This yields 
\begin{equation}
\label{eq:xi_slab2}
\xix(x)=
\begin{cases}
H~\dfrac{\cosh(\qs x)}{\cosh(\qs \Rs)}&\quad\text{S-modes}\\
\\
H~\dfrac{\sinh(\qs x)}{\sinh(\qs \Rs)}&\quad\text{P-modes}
\end{cases}\quad.
\end{equation}
Since in general $\omega$ and $q_{\rm s}$ are complex, $\xix(x)$ is complex and both its amplitude and its phase depend on $x$. Therefore, the amplitude and the phase of $\xix(x,z,t)$ from \Cref{eq:Fourier_xi} both depend on $x$ as well. Since the amplitude and phase of this function vary with $x$, fluid elements that began at different $x$ coordinates can cross at a finite time. This fluid crossing time is denoted $\tNL$ in \Cref{sec:body-results-critical}, and the magnitude of displacement at that time is denoted $\hNL$.


\section{Transition to Nonlinear Evolution in Body Modes}
\label{app:thNL}

\begin{figure*}
	\begin{tabular}{cc}
		\subfloat{
			\includegraphics[trim={0.25cm 1.1cm 1.0cm 0.6cm}, clip, width=0.4\textwidth]
			{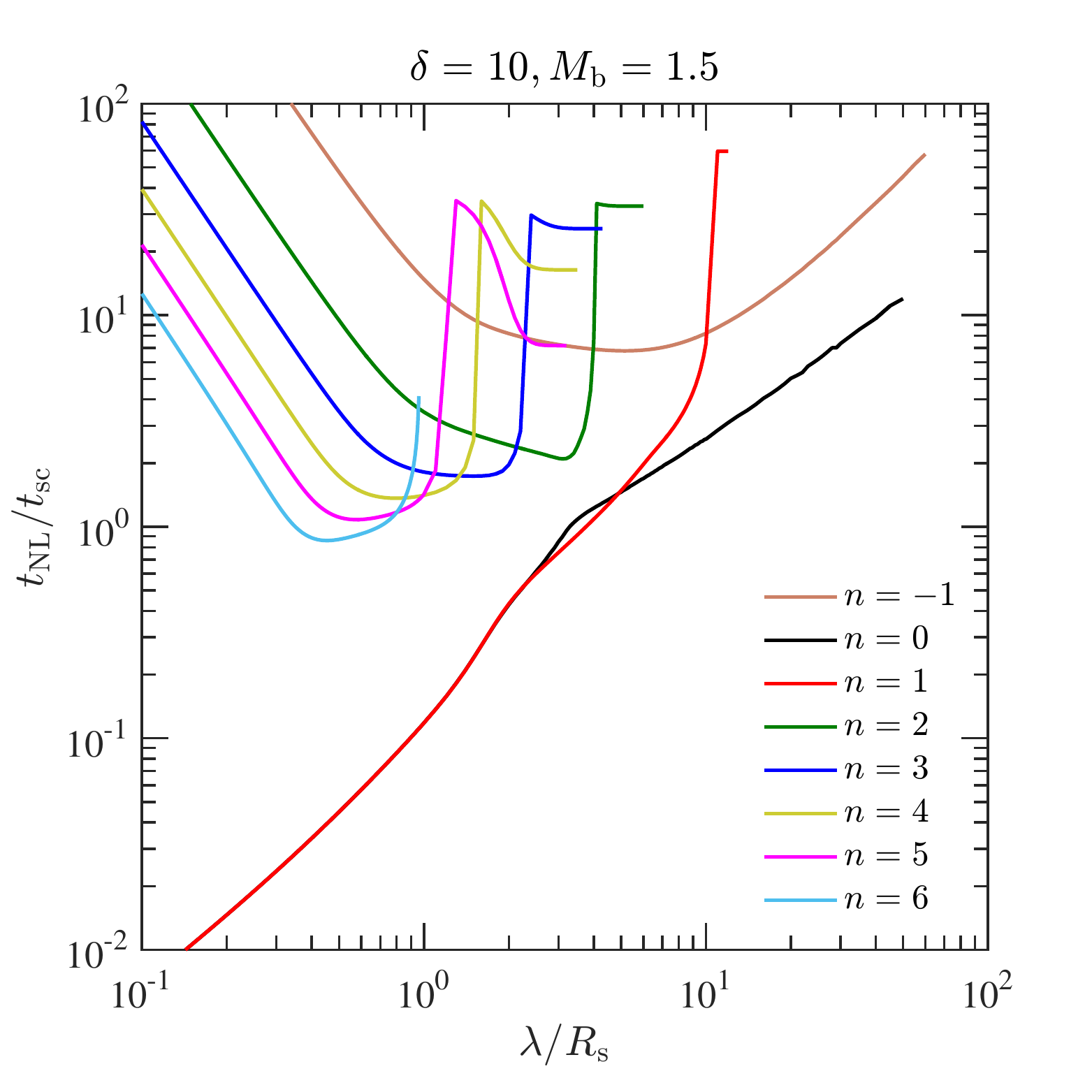}
		}&
		\subfloat{
			\includegraphics[trim={0.25cm 1.1cm 1.0cm 0.6cm}, clip, width=0.4\textwidth]
			{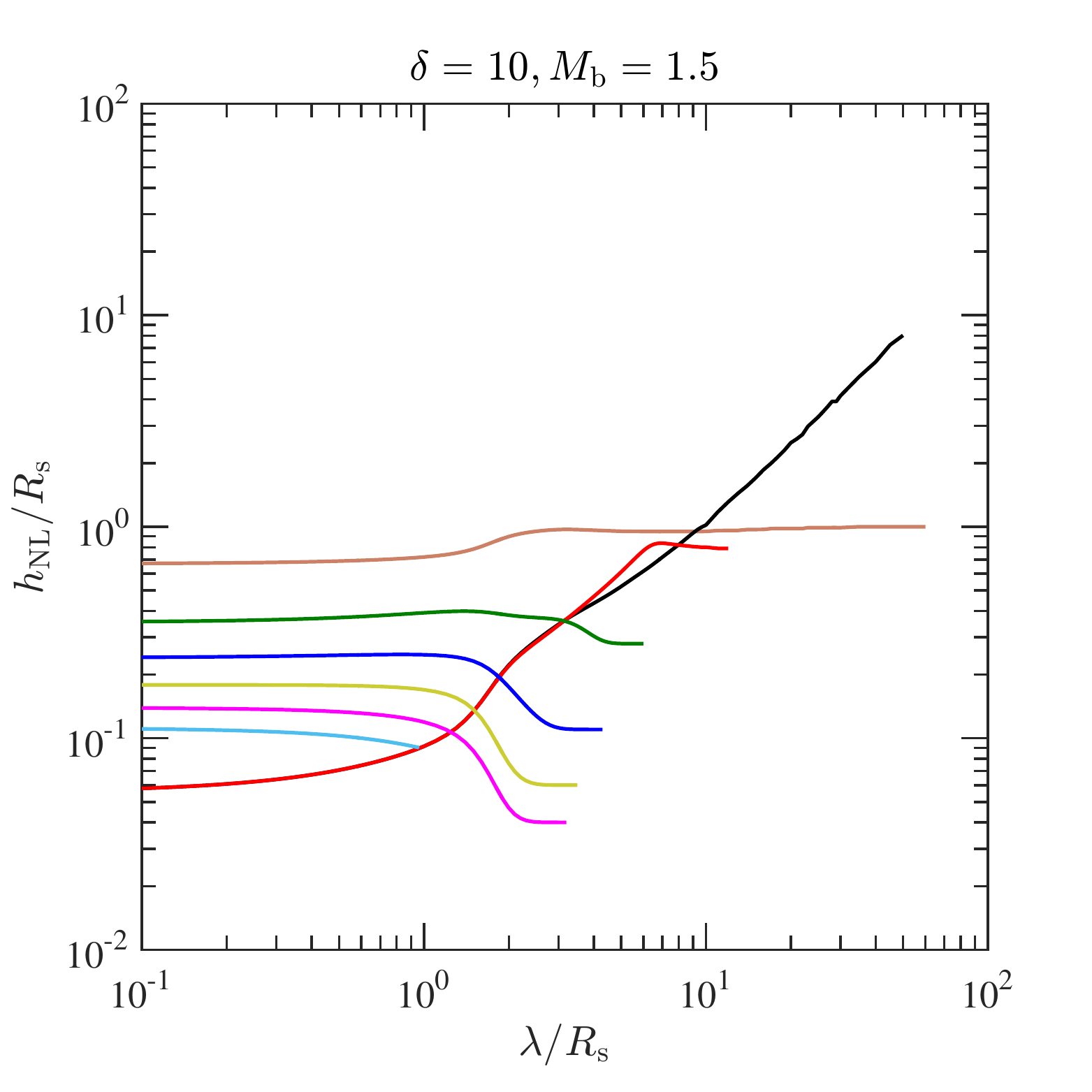}
		}\\
		\subfloat{
			\includegraphics[trim={0.25cm 1.1cm 1.0cm 0.6cm}, clip, width=0.4\textwidth]
			{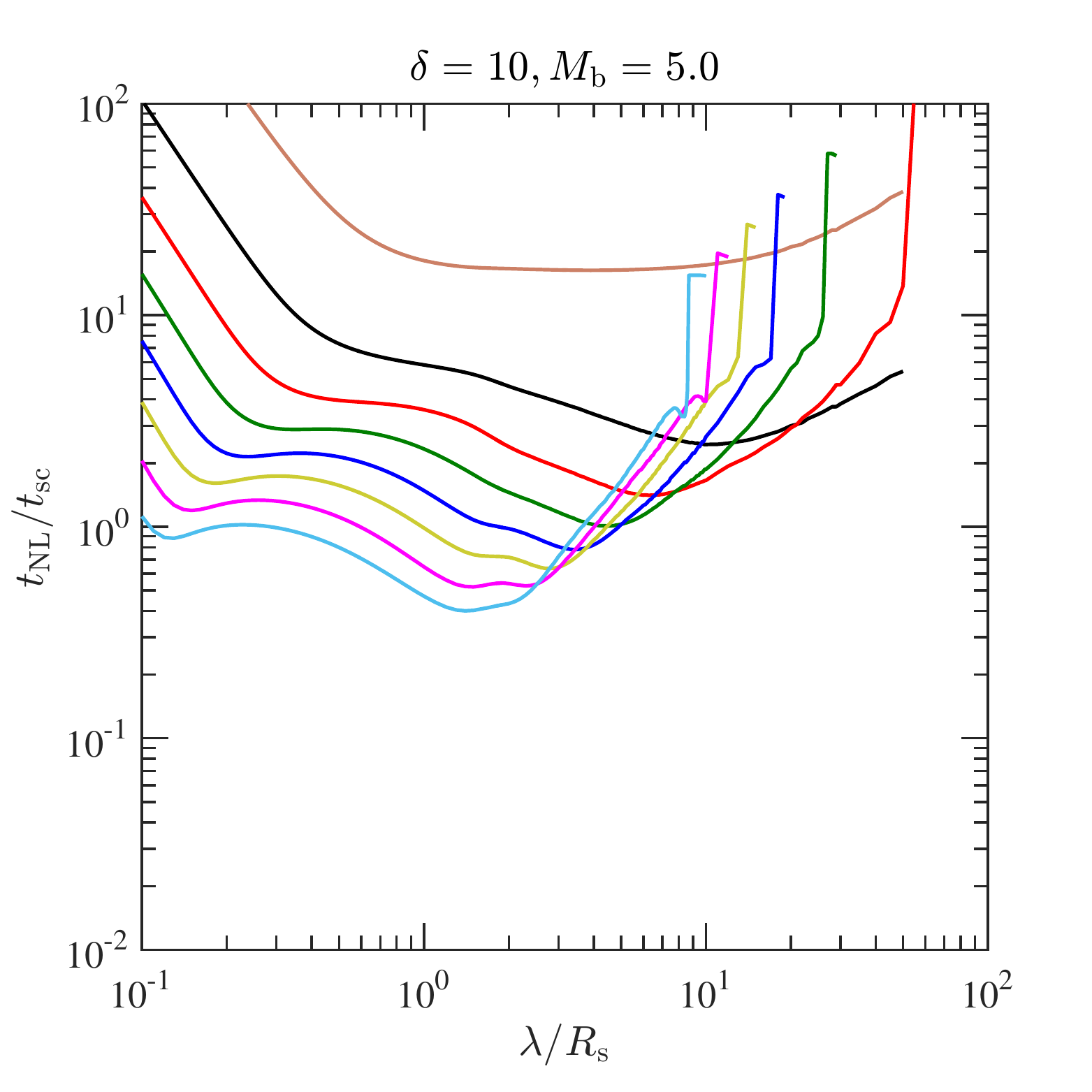}
		}&
		\subfloat{
			\includegraphics[trim={0.25cm 1.1cm 1.0cm 0.6cm}, clip, width=0.4\textwidth]
			{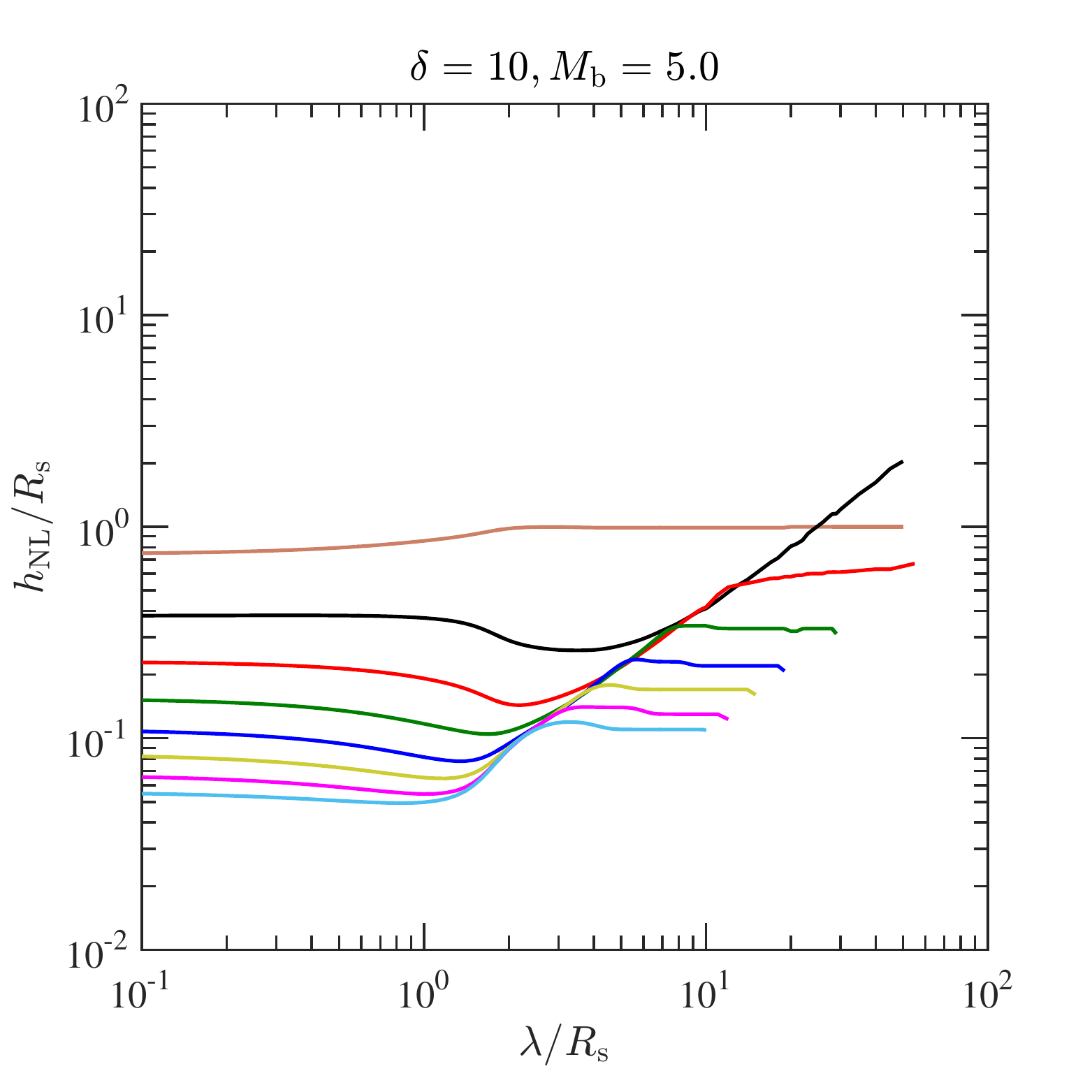}
		}\\
		\subfloat{
			\includegraphics[trim={0.25cm 0.25cm 1.0cm 0.6cm}, clip, width=0.4\textwidth]
			{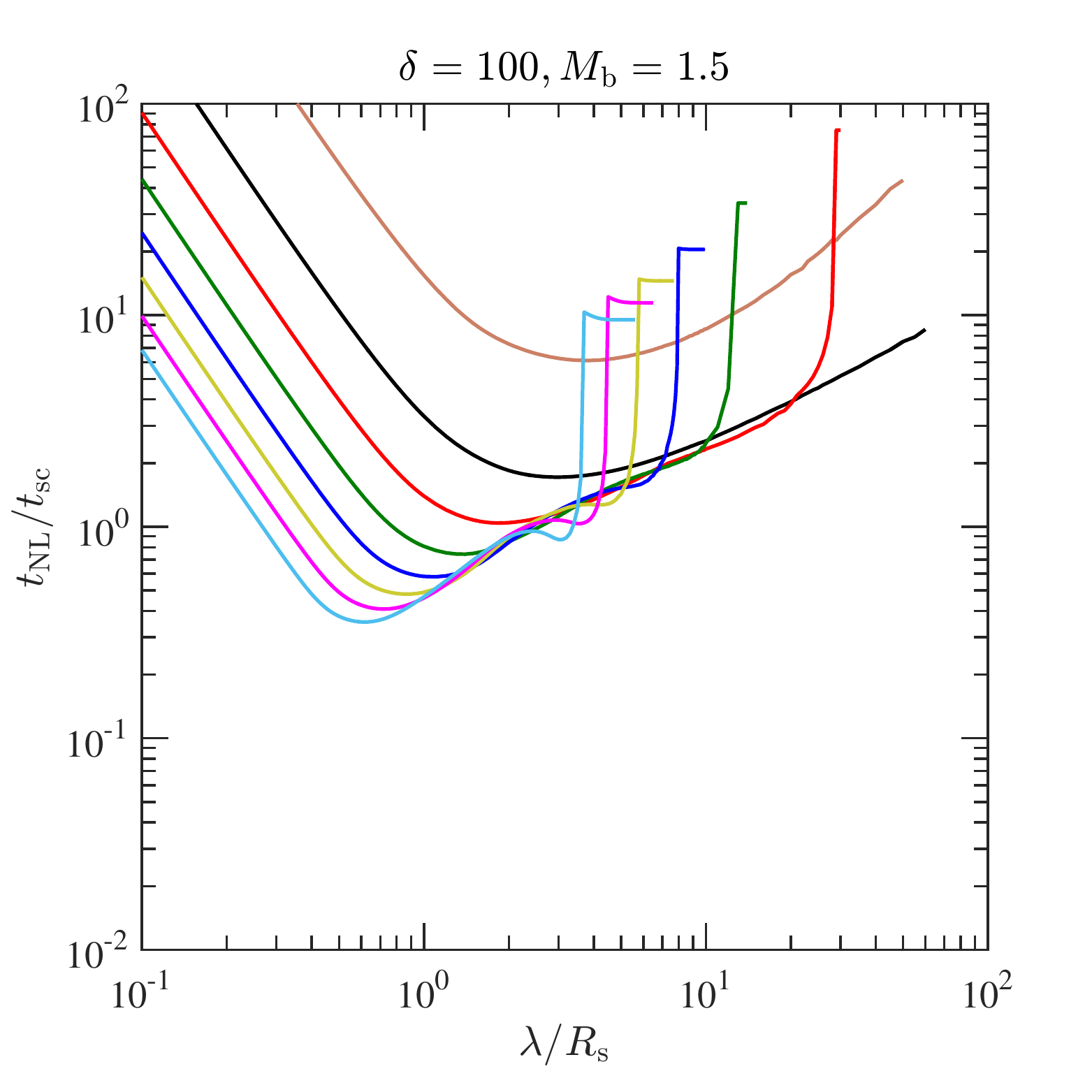}
		}&
		\subfloat{
			\includegraphics[trim={0.25cm 0.25cm 1.0cm 0.6cm}, clip, width=0.4\textwidth]
			{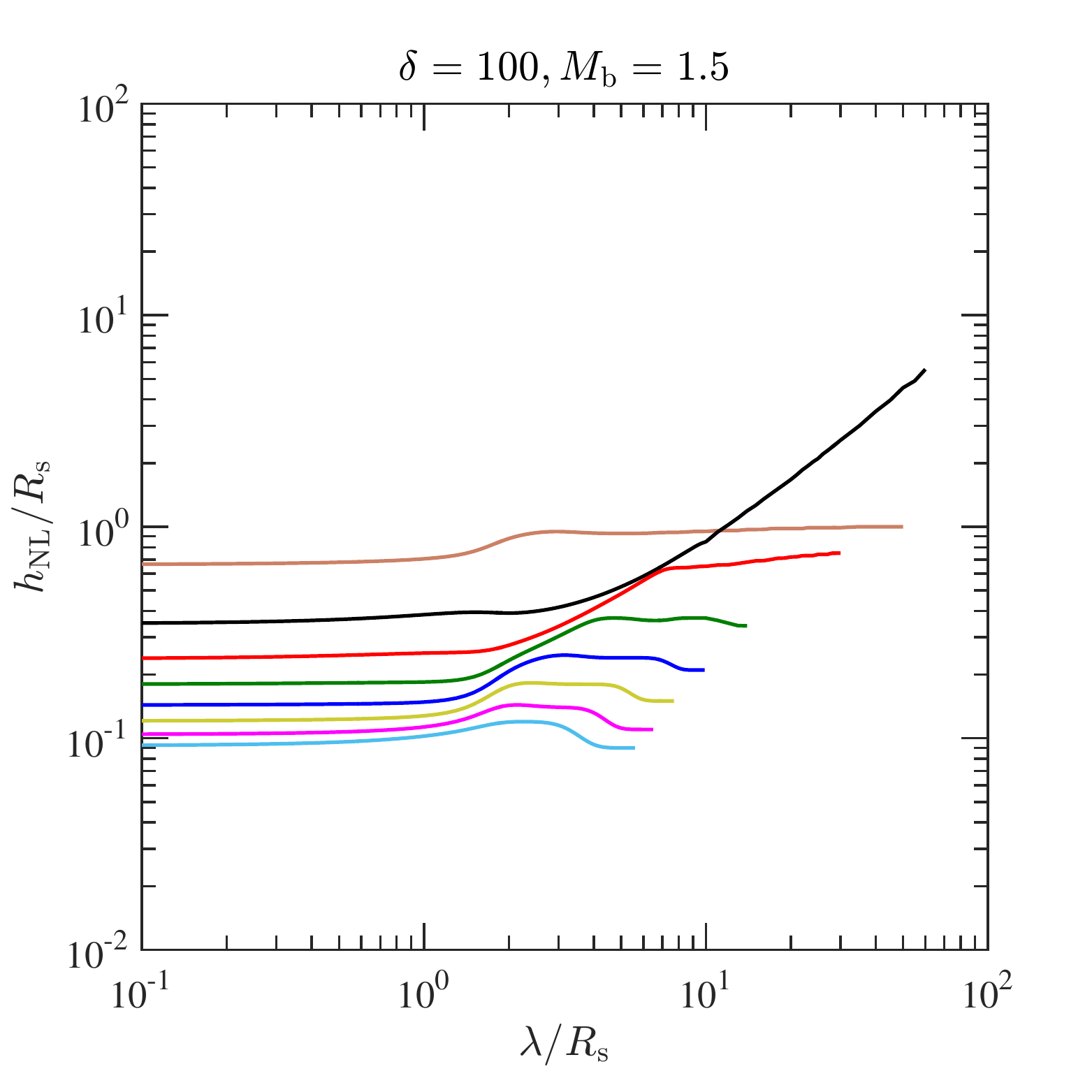}
		}
	\end{tabular}
	\caption{Same as \Cref{fig:thNL} for additional values of $\unpert$. In each row, the left panel shows $\tNL$ and the right panel shows $\hNL$ for some combination of $\unpert$. The legend in the top left panel refers to all panels. The initial perturbation amplitude is assumed to be $H=0.03\Rs$ for all wavelengths.}
	\label{fig:thNL-selected}
\end{figure*}

\Cref{fig:thNL-selected} shows the predicted amplitude and time of transition to nonlinearity for different values of $\unpert$, according to the analysis presented in \Cref{sec:body-results-critical}. The magnitude of displacement to the stream/background interface at the time of transition, $\hNL$, depends on the initial displacement, which is assumed to be $H=0.03\Rs$. Following the arguments laid out in \Cref{sec:body-results-critical}, \Cref{fig:thNL-selected} shows that a long wavelength S-mode is the critical perturbation mode  expected to cause the breakup of the stream $(\lambdacrit\gsim10\Rs,\ncrit=0)$ regardless of $\unpert$.

\section{Linear Growth Rates of Body Modes in Selected Cases}
\label{app:tkh-body}

\begin{figure*}
	\begin{tabular}{cc}
		\subfloat{
			\includegraphics[trim={0cm 0.25cm 0.5cm 0.5cm}, clip, height=0.35\textheight]
			{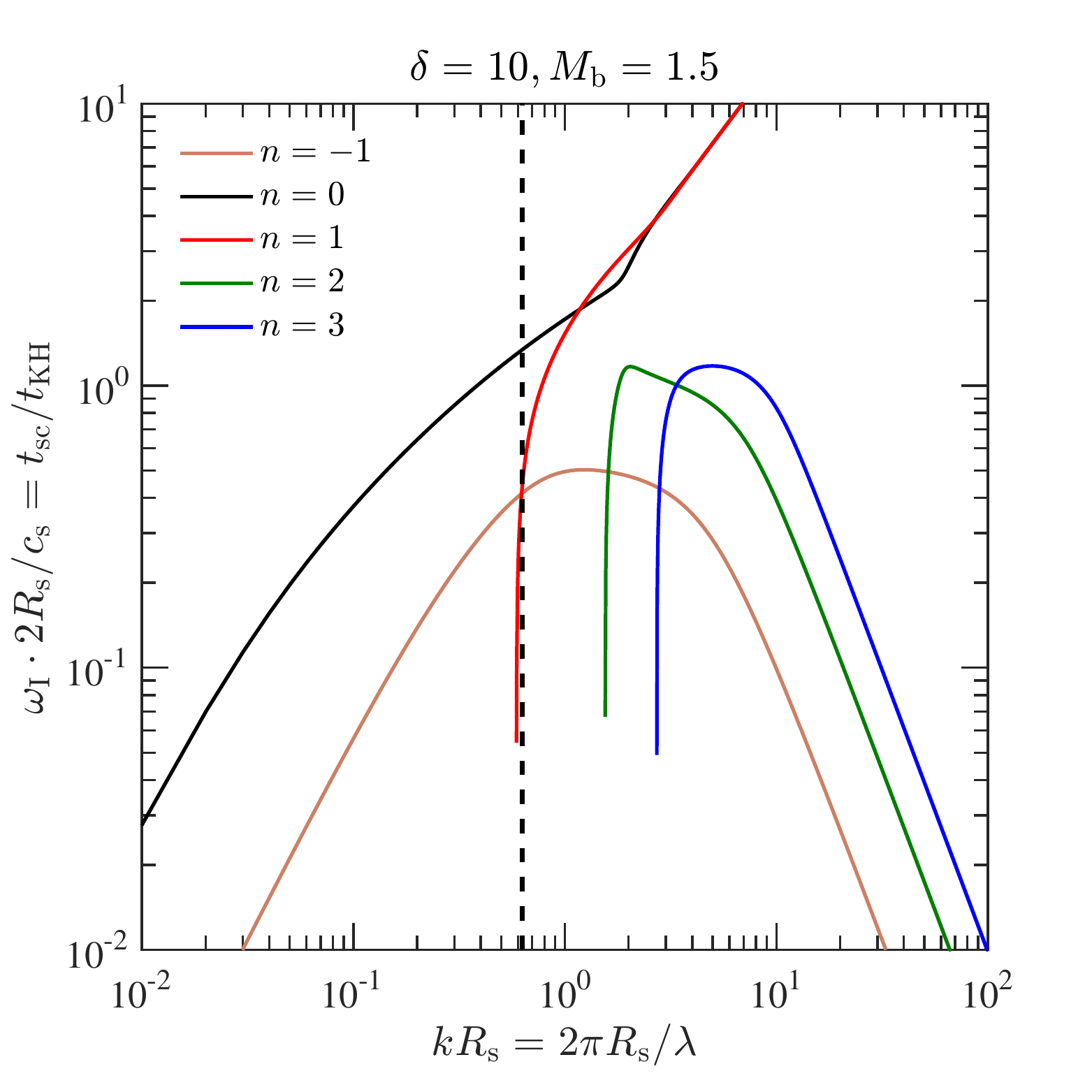}
		}&
		\subfloat{
			\includegraphics[trim={1.0cm 0.25cm 0.5cm 0.5cm}, clip, height=0.35\textheight]
			{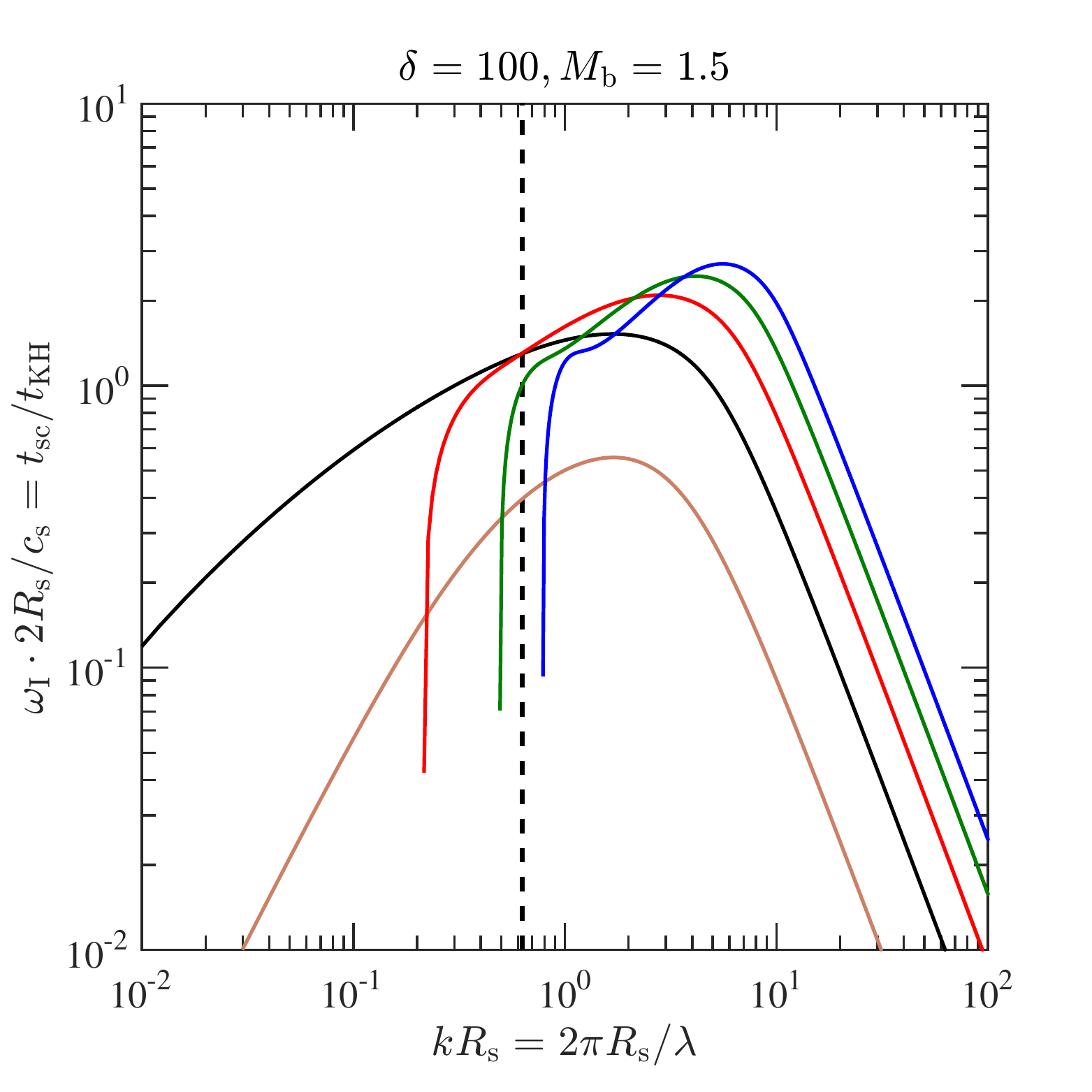}
		}\\
		\subfloat{
			\includegraphics[trim={0cm 0.25cm 0.5cm 0.5cm}, clip, height=0.35\textheight]
			{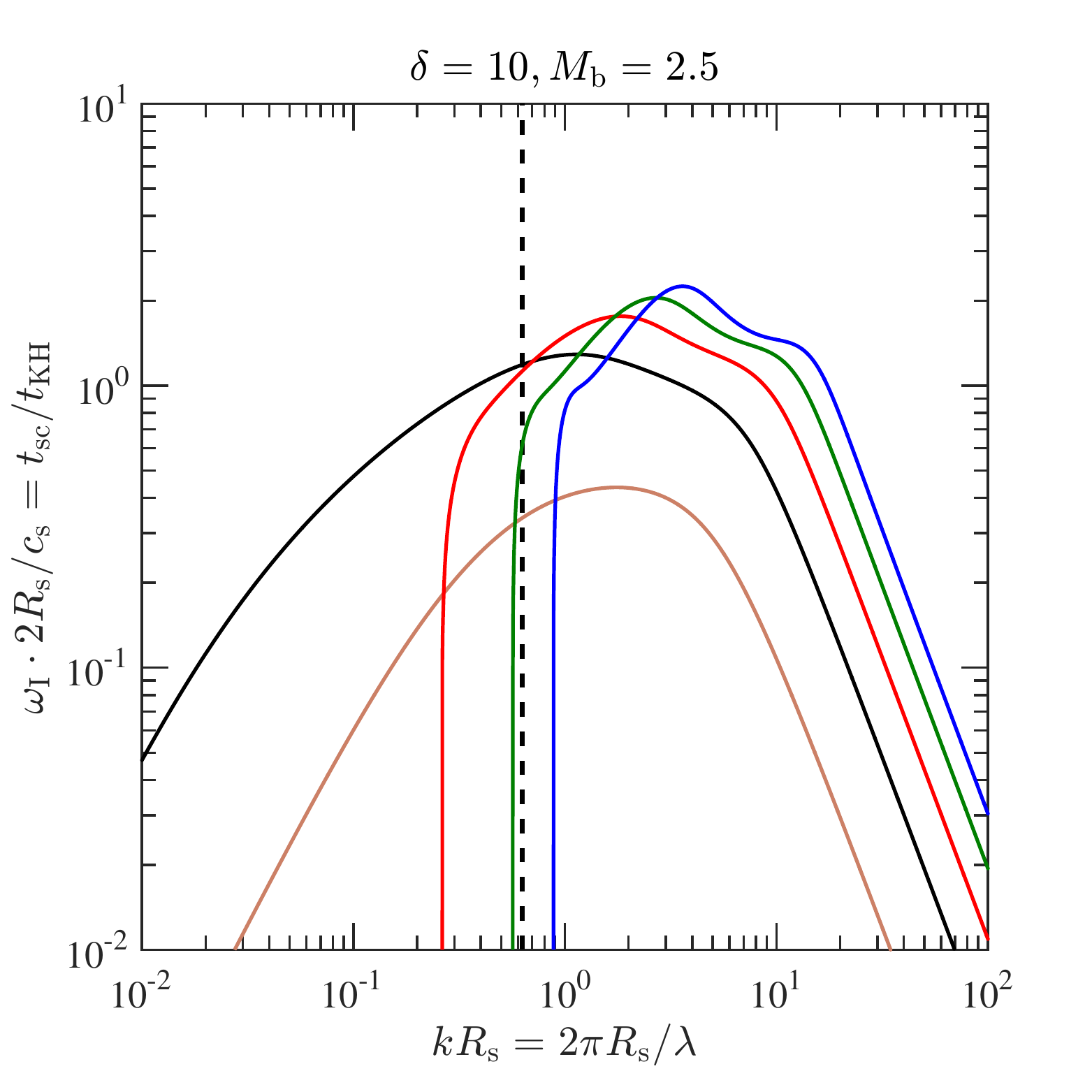}
		}&
		\subfloat{
			\includegraphics[trim={1.0cm 0.25cm 0.5cm 0.5cm}, clip, height=0.35\textheight]
			{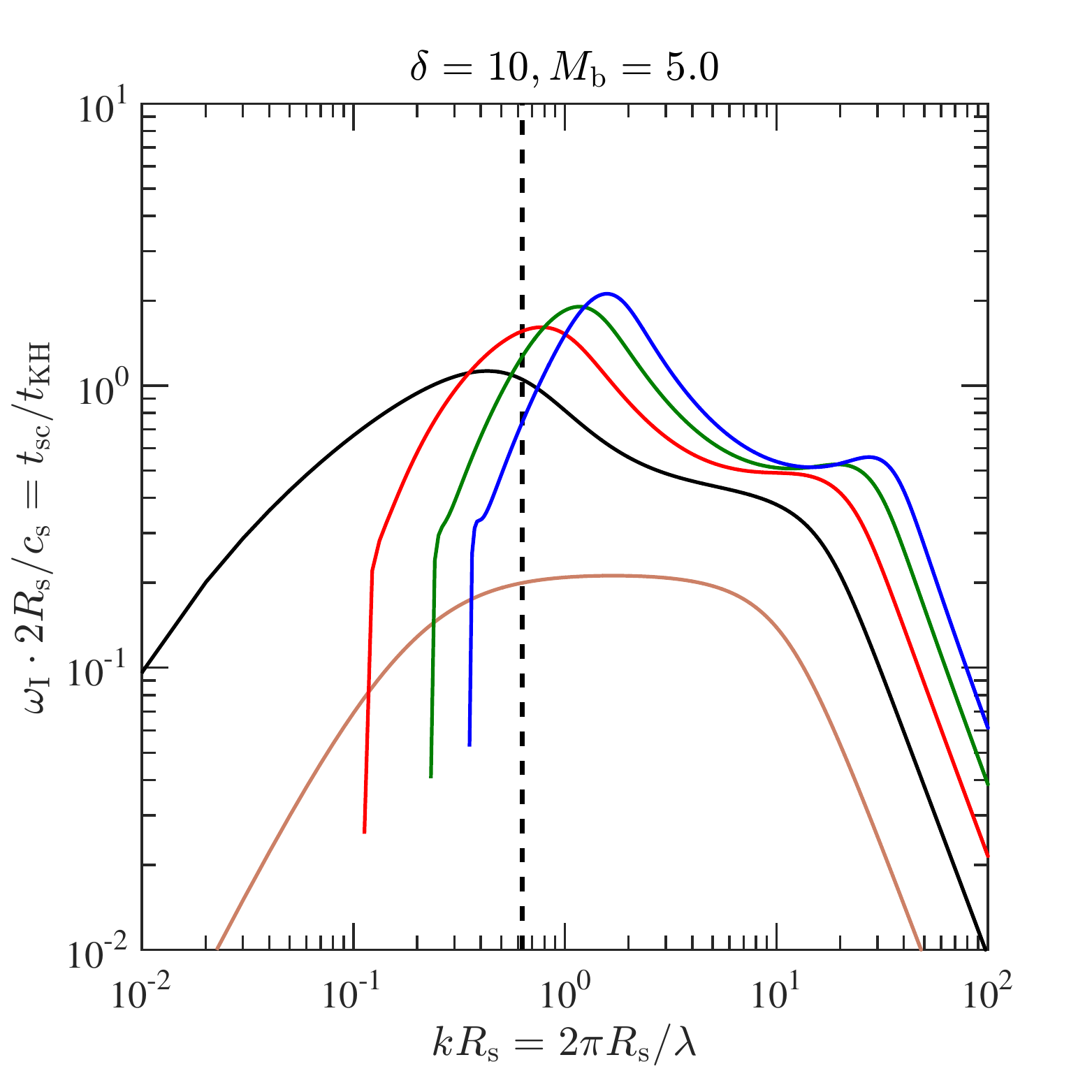}
		}
	\end{tabular}
	\caption{Linear growth rates of the first $5$ body modes in planar slab geometry with various combinations of $\unpert$. The vertical axis shows the growth rates, derived in \PIt, normalized by the inverse of the sound crossing time in the stream. The horizontal axis shows the wavenumber, normalized by the inverse of the stream radius. The solid lines of different colors correspond to different order modes. The legend in the top left panel refers to all panels. The vertical dashed black line marks $\lambda=10\Rs$. The critical perturbation mode (see \Cref{sec:body-results-critical,app:thNL}) typically corresponds to $(\ncrit=0,\lambdacrit\gsim10\Rs)$. }
	\label{fig:tkh-body-selected}
\end{figure*}

\Cref{fig:tkh-body-selected} shows the KH time of different order body modes for various combinations of $\unpert$. The KH time of the fundamental S-mode with $\lambda=10\Rs$, corresponding to the critical perturbation mode responsible for stream breakup (see \Cref{sec:body-results-critical,app:thNL}) is $1\tsc<\tkh<1.2\tsc$ for $1.5<\Mb<5.0$ and $10<\delta<100$.

\section{A Toy Model for the Inflow Velocity of Cold Streams Decelerated due to KHI}
\label{app:inflow-velocity-toy-model}

We present here a simple toy model describing the evolution of the inflow velocity of a cold stream as it penetrates through the halo CGM. The stream fluid is modeled as a rigid body falling radially into the halo under the influence of gravity, on the one hand, and a friction-like force associated with KHI, on the other hand. We limit the discussion to surface mode instability, since it gives rise to a simpler deceleration term than body mode instability (see \Cref{sec:surface-results-deceleration,sec:body-results-deceleration}). 

The gravitational acceleration given by
\begin{equation}
\label{eq:ag}
a_{\rm g} = -\frac{GM(r)}{r^2},
\end{equation}
where $r$ is the halo-centric radius and $M(r)$ is the total mass within a sphere of radius $r$, assumed to be distributed roughly spherisymmetrically. We approximate the mass distribution using the Singular Isothermal Sphere model, 
\begin{equation}
\label{eq:sis}
M(r) = \frac{\Mv}{\Rv}r.
\end{equation} 
Plugging in \Cref{eq:sis,eq:virial-equilibrium} into \Cref{eq:ag} gives 
\begin{equation}
\label{eq:ag-final}
a_{\rm g} = -\frac{\Vv^2}{r}.
\end{equation}

We assume that the deceleration of the stream due to sheer layer growth can be modeled using the term
\begin{equation}
\label{eq:aKH}
a_{\rm KH} = \frac{\alpha}{2(\delta+\sqrt{\delta})} \frac{v^2}{r_{\rm s}},
\end{equation}
which is a generalization of \Cref{eq:vdot-surface,eq:tausurf} where $v=v(t)$ is the time-dependent radial velocity of the stream and $r_{\rm s}=r_{\rm s}(r)$ is the stream radius at a given halo-centric radius. The stream is assumed to be roughly conical in shape, i.e.
\begin{equation}
\label{eq:rs}
r_{\rm s} = \frac{\Rs}{\Rv} r,
\end{equation}
where $\Rs$ is the stream radius at $r=\Rv$ (see \Cref{sec:application-parameters}). This assumption is equivalent to assuming that both the stream and the halo are isothermal with temperatures $\Tbs$.

Combining \Cref{eq:ag-final,eq:aKH} yields a differential equation for the stream velocity,
\begin{equation}
\label{eq:dvdt}
\frac{{\rm d}v}{{\rm d}t} = a_{\rm g} + a_{\rm KH} = 
-\frac{\Vv^2}{r} +\frac{\alpha\Rv}{2(\delta+\sqrt{\delta})\Rs} \frac{v^2}{r},
\end{equation}
which can be simplified to obtain
\begin{equation}
\label{eq:dvdt-simplified}
\frac{{\rm d}y}{{\rm d}x} = -\frac{2}{x} + \frac{By}{x},
\end{equation}
by substituting $y=v^2/\Vv^2$, $x=r/\Rv$ and
\begin{equation}
\label{eq:B-def}
B = \frac{\alpha\Rv}{(\delta+\sqrt{\delta})\Rs}.
\end{equation}
The solution to \Cref{eq:dvdt-simplified} is
\begin{equation}
y(x) = 
\begin{cases}
\dfrac{1}{B}(Cx^B+2) & B \neq 0 \\
\\
1-2\log(x) & B=0
\end{cases}
\end{equation}
where $C$ is a constant of integration and the case $B=0$ corresponds to free fall without friction. Applying the initial condition $v(\Rv)=-\Vv$ or $y(1)=1$ , we find the solution to \Cref{eq:dvdt},
\begin{equation}
\label{eq:v-sol}
v(r) = 
\begin{cases}
-\Vv \sqrt{\dfrac{(B-2)x^B+2}{B}} & B \neq 0 \\
\\
-\Vv \sqrt{1-2\log(x)} & B=0
\end{cases}
\end{equation}

\begin{figure}
	\includegraphics[trim={0.25cm 0.5cm 1cm 1cm}, clip, width=0.475\textwidth]{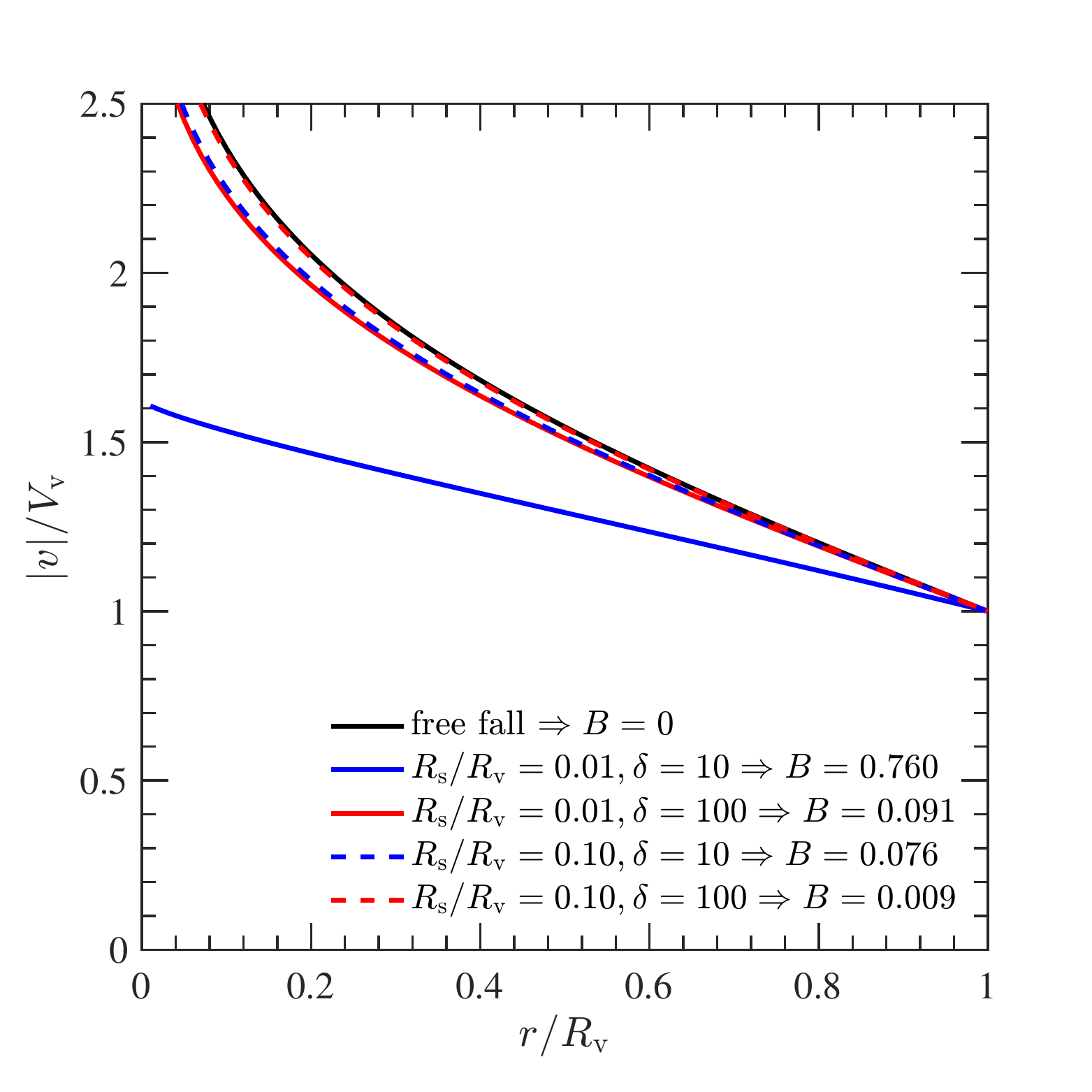}
	\caption{Stream radial velocity during infall into the virial halo according to solutions of a toy model including gravitational acceleration and a friction-like deceleration due to shear layer growth, \Cref{eq:dvdt,eq:v-sol}. The vertical axis shows the magnitude of the radial velocity, $v$, normalized by the virial velocity, $\Vv$. The horizontal axis shows the halo-centric radius, $r$, normalized by the virial radius, $\Rv$. The significance of the friction term is determined by the parameter $B=B(\alpha,\delta,\Rs/\Rv)$, defined in \Cref{eq:B-def}. The figure shows five cases with different $B$: free fall ($B=0$) and four cases with $B>0$ varying by $\delta$ and $\Rs/\Rv$. For simplicity, we use $\alpha=0.1$ throughout, as a reasonable approximation for $\unpert$ relevant to cosmic cold streams (see \Cref{sec:surface-results-density-mach}).}
	\label{fig:velocity-toy-model}
\end{figure}

\Cref{fig:velocity-toy-model} shows solutions to \Cref{eq:v-sol} for different values of $B$, quantifying the significance of the KH deceleration term, \Cref{eq:aKH}. In most cases relevant to cosmic cold streams, this term is negligible compared to the gravitational acceleration, \Cref{eq:ag}, so the resulting $v(r)$ profile barely differs from free fall. Only streams at the most dilute ($\delta=10$) and narrow ($\Rs/\Rv=0.01$) point in the allowed parameter range diverge considerably from free fall. This combination of parameters is particularly unlikely given \Cref{eq:RsRv-fiducial}, which suggests $\Rs/\Rv$ is inversely correlated with $\delta$.

\bsp	
\label{lastpage}
\end{document}